\documentstyle[12pt]{article}
\advance\oddsidemargin  by -1in
\advance\evensidemargin by -1in
\def\lsim{\mathrel{\rlap{\lower4pt\hbox{\hskip1pt$\sim$}}
   \raise1pt\hbox{$<$}}}         
\def\gsim{\mathrel{\rlap{\lower4pt\hbox{\hskip1pt$\sim$}}
   \raise1pt\hbox{$>$}}}         

\def\be{\begin{equation}}
\def\ee{\end{equation}}
\def\bq{\begin{eqnarray}}
\def\eq{\end{eqnarray}}

\mathchardef\alpha="710B
\mathchardef\beta="710C
\mathchardef\gamma="710D
\mathchardef\delta="710E
\mathchardef\epsilon="710F
\mathchardef\zeta="7110
\mathchardef\eta="7111
\mathchardef\theta="7112
\mathchardef\iota="7113
\mathchardef\kappa="7114
\mathchardef\lambda="7115
\mathchardef\mu="7116
\mathchardef\nu="7117
\mathchardef\xi="7118
\mathchardef\pi="7119
\mathchardef\rho="711A
\mathchardef\sigma="711B
\mathchardef\tau="711C
\mathchardef\upsilon="711D
\mathchardef\phi="711E
\mathchardef\chi="711F
\mathchardef\psi="7120
\mathchardef\omega="7121
\mathchardef\varepsilon="7122
\mathchardef\vartheta="7123
\mathchardef\varpi="7124
\mathchardef\varrho="7125
\mathchardef\varsigma="7126
\mathchardef\varphi="7127
\mathchardef\nabla="7272
\font\dozeb=cmmib10 scaled \magstep1
\font\dozesyb=cmbsy10 scaled \magstep1
\font\dezb=cmmib10
\def\pom {\hbox{I$\!$P}}

\def\si{\hbox{$\sum$$\!$$\int$}}

\begin{document}
\pagestyle{empty}
\vspace{2.0cm}

\begin{center}

\rightline{DFTT 57/98}

\vspace{1.0cm}

{\large \bf DIFFRACTION: PAST, PRESENT AND FUTURE\footnote{Lectures given at {\it Hadrons VI},
Florianopolis, Brazil, March 1998.}\\}

\vspace{1.0cm}

{\large  E.~Predazzi\\}

\vspace{1.0cm} 

{\it Dipartimento di Fisica Teorica, Universit\`a di Torino\\
and INFN, Sezione di Torino, 10125 Torino, Italy \medskip\\ }

\vspace{1.0cm}

\end{center}

\noindent{{\large \bf Abstract .- } Hadronic diffraction has 
become a hot and fashionable subject in recent years due to 
the great interest triggered by the HERA and Tevatron data. 
These data have helped to put the field in a different 
perspective paving the road to a hopefully more complete 
understanding than hitherto achieved. The forthcoming data
in the next few years from even higher energies (LHC)
promise to sustain this interest for a long time. It is, therefore, 
necessary to provide the younger generations with as complete 
as possible discussion of the main developments that have 
marked the growth of high energy diffractive physics in the 
past and to assess the present state of the art. For this reason, 
this part will be by far the largest. The analysis of the
relationship between conventional diffractive physics and the 
low-x physics from deep inelastic scattering will allow us also
to review the instruments which could help to understand the
developments we can expect from the future.}

\vfill\eject


\baselineskip 16 pt
\pagestyle{plain}
\bigskip

\bigskip

The plan of these lectures is the following:
\bigskip

{\large \bf Part I.- Conventional hadronic diffraction.} 

\begin {itemize}

\item {\bf I.1 What do we mean by hadronic diffraction?}
\item {\bf I.2 Some history.}
\item {\bf I.3 A summary of conventional hadronic diffraction.}
\end {itemize}
\medskip

{\large \bf Part II.- Elements of Deep Inelastic Scattering (DIS).}

\begin {itemize}

\item {\bf II.1 Basic kinematics.}
\item {\bf II.2 Basic properties of DIS.}
\end {itemize}
\medskip

{\large \bf Part III.- Modern hadronic diffraction.}

\begin {itemize}

\item {\bf III.1 Diffraction with hadrons.}
\item {\bf III.2 Diffraction at HERA.}
\item {\bf III.3 Concluding remarks and perspectives.}

\end {itemize}
\medskip

\bigskip

\vfill \eject

\centerline {\large \bf PART I.}
\medskip

\centerline {\large \bf  CONVENTIONAL HADRONIC DIFFRACTION} 
\bigskip

\noindent {\large \bf I.1 What do we mean by hadronic diffraction?}

\bigskip

The term, {\it diffraction} was introduced in nuclear high 
energy physics in the Fifties. Apparently, the very first to use
it were Landau and his school\footnote {Besides Landau,
the names associated with these early developments are those of 
Pomeranchuk, Feinberg, Akhiezer, Sitenko and Gribov etc. [1].}.
In this context, the term is used in strict analogy with what
was done for nearly two centuries in optics to describe the 
coherent phenomenon that occurs when a beam of light meets 
an obstacle or crosses a hole whose dimensions are comparable 
to its wavelength (so long as the wavelengths are much smaller 
than these dimensions, we have geometrical shadow).
To the extent that the propagation and the interaction of
extended objects like the hadrons are nothing but the 
absorption of their wave function caused by the many 
inelastic channels open at high energy, the use of the term
{\it diffraction} seems indeed appropriate.

Historically, the terminology comes, as we said, from optics
which, as a field, relies on approximations. Let us imagine a 
plane wave of wavelength $\lambda$ which hits (perpendicularly,
for simplicity) a screen with a hole of dimensions $R$ and 
let us suppose that the wave number $k= {{2 \pi} \over {\lambda}}$  
is sufficiently large that the {\it short wavelength condition}
$$kR >> 1\eqno(I.1.1)$$
is satisfied. 

If $\Sigma_0$ describes the hole on the screen, according to the
Huygens-Fresnel principle, each point becomes the center of
a spherical wave whose envelope will give the deflected wave. 
Let $\Sigma$ be the plane at a distance $D$ where we 
imagine to collect the image ({\it i.e.} the detector plane). 
Because of the varying distances to the point and varying
angles with respect to the original direction of the beam, the 
amplitudes and phases of the wavelets collected at each 
point will be different. As a consequence, 
cancellations and reinforcements can occur at different
points giving rise to the phenomenon of diffraction.
This propagation maps the value of this energy distribution
$T_0$ on $\Sigma_0)$ into its value $T$ at the
point $P(x, y, z)$ on the detector's plane. Mathematically,
this is given by the Fresnel-Kirchoff formula [2]
$$T(x,y,z) = {{-i}\over {2 \lambda}} {{e^{ik_0 r_0}}\over {r_0}} 
\int_{\Sigma} dS T_0 [1+ cos \theta] {{\exp{i \vec k \cdot \vec b}} \over 
{s}}\eqno(I.1.2)$$
where $\vec s$ is the distance of the point P from $\Sigma_0$ 
and $cos \theta$ is the inclination of this vector with respect to the 
normal to $\Sigma_0$. 

The problem is greatly simplified when the detector is
so distant that all rays from $\Sigma_0$ to the point 
$P(x,y,z)$ on $\Sigma$ can be considered parallel.
One talks of {\it Fraunhofer} or {\it Fresnel diffraction} 
according to whether the source is at a distance which can
or cannot be considered infinitely large. For the case at hand,
the large distance approximation will always be valid.

If the distance $D$ satisfies the {\it large distance condition} 
$$R/D << 1,\eqno(I.1.3)$$
we may expand the exponential ${{e^{i k s}} \over {s}}$ 
in power series of $ks$. The following various cases can occur:

\begin {itemize}
\item {i}) {\it Fraunhofer diffraction} when
$$kR^2/D<<1;\eqno(I.1.4a)$$
\item {ii}) {\it Fresnel diffraction} when
$$kR^2/D \approx 1;\eqno(I.1.4b)$$
\item {iii}) {\it geometrical optics} when
$$kR^2/D>>1.\eqno(I.1.4c)$$
\end {itemize}

The consequence of all this is that the parameter $kR^2/D$
is the one that dictates the optical regime.

In the Fraunhofer limit $(I.1.4a)$ (which we shall be consistently
assume), and turning to a terminology closer to that of particle
physics by introducing the {\it impact parameter} $\vec b$,
we shall rewrite eq. $(I.1.2)$ as
$$T(x,y,z) \approx {{k}\over {2 \pi i}} {{e^{i kr_0}} \over {r_0}}
\int_{\Sigma} d^2b S(\vec b) e^{i {\vec q}\cdot {\vec 
b}}\eqno(I.1.5)$$
where $\vec q$ is the two-dimensional momentum transfer
$$|\vec q| = k sin \theta\eqno(I.1.6)$$
and the Scattering matrix $S$ is expressed as
$$S(\vec b) \equiv 1- \Gamma(\vec b) \eqno (I.1.7)$$
in terms of the {\it profile function} of the target 
$\Gamma (\vec b)$. Equivalently,
inserting $(I.1.7)$ into $(I.1.5)$
one obtains the complete amplitude of which the term
that contains $1$ represents the unperturbed wave and the 
one that contains $\Gamma(\vec b)$ is the diffracted wave.

The factor multiplying the {\it outgoing spherical wave} is 
the physically relevant quantity {\it i.e.} the {\it scattering
amplitude} which we will write in the form
$$f(\vec q) = {{i k}\over {2 \pi }} \int {d^2b \,
\Gamma(\vec b) \, e^{i {\vec q}\cdot {\vec b}}},\eqno(I.1.8)$$
so that the scattering amplitude is given by the Fourier
Transform of the profile function and, viceversa,
we could also write
$$\Gamma(\vec b) = {{1}\over {2 \pi i k}} \int
{ d^2q f(\vec q) 
e^{- i {\vec q}\cdot {\vec b}}}.\eqno(I.1.9)$$

If, now, the profile function $\Gamma(\vec b)$ is spherically
symmetric, eq.$(I.1.8)$ can be written as the Bessel Transform
$$f(\vec q) = i k \int_0^{\infty} b db \Gamma(b)  
J_0(qb).\eqno(I.1.10)$$

Finally, if the profile function is a disk of radius $R$,
we obtain the so-called {\it black disk} form
$$f(\vec q) = i k R^2 {{J_1(qR)}\over {qR}}.\eqno(I.1.11)$$

The form $(I.1.10)$ can be integrated explicitly for a number
of cases but we shall not insist on these developments here.

Optics and hadronic physics may, at first sight, appear very 
distant fields. In the latter, the collision of very high energy 
particles produces an almost arbitrary number of large
varieties of different particles and this makes the situation 
considerably more confused than in optics where Huyghens 
principle dictates the solution to every problem. In practice, 
however, the two fields are very much alike; in both cases 
we have high wave numbers in the game and in both 
cases it is the ondulatory character of the phenomenon which 
is responsible for what we observe. And, what we observe are, 
precisely, series of diffractive maxima and minima which have
become familiar in nuclear physics first and then in hadronic
physics (see. 
Fig. 1). Later, (see Section I.3.8.10), we will see 
that the analogy is further strenghtened by the fact that the 
elastic scattering of two particles at high energy is indeed
dominated by diffraction near the forward direction in the
sense that it results from the {\it shadow of all inelastic 
channels open at these energies}. The same kind of 
representation $(I.1.2)$ will, therefore, be applicable to 
describe the physical situation also in hadronic high energy 
physics even though this representation is usually derived 
from quantum mechanical tools.

Thus, diffraction covers a large span of phenomena, 
from optics to nuclear and from nuclear to hadron physics
so long as the proper conditions $(I.1.1,4 )$ are obeyed.

\medskip
Fig. 1.  {\it Angular variation for diffraction in nuclear 
and in hadronic physics; i) elastic scattering of $1.7 \,
Gev$ protons on $^{40}Ca$ and $^{48}Ca$, ii) elastic 
scattering of $1.75 \, Gev$ protons on $^{208}Pb$ and 
iii) elastic $pp$ scattering at the ISR.}
\medskip

Analysing the optical limit and the diffraction of very 
high energy electromagnetic waves simulated by the
collision of perfectly conducting spheres, T. T. Wu [3] 
has come to the extraordinary conclusion that diffraction-like 
ideas applied to the realm of Maxwell equations {\it "...describe 
electromagnetic waves correctly over at least 18 orders of 
magnitude from the Edison-Hertz to the HERA wavelengths".}

The next task will be to {\it define} diffraction in purely 
particle physics terms. The first authors to give a 
definition of {\it hadronic diffraction} in perfectly 
modern terms were Good and Walker [4] who, in
1960, wrote: {\it  "...A phenomenon is predicted 
in which a high energy particle beam undergoing 
diffraction scattering from a nucleus will acquire components 
corresponding to various products of the virtual dissociations 
of the incident particle... These diffraction-produced systems 
would have a characteristic extremely narrow distribution in 
transverse momentum and would have the same quantum 
numbers of the initial particle..."}.

For the sake of definiteness, we will  say that {\it  every 
reaction in which no quantum numbers are exchanged  
between high energy colliding particles is 
dominated asymptotically by diffraction}. Turning things 
around, this implies that {\it diffraction dominates as
the energy increases whenever 
the particles (or ensembles of particles) diffused have the 
same quantum numbers of the incident particles}. 

It will be noticed that we have taken an indirect path to arrive
to a propositive definition of diffraction\footnote {We are proposing 
a definition of diffraction as if it were {\it sufficient} to demand 
no exchange of quantum numbers (other than those of the 
vacuum) for the process to be diffractive. In a way, things are 
the other way around: it is {\it necessary} that no exchange of 
quantum numbers (other than the vacuum) takes place for 
diffraction to be active in the process.}. The point is that it is 
essentially impossible to give a definition of diffraction entirely 
free of ambiguities. Our prescription has one disadvantage:
it does not allow us to recognize and eliminate possible  
contaminations of nondiffractive origin (like exchange of scalar
particles) ; the latter, however, asymptotically becomes weaker 
and weaker (as the c.m. energy increases, the ratio of the non 
diffractive and of the diffractive component vanishes). The 
advantage, however, is that it provides a simple and operational 
prescription to recognize if a reaction is dominated by diffraction 
at high energies. 

 Another advantage of our definition is that it covers all cases: 
 {\it i) elastic scattering} when exactly the same incident particles 
 come out after the collision, {\it ii) single diffraction} when {\it one} 
 of the incident particles comes out unscathed after the collision 
 while the other gives rise to a resonance (or to a bunch of final 
 particles) whose resulting quantum numbers are exactly those of 
 the other incident particle and, finally {\it iii) double diffraction} 
 when {\it each} incident particle gives rise to a resonance (or to a 
 bunch of final particles) with exactly the same quantum numbers
 of the two initial ones.
 
 According to our definition, the most general 
 diffractive reaction can then be represented as in Fig. 2
 $$ a \, + b \,  \rightarrow  a^* \, + b^* ,\eqno(I.1.12)$$
 where $a^*$ and $b^*$ have exactly the same quantum numbers 
 of the two incident hadrons
 $a$ and $b$ respectively (they may be any ensemble of 
 particles produced in the final state or may coincide with the 
 initial particles themselves). Owing to our definition, this 
 reaction is indeed diffractive since it corresponds to no 
 exchange of quantum numbers between the initial and 
 the final particles. As an example, Fig. 2 depicts the case of 
 {\it single diffraction}. 
 
\medskip
 Fig. 2. {\it Single diffraction $a+b \rightarrow a+b^*$.}
\medskip
 
One usually refers to a diffractive process of the kind 
$(I.1.12)$ by saying that it is do\-mi\-nated by the exchange 
 of a {\it Pomeron}\footnote {In honour of the Russian 
 physicist I.Y. Pomeranchuk, one of the founding fathers of 
hadronic diffractive physics.}. Many consider this term 
illdefined or misleading if not directly meaningless. In
our opinion it is, probably, not yet entirely understood
and the same entity 
may take different connotations in different contexts.
Possibly, it simply conceals a more profound mechanism
as we will try to illustrate later on (see Section I.3.8.10).
With our definitions, {\it Pomeron exchange} is
synonimous of {\it exchange of no quantum numbers}.

 As already mentioned, our definition of {\it diffraction} is 
 oversimplifying the matter a little; indeed, from this
 point of view, it does not allow us to distinguish true 
 diffraction from the exchange of scalar systems which 
 {\it a priori} are non-diffractive. Exchange of scalars, 
 however, becomes less and less important as the energy
increases.
 \bigskip

 In the rest of Chapter I we will first {\it i)} give a historical
 perspective of the developments of hadronic physics, next, {\it ii)}
we will give a very brief resum\'e of elastic and diffractive 
hadronic high energy data and, finally, {\it iii)}
 we will recall the main ideas and instruments which have 
been used to account for diffractive phenomena in high
 energy particle physics.
 \bigskip
 
\noindent {\large \bf I.2 - Some history.}
\bigskip

 \noindent {\bf I.2.1 Introduction.}
 \bigskip
 
>From the purely empirical or observational point of view, the
 distinctive features of diffractive hadronic reactions have been 
 recognized to be the following from the very earliest days:
 \begin {itemize}
 \item {i)} very steep angular (or momentum transfer) 
 distributions (as predicted in [4]). Such a feature is, qualitatively, 
 the simplest consequence of diffraction already in optics and we 
will see it is predicted by unitarity at high energy (hinting,
qualitatively, at a strict but somewhat mysterious relationship 
between diffraction and unitarity); 
 \item {ii)} total and integrated cross sections increase slowly with 
 energy\footnote {As a matter of fact it was only after the data 
 from the Serpukhov accelerator at $p_{lab} \approx 50-60 \, GeV$
 became available that total cross sections were recognized to 
 increase with energy; until then, the prevalent belief was that 
 they would, eventually, approach a constant at asymptotically 
 high energies (this was, wrongly, considered to be a prediction 
 of one of Pomeranchuk theorems, see Section I.3.7.2).};
 \item {iii)} the slopes of angular distributions shrink, {\it i.e.}, 
 they increase (slowly) with energy.
 \end {itemize}
 
Initially, the interest was focused on elastic reactions or, 
 in any case, on {\it exclusive} reactions {\it i.e.} on reactions 
 in which the kinematics of all particles in the final state is 
 fully reconstructed (this, typically, implies that one or at most
 two extra particles are produced in the final state and their
 kinematics is reconstructed).
 
 \bigskip
 \noindent {\bf I.2.2 The Sixties.}
 \bigskip
 
 We can, probably, take the early Sixties as the beginning of 
 diffraction in hadronic physics with the advent of the first high 
 energy accelerators. In these years, diffraction was a blend of 
 various fairly heterogeneous items. First, a number of rigorous 
 theorems like the optical theorem, the Pomeranchuk theorems, 
the Froissart-Martin theorem plus a variety of highly 
sophisticated theorems derived by several authors (among them,
we mention A. Martin, N. N. Khuri, H. Cornille, J. D. Bessis, T. Kinoshita 
and many others). We will discuss some of these theorems 
later on (Section I.3.7), but it is impossible to give 
the proper credit to all those who have contributed to this 
field. The interested reader is referred to few basic textbooks [5, 
especially 5c and 5d, 6, 7] where details and original references 
can be found. Next, general properties where used such as 
analyticity, crossing etc., enforced either exactly or within 
more or less simple approximations. Third, clever 
 representations of the scattering amplitude were devised to 
exhibit in an explicit and convenient form either some general 
properties of the theory or some special features of the data 
(such as the eikonal and the Watson Sommerfeld 
representations). In addition, various kinds of 
approximations and of intuitive or empirical properties were 
proposed, especially tailored to describe specific (or general) 
aspects of the problem such as duality (which 
inspired Veneziano to propose a celebrated formula 
[8] that was the beginning of a totally new laboratory
of theoretical physics), geometrical scaling etc.
Finally, the powerful techniques of dispersion relations 
 were employed often together with the most naive (in 
retrospective) intuitions to provide a framework in 
which to describe the data.
 
 Probably, the most longlasting development of this decade is the 
 introduction of {\it complex angular momenta} rediscovered 
by Regge [9] and applied to particle physics with the consequent 
use of the so-called Watson-Sommerfeld transform to represent 
the scattering amplitude [10]. This formulation seems the most 
useful to describe high energy data and we shall come back to 
this point given the present renewal of interest in such matter. It 
is in this context that, among other things, the {\it Pomeron} was 
first introduced as the {\it dominant or leading angular momentum 
trajectory i.e.} as the dominant singularity in the complex angular
momentum plane (see Section I.3.8). 
 
 To summarize, high energy physics in the {\it Sixties} was 
characterized by clean and clever mathematics but the physics 
issues were rather muddy.
 
>From the phenomenological point of view, basically all data on 
 angular distributions were easily reproduced with a fairly small 
 number of Regge trajectories (and of parameters). The big 
limitation came when it was realized that {\it polarization data}
 did not fit in the scheme.
 Spin, considered till that moment a {\it rather unessential 
 complication}, proved, well to the contrary and for the first time, 
 extremely subtle and difficult to cope with. This led to the first 
 attempt to incorporate unitarity corrections and, as a consequence,
 to the introduction of {\it complex angular momentum cuts} 
by Mandelstam [11]
 \bigskip
 
 \noindent {\bf I.2.3 The Seventies.}
 \bigskip
 
 As the energy increases, a larger and larger number of
 particles can be produced in the final state. Let us suppose
that $n$ final particles $c_i$ ($i =1, 2, ...n$) are produced in
 the collision of the two initial particles $a$ and $b$
 $$ a + b \rightarrow  c_1 + c_2 + .... + c_n. \eqno(I.2.1)$$
 We shall call this an {\it exclusive reaction} and it is easy to see
 that the number of independent scalar variables necessary to 
 describe completely the kinematics of such a final state is
 $3n-4$ {\it i.e.} increases very rapidly with the energy (see 
Section I.3.1).
 Partly due to these kinematical difficulties and partly 
 to the complexity of the dynamical situation, in this
decade the interest shifted gradually from exclusive 
 to {\it inclusive} reactions. The latter denote situations 
 in which, out of n particles produced in the final state,
 only one (or, perhaps, exceptionally, two) is {\it measured} 
 (meaning that its kinematics is completely reconstructed) and 
 all the others {\it are summed over}. We can denote such a  
physical situation by writing
 $$ a + b \rightarrow c_1 + X \eqno(I.2.2)$$
 where $X$ stays for all the particles whose kinematics is not 
 fully measured and reconstructed.
 
 Reaction $(I.2.2)$ is called {\it single inclusive} but we 
 may have {\it double inclusive reactions} (if one measures two 
 of the final particles) etc. 
 
 It is important to notice that eq. $(I.2.2)$ looks very much like
 the two-body reaction $(I.1.12)$ we started from in order to define
 a diffractive reaction. In $(I.2.2)$, $X$ is not a true particle (its 
 fourmomentum squared is not the mass squared like it would be 
for a {\it bona fide} particle). For that matter, however, also $a^*$ 
 and (or) $b^*$ in $(I.1.12)$ were not necessarily true particles 
 according to our general discussion. Thus, as a byproduct, we
 learn that an inclusive reaction can also be diffractive. For this, 
 according to our general prescription, it will be sufficient that 
 $c_1$ coincides with either $a$ or $b$. For instance,
 $$ a + b \rightarrow  a + X \eqno(I.2.3)$$
 is both diffractive and inclusive (such a case will turn out to
provide today's preferred path to analyse diffraction).
 
 Concerning inclusive reactions, a very remarkable extension
 of the optical theorem was proved by Mueller [12] through a 
 clever use of analyticity and crossing. This theorem will be
 discussed later on (see Section I.3.10) since it remains one of the 
most interesting tools to study diffraction in present days.
 
 We have already mentioned, that the number of variables 
 necessary to describe reaction $(I.2.1)$ is altogether $3n-4$. For 
 n=2, this gives the 2 independent variables 
 (the total energy and the scattering angle or the momentum 
 transfer) which are notoriously necessary to describe, for
 instance, an elastic reaction. By contrast, in a single inclusive 
 process, (at least) one of the final products ($X$) is not an on-shell 
 particle and one needs (at least) one extra variable. Reaction 
$(I.2.2)$, for example, would thus need a total of {\it three 
independent variables}.
 
 In the Seventies, however, diffractive physics in general 
lost ground; the main reason for this was the explosion of 
interest in the physics of Deep Inelastic Scattering (DIS) 
dominated by Bjorken scaling [13] whose main properties
we will very briefly recall in Part II. 
 
 It is both enlightening and unexpected that it will be, 
eventually, largely from the developments of DIS that 
diffraction will be rejuvenated (see Part III).
 \bigskip
 
 \noindent {\bf I.2.4 The Eighties.}
 \bigskip
 
 This decade witnessed a renewal of interest in diffractive 
physics due to several reasons among which:
 \begin {itemize}
 \item {i)} Donnachie and Landshoff proposal of an 
empirically simple phenomenology of high energy 
physics [14]. Though its merits have been overemphasized, 
it has significantly contributed to shifting the attention 
of high energy physicists back to diffraction;
 \item {ii)} a perturbative picture of the dominant 
diffractive component (the {\it Pomeron}) as made of two 
gluons [15]. This, eventually, led to the perturbative or 
BFKL Pomeron [16];
 \item {iii)} the first anticipation of the role of $ep$ physics in 
diffraction [17].
 \end {itemize}
 \bigskip
 
 \noindent {\bf I.2.5 The Nineties.}
 \bigskip
 
In the Nineties, diffraction made a grandiose comeback both from the 
theoretical and the experimental points of view. On the theoretical 
side, Bjorken pointed out a new exciting signature of diffractive 
 physics predicting {\it large rapidity gaps} as consequence 
 of it [18]. These rapidity gaps were immediately sought for, 
and found experimentally, both at the Tevatron and at HERA 
 [19, 20]. Diffraction, at the Tevatron is seen today not only
 through a variety of large rapidity gaps in jet physics but also
 via specific signatures of inclusive reactions some of which 
(like the $1/M_X^2$ behavior) will be discussed later on
(see Section I.3.9.2). Diffraction at HERA, started 
 when an {\it unexpected}\footnote {Not really so unexpected,
 a number of authors had indeed anticipated it [21].} sudden 
 growth at low Bjorken-$x$ was seen in the gluon component 
 of the structure function $F_2(x, Q^2)$ . 
 
 Most important, however, for the resurrection of diffraction 
in the Nineties was that, for the first time one managed to go 
beyond a purely qualitative understanding of origin and 
properties of the Pomeron 
 both perturbatively [16] and beyond. We shall return to some 
 of these issues because they remain at present, basically open 
 problems. Generally speaking, however, our present understanding
 of hadronic physics marks a great improvement over the past
 mostly because of the much larger flexibility of tools by which
 we can approach the problem. This is seen also in the spectrum
 of different variables that can be turned on and off in the various
 experimental set-ups. 
 
 Many are, however, the issues of present day's debates. They 
could be classified in various order but the basic question is 
{\it what (if any) is the precise relation between hard and soft
 diffraction?} Are there several Pomerons (as some believe [22])
 or is there only one (as others advocate [23])? Or, even more 
 fundamental, is the notion of {\it Pomeron} at all defined? It 
 is quite clear that, ultimately, a complete understanding of these 
 matters will require a much deeper mastering of higly complex 
 perturbative and, alas, nonperturbative techniques with which 
 to attack the problem. This will, presumably, be left for:
 \bigskip
 
 \noindent {\bf I.2.6 The Third Millenium.}
 \bigskip
  
 A full understanding of the origin and nature of the 
 Pomeron, {\it i.e.} of diffraction is, hopefully, the next step. 
 The difficulties to be surmonted, however, remain formidable 
 because of the intrinsecally nonperturbative nature of QCD  
 and because, probably, one will need a much better control
 of unitarity. A first step in this direction is, hopefully, 
content of the last paper of Ref. [16].
 
\bigskip
 \noindent {\bf  I.2.7 Brief discussion of the data.}
\bigskip
 
We will end this Section with a sketchy recollection of
the main aspects of high energy hadronic data (mostly elastic).
The reader interested in more details should consult Refs. [5-7].

Most of the elastic and diffractive hadronic data at high 
energy are concentrated in the small $t$ (or $cos \theta$) 
domain. This accumulation of data in the forward direction 
takes the name of {\it forward} or {\it diffraction peak} 
for the strict analogy with the optical
phenomenon illustrated earlier. Traditionally, the forward
peak, is parametrized (empirically) as
$${{d \sigma} \over {dt}} = A(s) e^{b(s) \, t}\eqno(I.2.4)$$
where $A(s)$ is called the {\it optical point} and $b(s)$ is
the {\it slope}; they all appear to increase gently with energy 
and they are, roughly, related via the elastic cross section 
(which one obtains integrating eq. $(I.2.4)$ by $b(s) \, 
\sigma_{el} \approx A(s)$. Their rate of growth is 
connected to that of {\it total and elastic cross sections} 
through the relation  
$$b(s) \ge {{\sigma^2_{tot}} \over 
{18 \pi \sigma_{el}}}\eqno(I.2.5)$$
This increase, known also as {\it shrinkage} of the diffraction 
peak is clearly exhibited in Fig. 3.

\medskip
 Fig. 3 . {\it $pp$ elastic forward high energy data (the 
continuous lines show how these data are reproduced by the 
lowest lying mesonic Regge trajectories, to be discussed in 
Section I.3.8.9).}
\medskip

As mentioned above, the {\it elastic cross section} is 
obtained integrating over $t$ the differential cross section.
At a {\it collider}, one has
$$\sigma_{el} = {{N_{el}} \over {\cal L}}\eqno(I.2.6)$$
where $N_{el}$ is the number of elastic events and $\cal L$ 
the {\it luminosity} of the collider. That $\sigma_{el} $ grows 
with energy compared with $\sigma_{tot}$ is by now an 
accepted fact but specific predictions concerning this point
are rather model dependent (see Fig. 4).

\medskip
Fig. 4  {\it The ratio ${{\sigma_{el}} \over { \sigma_{tot}}}$ for $p \bar 
p$ as a function of energy. The line is a guide to the eye.} 
\medskip

Similarly, one defines the {\it total cross sections} 
$\sigma_{tot}$ as
$$\sigma_{tot} = {{N_{el}+ N_{inel}} \over {\cal L}}
\eqno(I.2.7)$$
where $N_{inel}$ is, the number of inelastic events. 

A formally different way to write down the {\it total cross 
sections} makes use of the {\it optical theorem} (see below eq.
$(I.3.40)$) which we rewrite here as
$$\sigma_{tot}^2 = {{ 16 \pi} \over { 1 + \rho^2}} {{d \sigma} \over 
{dt}}|_{t=0}= {{ 16 \pi} \over { 1 + \rho^2}} {{d N_{el(t)}}\over 
{dt}}|_{t=0} \, {{1} \over {\cal L}}.\eqno(I.2.8)$$
Eliminating the luminosity $\cal L$ (which is not very
easy to measure accurately), we get
$$\sigma_{tot}^2 = {{ 16 \pi} \over { 1 + \rho^2}} {{d N_{el}/dt|_{t=0}} 
\over {N_{el}+ N_{inel}}}.\eqno(I.2.9)$$
In $(I.2.8,9)$, $\rho$ is the ratio of the real to the imaginary 
forward amplitude
$$\rho = {{Re f(s,0)} \over {Im f(s, 0)}}.\eqno(I.2.10)$$

For a long time it was either believed or assumed that total 
cross sections tend to a constant as the energy increases. It 
was only in the late Sixties, after the Serpukhov data became 
available that it was realized that they increase very slowly 
with energy. We will see that total cross sections cannot grow 
faster than a squared logarithm of the energy owing to a
celebrated result known as the Froissart bound 
(Section I.3.7.1). Experimentally, such a growth is indeed not 
incompatible with the data. Fig. 5, for instance, displays a 
fit to the data [5d]. What is 
instructive in this figure is the comparison of how a squared 
logarithmic behavior could be distinguished by a single power 
of $ln \, s$ once LHC data will be available. At present energies, 
however, a {\it transient power growth} (as suggested in Ref.
 [14] ) could not be ruled out by the data. The fit with the form 
suggested by the authors of Ref. [14] is shown in Fig. 6 for the 
same ($pp$ and $p \bar p$) total cross sections. For example, 
the case of $\sigma_{tot}(\bar p p)$ corresponds to
$$\sigma_{tot}(\bar p p) = 21.7 s^{0.0808} + 98.45 s^{-
0.45}.\eqno(I.2.11)$$

\medskip
Fig. 5 {\it Total $pp$ and $\bar p p$ cross sections up to the highest
energies with a $ln^{\gamma} s$ behavior. The best fit (solid line)
corresponds to $\gamma =2.2$. The dashed lines delimit the region
of uncertainty. The dotted line shows the extrapolation of the
$\gamma =1$ fit.}
\medskip

Fig. 6  {\it Total $pp$ and $\bar p p$ cross sections according to
the power fit eq. $(I.2.11)$.}
\medskip

As for the quantity $\rho$ defined in $(I.2.10)$,  the standard 
picture of high energy data is that this quantity should vanish
at increasing energies\footnote{As we will see when discussing
the optical theorem (Section I.3.5), the qualitative argument 
is that while unitarity keeps building up the forward imaginary 
part of the amplitude, this is not the case for the real part.}. 
Indeed, at the present highest energies, this ratio is about
10 \% (Fig. 7). As 
a matter of fact, analyticity of the scattering amplitude alone
({\it i.e. forward dispersion relations}) predicts that, as 
$s \rightarrow \infty$
$$\rho \equiv {{Re f_+(s,0)} \over {Im f_+(s,0)}} \rightarrow
{{const} \over {ln \, s}} \rightarrow 0.\eqno(I.2.12)$$

That such an asymptotic trend ($\rho \rightarrow 0$) is {\it not} 
yet visible in the data (Fig. 7) is still one of the cleanest
pieces of evidence that Asymptopia ( {\it i.e.} the utopic land where
all behaviors should be asymptotically simple) is not yet in 
sight.
\medskip

Fig. 7  {\it The quantity $\rho $ as function of energy with the
extrapolation from dispersion relations.}
\medskip

We will return time and again on the data we have just briefly
recalled either to discuss aspects related to their interpretation 
or to comment on the various points covered in our discussion.

 \bigskip
 
 \noindent {\large\bf I.3 A summary of conventional hadronic 
diffraction.}
 \bigskip
 
 Old-fashioned hadronic diffraction is a very complex 
mixture of a number of different ingredients ranging 
from rigorous theorems derived from analyticity, unitarity 
and crossing or directly from axiomatic field theory applied 
to powerful representations and models of the scattering 
amplitude (like the eikonal, the Watson Sommerfeld, the 
geometrical scaling model etc.) down to the use of 
intuitive arguments or empirical observations. It also 
relies on an ample collection of properly adjusted data. 
 
We will present, here, a discussion of many of these 
aspects. It is quite clear, however, that their choice and  
presentation is profoundly affected by the personal inclination of 
the present author. The reader is, therefore, urged to find
 alternative (and complementary) viewpoints in the 
existing literature (see, for instance, Refs. [5, 7]).
\bigskip
 
 \noindent {\bf I.3.1 Some kinematics.}
\bigskip
 
 Let us begin by reviewing a minimum of kinematics. This
 will not be confined, obviously, to the case of diffractive 
 reactions; for this reason, we will refer to a totally arbitrary
 reaction and use a terminology as general as possible.
 Ultimately, however, we will be interested in applying 
our general considerations to the specific case of reaction 
$(I.1.12)$.
 
 Let us consider the most general $2\rightarrow n$ 
 production process (in which all legs represent {\it 
 bona fide} particles)\footnote {Normally, one assumes, 
that {\it two particles come in and n go out}. The two 
that {\it come in} define the {\it initial state}. However, 
to the extent that, from the theoretical viewpoint,  an 
 {\it outgoing particle} is nothing but {\it an incoming 
antiparticle with opposite momentum}, {\it a priori}, 
any number $k$ of particles could be taken as incoming 
and the others as outgoing.}, 
$$p_1 + p_2 \rightarrow {p'}_1 + {p'}_2 +...+{p'}_n,
\eqno(I.3.1)$$
 where $p_i$ ($i=1, 2$) and ${p'}_i$ ($i=1, ... n $) are the 
 fourmomenta of the incoming and of the outgoing particles 
 respectively. For the process $(I.3.1)$, the number of 
 independent Lorentz invariant variables is $3n-4$. 
 To see how this comes about, let us begin by noticing 
that there are altogether $4(n+2)$ fourmomenta {\it 
components}, $8$ associated with the initial particles and 
 $4n$ with the outgoing ones. There are, however, several
 constraints acting on these $4(n+2)$ scalar quantities.  
First, we have $n+2$ {\it mass shell conditions}
 $$p_j^2 = m_j^2, (i=1,2); {p'}_i^2 = {m'}_i^2 (i=1, 2,...n).$$
 Next, there are $4$ constraints from energy-momentum 
 conservation (the total fourmomenta of the initial and of the
 final states must be equal)
 $$ p_1 + p_2 = {p'}_1 + {p'}_2 +...+{p'}_n,\eqno(I.3.2)$$
 Lastly, $6$ constraints are required to fix the frame of 
reference (the six {\it Euler angles} of a four-dimensional 
world). This gives a total of $n+12$ constraints for 
$4(n+2)$ quantities which brings the count of truly 
independent variables down to $3n-4$ as anticipated.
 
\noindent In the case of two-body reactions 
 $$ p_1 + p_2 = {p'}_1 + {p'}_2,\eqno(I.3.3)$$
 the count reduces to the familiar fact that two independent 
 variables are sufficient to describe the entire kinematics of 
 the process and the reference frames one usually introduces 
 are: i) the  CM system (characterized by $\vec p_1 + 
\vec p_2 = 0$), ii) the LAB system (where, for instance, 
$\vec p_2 = 0$) and, iii) the {\it brick-wall} (or Breit) 
system (defined by $\vec p_1 +  \vec {p'}_2$)\footnote{None 
of these systems would be appropriate for HERA where an 
asymmetric reaction takes place, 800 GeV protons hitting 
30 GeV electrons with a corresponding total CM energy 
of $\approx$314 GeV.}.
 
 We shall denote the momentum in the CM system by $p$ 
and the scattering angle by $\theta$. It is, however, 
customary to use the three scalar variables $s, t, u$ 
introduced long ago by Mandelstam, and defined by
 $$s\ =  (p_1\, +\, p_2)^2\eqno(I.3.4a)$$
 $$t\ =  (p_1\, -\, {p'}_1)^2\eqno(I.3.4b)$$
 $$u\ =  (p_1\, -\, {p'}_2)^2\eqno(I.3.4c)$$
 where the identity
 $$s\, +\, t\, +\, u\, = m_1^2 + m_2^2 + {m'}_1^2 +
 {m'}_2^2\eqno(I.3.5)$$
 brings back the count of independent variables to 2. 
 
 Assuming (for simplicity) that all the particles have 
equal masses\footnote {Technically, this would be the 
case of {\it elastic scattering} of {\it identical} particles. 
At high energies, however,  mass differences become 
irrelevant so long as all particles are on their mass 
shell. Thus, this assumption is rather unessential.}, 
the relation between the two sets of variables 
 $[p, \theta]$ and $[s, t, u]$ would be
 $$s\, =\, 4( p^2 +  m^2 ) \eqno(I.3.6a)$$
 $$t\, =\, - 2  p^2 ( 1 - cos\, \theta )\eqno(I.3.6b)$$
 $$u\, =\, - 2  p^2 ( 1 + cos\, \theta )\eqno(I.3.6c)$$
 so that $s$ is the (squared) CM  energy of the reaction 
$(I.3.3)$; $t$ and $u$ are (minus) the (squared) fourmomenta 
transferred from particle $1$ to particle $1'$ or to particle 
$2'$ respectively. 
 
Let us, now, consider a case in which not all legs represent 
true particles (like $(I.1.12)$ when either $a^*$ and/or $b^*$ 
may represent a resonance or an arbitrary number of 
particles)\footnote{ In this case, $m^2_{a^*}$ and/or 
$m^2_{b^*}$ do not represent physical masses and, 
therefore, are not constant.}. Formally, we can still, of 
course, define
 $$s\ =  (p_{a}\, +\, p_{b})^2\eqno(I.3.7a)$$
 $$t\ =  (p_{a}\, -\, p_{a^*})^2\eqno(I.3.7b)$$
 $$u\ =  (p_{a}\, -\, p_{b^*})^2\eqno(I.3.7c)$$
 with the identity
 $$s\, +\, t\, +\, u\, = m_a^2 + m_b^2 + m_{a^*}^2 +
 m_{b^*}^2\eqno(I.3.8)$$
 In this case, however, for any leg which does not describe 
 a true particle we have one (mass shell) constraint less and, 
therefore, one additional variable in the game. As an example,
 in the case of {\it single diffraction} ($a^*$, say, 
is the same incoming {\it real} particle $a$ and $b^*$ is a
pseudoparticle), in addition to 
the previously mentioned two variables, one has one 
additional variable to introduce which can, for instance, 
be the {\it missing mass}
 $$M_{b^*}^2 \equiv M_X^2 = (p_a+p_b-p_{a^*})^2.
\eqno(I.3.9)$$
 We will come back to discussing this physical situation.
 
 To simplify our discussion, in the next Section we will 
 confine ourselves to the fictitious case of identical isospin 
 zero particles and we will limit our considerations to the
 bare minimum necessary for our future developments
 referring the reader interested in more details to 
classical textbooks [24]. 
 \bigskip
 
 \noindent{\bf I.3.2 The S matrix.}
 \bigskip
 
 The {\it Scattering } matrix is, by definition, the operator 
which transforms the initial or {\it ingoing} state (a plane 
wave at time $t = -\infty$ where all the particles can be 
considered free and non interacting) into the corresponding 
final or {\it outgoing} state (a plane wave at time $t = +\infty$ 
where the particles are once again free and non-interacting) or
 $$|\psi _f> = S |\psi_i>.$$
 The general properties one assumes for $S$ are
 
 \begin {itemize}
 
 \item{i)} {\it relativistic invariance,}
 \item{i))} {\it unitarity,}
 \item{iii)} {\it crossing.}
 
 \end {itemize}

The first of these properties is self-evident; the second
and third will be discussed soon. 

 \noindent To these, one often adds
 \begin {itemize}
 
 \item{iv)} {\it analyticity}.
 \end {itemize}
 
 Part of this Chapter will be devoted to discussing
 these properties.
 
 The $S$ matrix and the {\it Transition matrix} are related by
 $$ S \, = \, 1 \, + \, iT \eqno(I.3.10)$$
 or, denoting by $i$ and $f$ the ensemble of discrete
indices and continuous variables 
characterizing the {\it initial} and the {\it final} state,
 $$S_{if} = \delta_{if}+(2 \pi)^4 i \delta^{(4)} (p_1+p_2- 
\sum_{i=1}^n {p'}_i) {{T_{if}} \over {n_{p_1} n_{p_2} 
n_{{p'}_1}...n_{{p'}_n}}}\eqno(I.3.11)$$
where the factors $n_{p_k}$ are given by the normalization 
of the free particle states and are introduced so that the 
$T_{if}$ matrix elements be relativistically invariant. 
With the normalization
$$<\vec p' | \vec p> = \delta (\vec p' - \vec p)\eqno(I.3.12a)$$
for spinless particles and
$$<\vec p' , s' | \vec p , s> = \delta (\vec p' - \vec p) 
\delta_{ss'}\eqno(I.3.12b)$$
for fermions, one finds
$$n_p = (2 \pi)^{{{3} \over {2}}} \sqrt {2E}\eqno(I.3.13a)$$
fos spinless bosons and 
$$n_p = (2 \pi)^{{{3} \over {2}}} \sqrt {E/m}\eqno(I.3.13b)$$
for massive fermions.
 
 Given the transition matrix elements, the differential cross 
section for reaction $(I.3.1)$ is 
 $$ d\sigma_{if} = (2 \pi)^4  \delta^{(4)} (p_1+p_2-
\sum_{i=1}^n {p'}_i) d^3 {p'}_1 ... d^3 {p'}_n  
{{|T_{if}|^2} \over {\Phi n^2_{{p'}_1}...n^2_{{p'}_n}}}
\eqno(I.3.14)$$
where $\Phi$ is the incoming flux defined as
$$\Phi = n_1^2  n_2^2 |\vec v_1 - \vec v_2| /(2 \pi)^6. 
\eqno(I.3.15)$$
Other choices of normalization are frequently used in the 
literature; for instance, instead of $(I.3.12)$ one often uses
 $$<\vec p' | \vec p> = 2E (2\pi)^3 \delta (\vec p' - \vec 
 p)\eqno(I.3.16a)$$
for spinless particles and
$$<\vec p' , s' | \vec p , s> = {{m} \over {E}} (2 \pi)^3 
\delta (\vec p' -  \vec p) \delta_{ss'}\eqno(I.3.16b)$$
 for fermions. In this case, all $n_p =1$.
 
 An alternative definition of the cross section involving 
the differentials of the fourmomenta is 
 $$d\sigma_{if} = {{\delta^{(4)} (p_1+p_2-\sum_{i=1}^n {p'}_i) } 
\over {2 (2\pi)^{3n-4} \lambda(s, m_1^2, m_2^2)}} 
\prod_{r=1}^n [\delta ({p'}_r^2 - {m'}_r^2) \theta ({{p'}_r}_0) \, d^{(4)}{p'}_r]  
|T_{if}|^2 \eqno(I.3.17)$$
 where 
 $$\lambda (x,y,z) = [x^2+y^2+z^2-2xy-2yz-2zx]^{1/2}.
\eqno(I.3.18)$$
In $(I.3.17)$ the step functions $\theta ({{p'}_r}_0)$ 
guarantee that each particle has positive energy while 
the $\delta ({p'}_r^2 - {m'}_r^2)$'s make sure that each 
final particle is on its mass shell.
 
An especially important case is the differential cross 
section for the elastic scattering (between two identical 
spinless particles of mass $\mu$, for simplicity). From 
eq. $(I.3.14)$ with $n=2$ we find
$$d\sigma = {{ (2 \pi)^4 \, \delta^{(4)}(p_1 + p_2 - 
{p'}_1 -{p'}_2) \, |T_{22}|^2} \over 
{\Phi \, (2 \pi)^3 \, 2{E'}_1 \, (2 \pi)^3 \, 2{E'}_2}}
 d^3 {p'}_1 d^3 {p'}_2. \eqno(I.3.19)$$
 The flux $\Phi$ is an invariant, so that we can evaluate 
it in any reference frame. In general, one has
 $$\Phi = 2 \lambda (s, \mu ^2, \mu ^2).\eqno(I.3.20)$$
 Using the vector part of the energy-momentum 
conservation $\delta^{(4)}(p_1 + p_2 - {p'}_1 -{p'}_2) $ 
to integrate over $d^3 {p'}_2$ and integrating over the 
{\it modulus} of $\vec {p'}_1$, we get the {\it differential 
cross section} 
 $${{d \sigma} \over {d \Omega}}=
{{(2 \pi)^4} \over {8 \, \lambda (s, \mu ^2, \mu ^2) \, (2 \pi)^6}} 
\int_0^{\infty} {{|\vec {p'}_1|  d|\vec {p'}_1| } \over
 {{E'}_1 {E'}_2}} \, \delta (E_1 + E_2 - {E'}_1 - {E'}_2) 
 |T_{22}|^2.\eqno(I.3.21)$$
We can calculate $(I.3.21)$ in any reference frame;  
in the CM system, for instance, ($\vec p_1=-\vec p_2 \, ; 
\, \vec {p'}_1=-\vec {p'}_2 $), using $| \vec p| = p$ and 
$| \vec p'| = p'$ for short, we have
 $$\delta (E_1 + E_2 - {E'}_1 - {E'}_2) \, = \,  
\delta (\sqrt {p^2 + \mu ^2} +\sqrt {p^2 + \mu ^2} - 
\sqrt {{p'}^2 + \mu ^2} - \sqrt {{p'}^2 + \mu ^2})
 ={{{E'}_2} \over {2 p'}} \, \delta (p-p')$$
 which, inserted into $(I.3.21)$ gives
 $$ {{d \sigma} \over {d \Omega}} =  {{|T_{22}|^2} \over 
{64 \ \pi ^2 \, s}}  \eqno(I.3.22)$$
 where we have also used  $p = {{\lambda (s, \mu ^2, \mu ^2)} 
\over {2 \, \sqrt s}}$ and $2 {E'}_2 = \sqrt s$.
 
 Collecting everything together and denoting by $B$ a 
(spinless) boson of mass $\mu$ and by $F$ a spin 
$1/2$ fermion of mass $m$, we have the following 
various cases
$$ {{d \sigma} \over {d \Omega}} =  
{{|T_{22}|^2} \over  {(2 \pi )^2 \, 16 \, s}}; 
 \quad \quad \quad [B+B \rightarrow B+B] \eqno(I.3.23a)$$
$$ {{d \sigma} \over {d \Omega}} =  
{{|T_{22}|^2} \over  {(2 \pi )^2}} \, \left [ {{s-4m^2} 
\over {s-4 \mu ^2}} \right ] ^{1/2}  \, {{m^2} \over {4 s}};  
 \quad \quad [B+B \rightarrow F+\bar F]\eqno(I.3.23b)$$
$$ {{d \sigma} \over {d \Omega}} =  
{{|T_{22}|^2} \over  {(2 \pi )^2}} \, {{m^2} \over { 4s}}; 
\quad \quad [B+F \rightarrow B+F] \eqno(I.3.23c)$$
$$ {{d \sigma} \over {d \Omega}} =  
{{|T_{22}|^2} \over  {(2 \pi )^2}} \, {{m^4} \over {s}}; 
\quad \quad [F+F (or \bar F) \rightarrow F+F (or \bar F)]. 
 \eqno(I.3.23d)$$
 
More concisely, we can write
$$ {{d \sigma} \over {d \Omega}} =  {{|T_{22}|^2} \over  {(2 \pi )^2}} \,
{{p} \over { \Phi}}\eqno(I.3.24)$$
where for the various cases, the flux $\Phi$ is given by
 $$ \Phi = 4 \, p\, \sqrt s; \quad \quad [B+B \rightarrow B+B] 
\eqno(I.3.25a)$$ 
 $$\Phi = 2 \, {{p} \over {m}}\, \sqrt s; \quad \quad [B+F 
\rightarrow B+F] 
 \eqno(I.3.25b)$$
 $$\Phi = {{p} \over {m^2}}\, \sqrt s; \quad [F+F (or \bar F) 
\rightarrow F+F (or \bar F)]. \eqno(I.3.25c)$$
 
 In practice, we shall often prefer to introduce the {\it 
scattering amplitude} $f(p, \theta)$ related to the $T_{22}$ 
matrix element by
$$T_{22} = 2 \pi  {{\Phi}\over {p}} \, f(p, \theta)
\eqno(I.3.26)$$
 with the simplifying relation
 $${{d \sigma} \over {d \Omega}} =  |f(p, \theta)|^2.
\eqno(I.3.27)$$
 
 \bigskip
 \bigskip
 \noindent {\bf I.3.3 Partial wave expansion and unitarity.}
 
 \bigskip
 
 If we are dealing with spinless particles, rotational 
invariance insures that the scattering amplitude can be 
decomposed in {\it partial waves, i.e.} that the angular 
dependence must satisfy the following decomposition
 $$f(p,\theta) =  \sum_{\ell =0}^{\infty} \,
 (2 \ell  +1) \, a_{\ell}(p) P_{\ell }(cos \theta)\eqno(I.3.28)$$
 where $P_{\ell}(cos \theta)$ are the usual Legendre 
polynomials and $a_{\ell}(p)$ take the name of {\it partial 
waves}. For particles with spin, there are various degrees 
of complications which we shall not get into [25].
 
 Expanding the S-matrix elements in partial waves, one 
identifies $a_{\ell}(p) $ with the corresponding {\it 
partial wave S-matrix elements} ($S_{\ell}(p) = 
e^{2i\delta_{\ell}(p)} $)
 $$  a_{\ell}(p)  = {{S_{\ell}(p) -1} \over {2ip}} =
{{e^{2i\delta_{\ell}(p)} - 1} \over {2ip}}\eqno(I.3.29)$$
 where $\delta_{\ell}(p) $ are known as the {\it phase shifts}.
 
 It is immediate to check that, for {\it elastic collisions i.e.} 
for collisions {\it below} the threshold for production of 
new particles, the {\it unitarity of the S-matrix} 
 $$SS^{\dagger} = S^{\dagger}S=1\eqno(I.3.30)$$ 
(which is the same as the {\it conservation of probability}), 
translates into the condition that $\delta_{\ell}(p)$ be {\it 
real}. In turn, this {\it elastic unitarity} can be written, in 
terms of the partial waves $a_{\ell }(p)$ as
 $$Im a_{\ell}(p) = p |a_{\ell}(p) |^2.\eqno(I.3.31)$$
 
Notice that eq.$(I.3.30)$ is {\it automatically satisfied} by
the form $(I.3.29)$ so long as $\delta_{\ell}(p)$ is real. 
Another form which satisfies identically the elastic 
unitarity condition $(I.3.30)$ is
$$a_{\ell}(p)  = {{1} \over {g_{\ell}(p) - ip}}\eqno(I.3.32)$$
where $g_{\ell}(p) $ must be real but is otherwise arbitrary.
 
{\it Above threshold}, $\delta_{\ell}(p)$ is no longer real 
but develops a {\it positive} imaginary part\footnote {A 
negative imaginary part of $\delta_{\ell}(p)$ would imply 
a partial wave scattering amplitude larger than one in 
violation of $(I.3.30)$.}. We shall also rewrite $S_{\ell}(p)$ 
as
$$S_{\ell}(p) =\eta_{\ell}(p) e^{2i\alpha_{\ell}(p)}
\eqno(I.3.33)$$ 
 where $\alpha_{\ell}(p)$ and $\eta_{\ell}(p) $  are real. 
$\eta_{\ell}(p) $ are known as the {\it absorption 
coefficients} and obey
$$0 \le \eta_{\ell}(p)  \le 1.\eqno(I.3.34)$$
The elastic unitarity condition over the partial waves 
$(I.3.31)$ is now replaced by
$$Im a_{\ell}(p) - p |a_{\ell}(p) |^2 =  
{{1 - {\eta_{\ell}(p)}^2} \over {4p}} \eqno(I.3.35)$$
where the {\it rhs} reduces to zero in the {\it elastic} limit 
$\eta_{\ell}(p) =1$.
 
 \bigskip
 \noindent {\bf I.3.4 Elastic, total and inelastic cross sections. 
The optical theorem.}
 
\bigskip
The elastic cross section is obtained integrating over the solid 
angle the differential cross section. Inserting the partial wave
expansion $(I.3.28)$ into $(I.3.24)$ and using the orthonormality 
of Legendre polynomials
 $$\int_{-1}^{+1} dx P_{\ell}(x) P_{\ell '}(x) = {{2} \over {(2 \ell +1)}}
 \delta_{\ell, \ell '}\eqno(I.3.36)$$ 
we get
 $$\sigma_{el} (p) = \int d\Omega {{d \sigma} \over {d \Omega}} =
 \sum_{\ell} \sum_{\ell '} a^*_{\ell}(p)  a_{\ell '}(p) (2 \ell +1) 
 (2 \ell ' +1) \, 2 \pi \, \int_{-1}^{+1}  d cos \theta \, 
P_{\ell}(cos \theta) P_{\ell '}(cos \theta)$$
 $$ = 4 \pi  \, \sum_{\ell} (2 \ell +1) 
 | a_{\ell}|^2 =  {{\pi} \over {p^2}} \sum_{\ell} (2 \ell +1)  
[{\eta_{\ell}}^2 
 sin^2  2 \alpha_{\ell}(p) + (1 - \eta_{\ell} cos 
2\alpha_{\ell})^2]\eqno(I.3.37)$$ 
 where the last term in square bracketts vanishes below the 
threshold for production. Above this threshold, one defines 
also the {\it absorption cross section} $\sigma_{abs}$ as
$$\sigma_{abs} \equiv {{\pi} \over {p^2}} \, \sum_{\ell =0}^{\infty}
 (2 \ell +1) [1 - |S_{\ell}(p)|^2] = {{\pi} \over {p^2}} \, 
\sum_{\ell=0}^{\infty} (2 \ell  +1) [1 - |\eta_{\ell }(p)|^2]
\eqno(I.3.38)$$
 (which, of course, reduces to zero below threshold where
 all  $\eta_{\ell }(p) \equiv1$). Finally, we have, by definition
 $$\sigma_{tot} = \sigma_{el} + \sigma_{abs} = 
{{2 \pi} \over {p^2}} \sum_{\ell =0}^{\infty} 
(2 \ell  + 1) ( 1 - \eta_{\ell } cos 2  \alpha_{\ell }).
\eqno(I.3.39)$$
 
 By comparison between $(I.3.39)$ and $(I.3.28, 29)$ and $(I.3.33)$ 
and using $P_{\ell}(1)=1$, we find the most remarkable sum rule
 $$Im f(p,0^{\circ}) = {{p} \over { 4 \pi}} \, 
\sigma_{tot}\eqno(I.3.40)$$
 which is known as the {\it optical theorem} and is a direct 
 consequence of unitarity (as we shall see in the next Section).
 
 \bigskip
 \noindent {\bf I.3.5 Unitarity, the optical theorem and elastic 
unitarity.}
 \bigskip

 If we denote by the symbol \si the integration over all 
continuous variables (the momenta of the various particles) 
and the sum over all discrete quantum numbers, the 
completeness of the physical states inserted into the 
unitarity condition $(I.3.30)$ bracketed between the 
initial ($i$) and the final ($f$) states gives
 $$\si S_{nf} S^*_{ni} = \delta_{if}\eqno(I.3.41)$$
 or, introducing the T-matrix elements $(I.3.10)$  
 $$S_{nf} = \delta_{nf} + (2 \pi)^4 i \delta^{(4)}(p_n - p_f)
 {{T_{nf} \over {\prod_n {n_n}{n_f}}}}\eqno(I.3.42)$$
 and, inserting them into the unitarity of the $T$-matrix 
$$TT^{\dagger}=i(T^{\dagger} - T),$$ 
after a little algebra we find
 $$i(T^*_{fi}- T_{if}) = \si (2 \pi)^4 \, \delta^{(4)}(p_i-p_f)
 {{T^*_{ni} T_{nf}} \over {\prod {n_n}^2}}.\eqno(I.3.43)$$
Rewriting the left hand side as
 $$T_{if}-T^*_{fi}= 2i Im T_{if},$$
in the special case when {\it the final state is
 identical to the initial one} ($|f> \equiv |i>$) eq.$(I.3.43)$ 
gives
 $$Im T_{ii} = {{(2 \pi)^4} \over {2}} \si_n  \delta^{(4)}(p_i-p_n)
 {{|T_{ni}|^2} \over { \prod n_n^2}}.\eqno(I.3.44)$$
The latter, up to (a flux) factor, is nothing but the 
definition itself of the total cross section $\sigma_{tot}$. 
Putting everything together, the precise relation is
 $$Im T_{ii} = {{\Phi} \over {2}} \, \sigma_{tot}.
\eqno(I.3.45)$$
 Using the previous definition $(I.3.25)$ for the flux $\Phi$ and
 the expression $(I.3.26)$ between the scattering amplitude and
 the two-body transition matrix element, we find, once again
 the {\it optical theorem}
 $$Im f(p,0^{\circ}) = {{p} \over { 4 \pi}} \, \sigma_{tot}
\eqno(I.3.40)$$
 which, in the present derivation is a direct consequence 
of unitarity (as anticipated).

One point is worth stressing right away about this sum rule. 
The fact that $\sigma_{tot}$ is the sum over an increasingly 
larger number of terms (as $s \rightarrow \infty$ more 
and more production channels contribute), suggests that the 
{\it imaginary part of the scattering amplitude could even 
increase with the energy}\footnote{This will depend on whether
the increase in the opening of new production channels will
or not exceed the smallness of each individual contribution and
the fact that each one individually will tend to zero as the
energy increases.}. More 
properly, given that no such constraint exists over the {\it 
real part} of the scattering amplitude, one can reasonably 
expect that the {\it scattering amplitude should 
become predominantly imaginary} as the energy increases.  
When comparing with the data, we saw that this is 
indeed the case (see Fig. 7).
 
 Later, we shall return to the previous expressions 
$(I.3.43, 44)$. For the time being, let us just notice that 
these equations, though written in a symbolic way 
(remember the meaning of \si), imply, first of all,
 that an {\it extended optical theorem} can, in principle, 
be written for any transition $i \rightarrow f$ 
when $|f> \equiv |i>$ and not just for the special case 
($2 \rightarrow 2$). Even though this may never become 
of any practical use (other than in the applicationof Section
I.3.10.1), it has very far reaching theoretical 
implications on the strict connection between different 
matrix elements induced by the unitarity condition. 

A crucial point worth noticing is that the left hand side of 
$(I.3.44)$ is always real (and positive) and so is {\it each 
coefficient} on its right hand side (which is proportional to 
$|T_{ni}|^2$). By contrast, this {\it is not} the case for the 
individual terms which appear under the sum on the right 
hand side of $(I.3.43)$ (these are proportional to $T^*_{ni} 
T_{nf}$ which become real and positive precisely as $|f> 
\rightarrow |i>$). The {\it whole sum} in $(I.3.43)$, however, is 
anyway real (because so is the left hand side of the equation). 
As a consequence, all the phases of the individual terms in 
this sum must mutually compensate and the reality of the
right hand side must thus result from elaborate cancellations 
between all the terms and not from the simple algebra of 
$(I.3.44)$. This argument illustrates
the complexity of the unitarity condition when considered in
its full glory and explains why it has not been possible to
come to terms with it so far.
 
 Let us now see yet another way of writing things by 
considering first the two body intermediate state 
contribution to the complete unitarity condition 
$(I.3.44)$ applied to a $2 \rightarrow 2$ process
 ({\it i.e.} to a process in which the final state $|f>$ not 
only is identical to the initial one $|i>$ but the latter is 
a two-body state; in essence, this means that we are 
going to confine ourselves to the {\it elastic transition 
amplitude}). Referring to Fig. 8, we use the Mandelstam 
variables 
$$s\ =  (p_1\, +\, p_2)^2= ({p'}_1\, +\, {p'}_2)^2= (k_1 \, + \, 
k_2)^2\eqno(I.3.46a)$$
$$t\ =  (p_1\, -\, {p'}_1)^2=-2p^2 (1 - cos \theta)
\eqno(I.3.46b)$$
$$t_1\ =  (p_1\, -\, k_1)^2=-2p^2 (1 - cos \theta_1)
\eqno(I.3.46c)$$
$$t_2\ =  ({p'}_1 \, - \, k_1)^2=-2p^2 (1 - cos \theta_2)
\eqno(I.3.46d)$$
where $\theta, \theta_1$ and $\theta_2$ are the (CM) 
angles between $(\vec p_1, \vec{p'}_1), 
(\vec p_1, \vec k_1)$ and  $(\vec {p'}_1, \vec k_1)$ 
respectively.
 
\medskip
Fig.8 {\it Kinematics of the 2-body intermediate state
in a $2 \rightarrow 2$ reaction.}
\medskip

In the two-body intermediate state approximation and 
explicitating the symbolic unitarity equation $(I.3.43)$, 
we have
$$Im T(s,t) = {{1} \over {2}} \int d^3k_1 \int d^3k_2 
{{(2 \pi)^4 \delta^{(4)}(p_1 + p_2 - k_1 - k_2)} 
\over {(2 \pi)^6 \, 2E_{k_1} \, 2E_{k_2}}} 
T(s, t_1) T^*(s, t_2)=$$
 $$= {{1} \over {2 \, (2 \pi)^2}} \, \int d^4k_1 
 \int d^4k_2 \, \delta (k_1^2-\mu^2) \, \delta (k_2^2-\mu^2) 
 \delta^{(4)}(p_1 + p_2 - k_1 - k_2) T(s, t_1) T^*(s, t_2)=$$
 $$ = {{1} \over {8  \pi^2}} \, \int d^4k_1 \, 
\delta [ (p_1 + p_2- k_1)^2 - \mu^2] \, \delta (k_1^2-\mu^2) \,
 T(s, t_1) T^*(s, t_2)=$$
 $$ = {{1} \over {8 \pi^2}} \, \int d^4k_1 
 \, \delta (k_1^2-\mu^2) \, \delta [ (p_1 + p_2)^2 + k_1^2 - 2 
 k_1(p_1+p_2) - \mu^2 ] \, T(s, t_1) T^*(s, t_2).$$
 We now evaluate the above equation in the CM system where 
$\vec p_1 + \vec p_2 = 0$ and, therefore, $s\ =  (p_1\, +\, p_2)^2 = 
(p_1+p_2)_0^2$  so  that we have
 $$Im T(s,t) = {{1} \over {8 \pi^2}} \, \int d^4k_1 \, 
\delta (k_1^2-\mu^2) 
 \, \delta [s - 2 k_{10} \sqrt s] \, T(s, t_1) T^*(s, t_2) =$$
 $$ = {{1} \over {8 \pi^2}} \, \int d^3k_1 dk_{10} \, 
\delta (k_{10}^2 - \vec k_1^2-\mu^2) {{\delta ({{1} \over {2}} 
\sqrt s - k_{10})} \over {2 \sqrt s}}  T(s, t_1)  T^*(s, t_2)=$$
 $$ = {{1} \over {16 \pi^2 \sqrt s }} \int d^3k_1 \delta ({{s} \over 
 {4}} - \vec k_1^2 - \mu^2) \, T(s, t_1) T^*(s, t_2) =$$
 $$ = {{1} \over {16 \pi^2 \sqrt s }} 
\int d|\vec k_1| {\vec k_1}^2 \, d{\Omega _{k_1}}  \delta ({{s} 
 \over {4}} - \vec k_1^2 - \mu^2) \, T(s, t_1) T^*(s, t_2) =$$
 $$ = \, {{1} \over 
 {32 \pi^2 \sqrt s }} \sqrt{{{s} \over {4}} - \mu^2} \, \int 
d{\Omega_{k_1}} \, T(s, t_1) T^*(s, t_2)$$
or
$$Im T(s,t) \, = \, {{1} \over {32 \pi^2 \sqrt s }} \sqrt{{{s} \over {4}} - 
 \mu^2} \int d{\Omega _{k_1}} \, T(s, t_1) T^*(s, t_2).
\eqno (I.3.47a) $$
Rewritten in terms of $\theta, \theta_1, \theta_2$ we have
$$Im T(p, cos \theta) \, = \, {{p} \over {32 \pi^2 \sqrt s }} 
\int d{\Omega_{k_1}} \, T(p, cos \theta_1) T^*(p, cos 
\theta_2).\eqno(I.3.47b) $$
 If the above equation is projected in partial waves
 $$T(p, cos \theta) = \sum_{\ell = 0}^{\infty} T_{\ell }(p) P_{\ell }(cos 
 \theta),$$
using repeatedly this expansion and inverting freely orders of 
 summations and of integrations, eq. $(I.3.47b)$ becomes
 $$Im T_{\ell }(p) = {{1} \over {2}} \int_{-1}^{+1} d cos \theta 
P_{\ell}(cos 
 \theta) Im T(p, cos \theta)=$$
 $$ = \,{{p} \over {64 \pi^2 \sqrt s }} 
 \sum_{\ell_1, \ell_2} \, (2 \ell_1 +1) (2 \ell_2 +1) T_{\ell_1}(s)
 T^*_{\ell_2}(s) \, \int_{-1}^{+1} $$
$$d cos \theta \, d \Omega_{k_1} P_{\ell }(cos \theta) P_{\ell_1}(cos 
\theta_1) P_{\ell_2}(cos \theta_2).\eqno(I.3.47c) $$
 The angular integrations in $(I.3.47c) $ can be completely 
performed recalling that
 $$cos \theta_1 = cos \theta \, cos \theta_2 + sin \theta sin \theta_2
 cos \phi$$
and using (recall $(I.3.46)$)
$$\int d\Omega_{k_1} P_{\ell}(\vec p_1 \cdot \vec k_1) \,
P_{\ell'}(\vec p_2 \cdot \vec k_1) ={{4\pi} \over {2 \ell +1}}
P_{\ell}(\vec p_1 \cdot \vec p_2) \delta_{\ell \ell'}$$
or which amounts to the same, using directly the 
orthonormality of Legendre polynomials. For this,
we write
$$P_{\ell_1}(cos \theta_1) = P_{\ell_1} (cos \theta \, cos \theta_2 + 
sin \theta sin \theta_2 cos \phi)=$$
$$= P_{\ell_1}(cos \theta) \, P_{\ell_1}(cos \theta_2) + 2 
\sum_{m=1}^{\infty} {{\Gamma (\ell_1 -m +1)} \over {\Gamma 
(\ell_1 + m +1)}}$$
$$ P_{\ell_1}^m(cos \theta) \, P_{\ell_1}^m(cos \theta_2) \, cos (m\, \phi).$$
Integrating over $\phi$ and $\theta_2$ using 
$$\int_0^{2 \pi}  d\phi \, cos (m\, \phi) = 0$$
and of $(I.3.36)$, eq. $(I.3.47c) $ becomes, finally
 $$Im T_{\ell }(p)  \, = \, {{1} \over {8 \pi \sqrt s }} 
\sqrt{{{s} \over {4}} - \mu^2} \, |T_{\ell }(p) |^2. \eqno(I.3.48)$$
Using the relationship between the T-matrix elements and 
 the scattering amplitude $f(p, \theta)$ (eq $(I.3.26)$) and 
the partial wave expansion $(I.3.28)$ we get back the partial
 wave unitarity $(I.3.35)$ or, below threshold, $(I.3.31)$.
 
 Before closing this section, let us notice that the {\it complete}
 $2 \rightarrow 2$ unitarity equation ({\it i.e.} not confined to a
 two-body intermediate contribution, like $(I.3.47)$) can be,
formally, written as
 $$Im T(s,t) \, = \, E(s, t) \, + \, I(s, t)\eqno(I.3.49)$$
 where $E(s, t)$ is the two-body intermediate state contribution
to the unitarity condition for the $2 \rightarrow 2$ process 
which we have just evaluated. About 
$I(s,t)$ we know basically nothing (unless we resort to models) 
except for one, very important, property: $I(s, t)$ is inherently 
{\it positive} (or, more correctly, {\it non-negative}) 
(this has the same content of $(I.3.35)$ of which $I(s,t)$ would the
right hand side.
 
 This {\it positivity property} is about the only simple statement 
that can be made concerning {\it unitarity} in its full generality. 
We shall encounter, later on, several properties that can be 
derived from unitarity but most of these will follow from making 
other specific assumptions.
 
 \bigskip
 \noindent {\bf I.3.6 Crossing (the substitution law) and relativistic 
 invariance. Analyticity.}
 \bigskip
 
 It is an important property of {\it relativistic field theory} 
that in a given reaction, an {\it incoming particle} of 
momentum $p$ can be viewed as an {\it outgoing antiparticle} 
of momentum $-p$\footnote{Crossing is proved within 
perturbative field theory (Feynman diagrams) and is 
postulated within the S-matrix approach.}. To analyze the 
consequences of this invariance, let us go back to reaction 
$(I.3.3)$
 $$ p_1 + p_2 = {p'}_1 + {p'}_2,\eqno(I.3.3)$$
 for which we have introduced earlier the three scalar 
Mandelstam variables $s, t, u$ $(I.3.4).$

 These variables, in the CM of the two incoming particles 
$\vec p_1 + \vec p_2 = 0$ ({\it i.e.} $|\vec p_1| = |\vec p_2| = p)$ 
and in the simplyfying case of equal mass particles, become
 $$s\, =\, 4( p^2 +  m^2 )\, =\, E_{C.M.}^2 \eqno(I.3.6a)$$
 $$t\, =\, - 2  p^2 ( 1 - cos\, \theta )\eqno(I.3.6b)$$
and
 $$u\, =\, - 2  p^2 ( 1 + cos\, \theta ).\eqno(I.3.6c)$$
 What is important to notice is that the kinematical limits
$$p \geq 0, \quad \quad -1 \leq \cos \, \theta \leq 1,$$ 
translate into the following physical ranges of the Mandelstam 
variables 
 $$s \geq 4 m^2 \eqno(I.3.51a)$$
 $$-4p^2 \leq t \leq 0 \eqno(I.3.51b)$$
 $$-4p^2 \leq u \leq 0 \eqno(I.3.51c)$$
 {\it i.e.}, $s$ is the square of the total CM energy (while $t$ 
and $u$ are definite non-positive and are the momentum 
transfers between the appropriate particles). We shall refer 
to this case as to the {\it s-channel} of reaction\footnote{For  
more details, see Refs. [5,7].} $(I.3.3)$. Just to be specific, 
imagine that the reaction at hand is\footnote{Ignore the 
fact that pions and protons have different masses, this is
 immaterial for our discussion.}
 $$\pi^-(p_1)  \, p(p_2) \rightarrow \pi^-({p'}_1) \, 
 p({p'}_2)\eqno(I.3.52a)$$
 or, by {\it time reversal},
 $$  \pi^-({p'}_1)  \, p({p'}_2) \rightarrow  
\pi^-(p_1) \, p(p_2).\eqno(I.3.52b)$$
 
 If the theory is crossing invariant, {\it the same amplitude} 
which describes the process $(I.3.52a)$ describes also the 
{\it crossed} reaction
$$\bar p(-p_2) \, p({p'}_2)\rightarrow \pi^-(p_1)  \, \pi^+(-{p'}_1) 
\eqno(I.3.52c)$$
 (together with its time reversed) {\it i.e.} $p \bar p$ 
annihilation into a pair $\pi^- \pi^+$. In this case, however, 
if one goes into the proper
 ($p \bar p$) CM, the variable which would be {\it positive definite 
 i.e.} the variable which would be the square of the energy, would 
 now be $t$ while $s$ and $u$ would be {\it definite non-positive} 
 and would represent the momentum transfers between the
appropriate particles. One refers to this second class of reactions 
as to the {\it t-channel} of reaction $(I.3.3)$ {\it i.e.} as the
{\it t-channel} counterpart of $(I.3.52a)$.
 
 Similarly, always because of crossing, {\it the same amplitude} 
 which describes reactions $(I.3.52a, b)$ will describe also 
 $$\pi^-(p_1)  \, \bar p(-{p'}_2) \rightarrow  \pi^-({p'}_1)
\bar p(-p_2)\eqno(I.3.52d)$$
 or, (which is again the same)
 $$\pi^+(-{p'}_1)  \, p(p_2) \rightarrow  \pi^+(-p_1) 
 p({p'}_2).\eqno(I.3.52e)$$
 In this case, if one goes into the appropriate ($\pi^-(p_1)  \, 
\bar p(-{p'}_2) $) CM, the variable which would now be 
 {\it positive definite i.e.} the corresponding squared energy, 
 would be $u$ while $s$ and $t$ would now be {\it definite 
 non-positive} and would represent the momentum transfers of the 
 third (and last) channel which we will call the {\it u-channel}.
 
 Thus, dictated by relativistic invariance, the law of crossing 
 (known also as the {\it substitution law}), relates three
 different reactions (six, actually, taking time reversal into 
 account): the same amplitude (or appropriate combinations of 
amplitudes in the case of particles other than scalars {\it
i.e. } endowed with realistic  quantum numbers like spin, 
isospin etc.) will thus describe the {\it different
 processes} which are obtained one from the other by crossing.
 The key point, however, is that, as we have seen, the kinematical
 domains in which this amplitude, let us call it T(s, t, u), will 
 describe the {\it s-channel} or the {t-channel} or the {\it 
 u-channel} are {\it different and non overlapping.}
 
 What all this means is very simple: if the scattering 
amplitude has decent analyticity properties, the same 
function (or combination of functions) of the Mandelstam 
variables $s, t, u$ will describe the three different physical 
reactions $(I.3.52a, b, d)$ in {\it different domains of these 
 variables}. If, therefore, we knew the analytic properties of
 the scattering amplitude in any given channel, we could, in 
 principle, continue it to the other channels. As a consequence,
from the {\it analyticity properties} of the scattering 
amplitude(s) one can, in principle, perform the appropriate 
analytic continuations and describe all the reactions 
connected by crossing.
 
Everything becomes considerably more complex when we 
discuss {\it real particles, (i.e. } not the hypothetical
scalar identical bosons considered so far). In this case, 
there are a number of different amplitudes in each channel
and they are related by appropriate crossing relations
obtained via a crossing matrix obtained by Clebsch Gordan
coefficients [5, 7]. For example, in the case of reaction(s)
$(I.3.52)$, the {\it s-channel} would a mixture of $I=1/2$ 
and $I=3/2$ amplitudes ($I$ being the isospin) related to 
the $I=0$ and $I=1$ amplitudes which the {\it t-channel} 
would be made off (similarly for the {\it u-channel}). The
technical discussion of this situation is relatively involved
and we will not tackle this issue here any further.

 The key ingredient used to investigate the analyticity 
properties of the amplitudes have been the {\it dispersion 
relations} of which, in particular, the so-called {\it double 
dispersion relation or Mandelstam representation} has 
been probably the most ambitious conjecture. 
 
 This, chopping out all the frills, was the program of {\it 
 Analyticity, Crossing and Unitarity} of the Sixties to 
determine the S-Matrix elements. We will not enter into 
 this discussion which can be considered nowadays largely 
 obsolete; this approach, however, was quite successful in 
many ways and very instructive. In what follows, some of 
these considerations will come back from time to 
time\footnote{Here we will also ignore that $s-u$ crossing 
recommends the use of the quantity $s \, e^{-i {{\pi} \over 
{2}}}$ instead of $s$ as more suited to enforce analyticity. 
In what follows we will never get into problems for which 
this distinction could matter but it is advisable to keep it 
in mind.}.
 
 \bigskip
 \noindent{\bf I.3.7  Rigorous theorems.}
 \bigskip

In this Section we shall discuss in the simplest possible terms
a few of the many rigorous theorems which have left a mark in the 
development of particle physics. Among these, most prominent
are the Froissart bound (Sect. I.3.7.1) and the Pomeranchuk 
theorems  (Sect. I.3.7.2). We shall only give intuitive proofs
of these theorems which can be considered consequence of 
unitarity and analyticity. More complete derivations are
found in the literature [5, 7, 27, 28]. 

\bigskip
\noindent{\it I.3.7.1 The Froissart theorem.}
 \medskip

The Froissart [26] (or, perhaps, more correctly, the 
Froissart-Martin theorem [27]) states that {\it total cross 
sections can not grow faster than $(ln \, s)^2$}, or, more 
quantitatively, as $s \rightarrow \infty$,
$$\sigma_{tot} \le {{\pi} \over {\mu^2}} \, ln^2s\eqno(I.3.53a)$$
where $\mu$ is the pion mass. The coefficient in front of 
$(I.3.53)$ is a very large number ($\approx 60 \, mb$) but 
arguments have been given that, in practical cases, it could be 
considerably smaller. The point remains that this bound puts 
a strict limit to the rate of growth with energy of {\it any} 
total cross section. The catch is, logarithms are very slowly 
growing functions which could be simulated by slowly 
growing functions over a very large interval of variation 
of their variables. This, for instance, is, no doubt, what happens 
with the empirical power form $(I.2.11)$ proposed in Ref. [14]. 
Although it provides a good qualitative (not quantitative,
though) fit to the data up to a very large energy interval 
(Fig. 6), it will eventually lead to violation of unitarity. 
However, while this violation is known to occur at energies 
around $3 TeV$, the actual violation of 
the bound $(I.3.53)$ by $(I.2.11)$ would occur only at 
unattainably and unrealistic large energies.

There are various degrees of sophistication at which the bound
$(I.3.53)$ could be proved\footnote{In its fullest generality it has
been proved  in axiomatic field theory [27a]} and follows from
the {\it positivity} of the imaginary part of the scattering 
amplitude (therefore, in essence, of unitarity) and of a fixed $s$ 
dispersion relation (therefore of analyticity) of the (N times) 
subtracted form
$$T(s, cos \theta) = {{1} \over {\pi}} (cos \theta)^N \int_{x_1}^{\infty}
dx {{A_t(s,x)} \over {x^N \, (x-cos \theta)}}+$$
$${{1} \over {\pi}} (cos \theta)^N \int_{x_2}^{\infty}
dx {{A_u(s,x)} \over {x^N \, (x+cos \theta)}}+{{1} \over {\pi}} 
\sum_{r=o}^{N-1} C_r(s) \, (cos \theta)^r \eqno(I.3.54)$$
where $x_1=1 + {{t_0} \over {2 p^2}}$ and $x_2=1 + {{u_0} \over {2 
p^2}}$\footnote{$t_0$ and $u_0$ are the thresholds of the physical cuts
in the $t-$ and $u-$channels.} and $A_t$, $A_u$ are the absorptive parts 
of the amplitude coming from the $t$ and $u$ channel discontinuities 
respectively (it is over these quantities that positivity is imposed).

Here, we shall limit ourselves to the simplest and most intuitive 
derivation of $(I.3.53)$ warning the reader that this derivation
could be criticized on several grounds but, most of all, because
we assume that the target acts 
as an absorptive Yukawa-like potential {\it i.e.}, neglecting an 
unessential $1/r$ factor, as
$$P(r) \sim g e^{-\chi r}$$
where $r$ is the distance from its mean position.
Next, if for an incoming particle of energy (squared) $s$, the 
probability of interaction is bounded by some power $s^N$ 
(where $N$ is fixed), the (energy dependent) 
probability of an interaction at some distance $r$  
between the center of the target and the 
incident particle with energy (squared) s will satisfy the 
inequality
$$P(s, r) < g \, s^N \, e^{-\chi r}.$$
This probability of interaction will become negligible for 
$r > r_0$ where,
$$r_0 \approx {{N} \over {\chi}} \, ln s.\eqno(I.3.55)$$
We thus have the following upper bound for $\sigma_{tot} \propto r_0^2$
$$\sigma_{tot} \le C ln^2s\eqno(I.3.53b)$$
which is nothing but $(I.3.53a)$ as far as its energy 
dependence is concerned.

For a rigorous derivation of $(I.3.53)$, see Ref.[27a)] where 
the tools used are the positivity of the imaginary parts of 
the partial wave amplitudes {\it i.e.} the unitarity 
condition $(I.3.35)$ rewritten in the form
$$0 \le p |a_{\ell}(p) |^2   \le Im a_{\ell}(p)\le 1\eqno(I.3.56)$$
and where the proof proceeds by showing that the partial 
wave amplitudes decrease exponentially with $\ell$ and 
become negligibe when $s$ is large and
$$\ell > L = Const \sqrt s  \, ln s.\eqno(I.3.57)$$

A wealth of rigorous theorems has been established by a 
number of authors. This is not the place for a detailed
analysis of these results for which the interested reader is 
referred to the existing literature (see Ref.[27], in particular
[27 b)] and [27 c)]). As an illustration, we recall a relation (Ref. 
[27 d)]) which we have encountered earlier (eq. $(I.2.5)$) 
$$b(s) \ge {{\sigma^2_{tot}} \over 
{18 \pi \sigma_{el}}}.\eqno(I.2.5)$$

\bigskip
\noindent{\it I.3.7.2 The Pomeranchuk theorems.}
 \medskip

Back in 1956 it was argued [28] 
that the smallness and apparent vanishing of charge exchange 
reactions at high energies, together with isospin conservation 
suggests a variety of relationships relating the amplitudes 
and the cross sections of different processes which are
collectively known as {\it  Pomeranchuk theorems}.

There are several classes of Pomeranchuk theorems and, 
again, they can be proved in the largest generalitiy making 
use of dispersion relations, crossing and unitarity. For instance, 
writing the ({\it once subtracted}) forward dispersion relation  
giving the {\it dispersive} ({\it i.e.} real) parts of the processes
$AB \rightarrow AB$  and $\bar A B \rightarrow \bar A B$ 
($D(E)$ and $\bar D(E)$ respectively) in terms of their 
{\it absorptive} ({\it i.e.} imaginary) parts ($A(E)$ and 
$\bar A (E)$ respectively) and taking the asymptotic limit 
one gets\footnote{The reason the dispersion relation has to 
be subtracted is to accomodate growing total cross sections 
and the reason it suffices one subtraction is that this growth 
can at most be some power of logarithm of the energy $E$ 
according to the bound $(I.3.53)$ just discussed.}, we get
$$D_{as}(E) \approx E A(E_0) + {{E} \over {4 \pi^2}} \,
 ln[{{E} \over {4 \pi^2}}] [\bar \sigma (\infty) - \sigma 
(\infty)]\eqno(I.3.58a)$$
and
$$\bar D_{as}(E) \approx E \bar A(E_0) + {{E} \over {4 \pi^2}} \, 
ln[{{E} \over {4 \pi^2}}] [\sigma (\infty) - \bar \sigma (\infty)]
\eqno(I.3.58b)$$
where the suffix $as$ means {asymptotic}.
The above relations show that, to avoid that the real part 
grows faster than the imaginary part ({\it i.e.} than the total 
cross section), the asymptotic following equality ({\it 
Pomeranchuk theorem} must hold
$$\lim_{s \rightarrow \infty} \sigma_{tot}(AB) =
\lim_{s \rightarrow \infty} \sigma_{tot}(\bar A B).\eqno(I.3.59)$$

It is perhaps worth showing how another class of Pomeranchuk theorems
can be similarly proved in an intuitive way. 

Consider the ensemble of all $N \bar N $ reactions
\begin{itemize}
\item {1)}  $\bar p p \rightarrow \bar p p$
\item {2)}  $\bar p n \rightarrow \bar p n$
\item {3)}  $\bar p p \rightarrow \bar n n$
\item {4)}  $\bar n n \rightarrow \bar n n$
\item {5)}  $\bar n p \rightarrow \bar n p.$
\end {itemize}
Given that $(p, n)$ are an isospin doublet ($I={{1} \over {2}}$) with 
the third components given by [$+{{1} \over {2}}, -{{1} \over {2}}$] 
while the isodoublet $(\bar p, \bar n)$ have third components
[$-{{1} \over {2}}, +{{1} \over {2}}$], denoting by $I_0$ and $I=1$
the relative amplitudes, the isospin decomposition for the five 
reactions $N \bar N $ listed above is easily found to be
$${{d \sigma_1} \over {d \Omega}} = {{d \sigma_4} \over {d \Omega}} =
{{1} \over {4}} |I_0 \, + \, I_1|^2$$
 $${{d \sigma_2} \over {d \Omega}} = {{d \sigma_5} \over {d \Omega}} =
| I_1|^2,$$
$${{d \sigma_3} \over {d \Omega}} = {{1} \over {4}} |I_1 \, - \, I_0|^2.$$
Notice now that $\bar p p \rightarrow \bar p p$ (or 
$\bar n n \rightarrow \bar n n$) is an elastic reaction to 
which {\it all} annihilation channels contribute (of which
$\bar p p \rightarrow \bar n n$ is one of the many and 
many that open progressively as the energy increases). As a 
consequence, we expect that, at increasing energy, $|I_1 - I_0| 
<< |I_1 + I_0|$ {\it i.e.} that
$$lim_{s \rightarrow \infty} I_1 = I_0.\eqno(I.3.60)$$ 
The consequence of $(I.3.60)$ is that, asymptotically, all 
total cross sections for $N \bar N$ become equal 
$$\sigma_{tot} (\bar p p)=\sigma_{tot}(\bar n p)=
\sigma_{tot}(\bar n n).\eqno(I.3.61)$$
In a perfectly similar (intuitive) way one {\it proves} that,
asymptotically
$$\sigma_{tot} (\pi^+ p)=\sigma_{tot}(\pi^- p)$$
which is a special case of eq. $(I.3.59)$.

 We will not insist further on this subject, let us just quote
a few very general rigorous theorems that one can prove 
concerning the $\bar N N$ reactions. First of all, we
define the crossing even and crossing odd amplitudes
$$F^{\pm} \equiv {{1} \over {2}} [F(p \bar p) 
\pm F(pp)].$$
Using these, it has been proved [27] that at infinite energies
$$\sigma_{tot}(\bar p p) = \sigma_{tot}(p p)\eqno(I.3.62)$$
provided either
\begin {itemize}
\item{i)} $lim_{E \rightarrow \infty} {{F^-} \over {E \, ln E}}=0$ 
~~~[Martin]

or
\item{ii)} $lim_{E \rightarrow \infty} {{Re F^-} \over 
{Im F^- \, ln E}}=0$ ~~~[Khuri-Kinoshita]

or
\item{iii)} $Re F^- \times Im F^- >0$ (for some $E_0$ such 
that $E>E_0$ ~~~[Fischer]
\end {itemize}
where $E= s/2m$ (essentially), is the lab energy.

Many other theorems and results have been obtained in
the past. The interested reader should consult the existing
literature (Ref. [27]).

 \bigskip
\noindent{\bf I.3.8 The introduction of complex angular 
momenta.}
 \bigskip
 
 One of the most interesting outcomes of the analyticity 
program mentioned previously (and the one which has 
been not only the most fruitful from the phenomenological 
point of view but also for its various developments), is the 
continuation to complex angular momenta. This technique 
is fairly old since it was first applied to the study of the 
propagation of electromagnetic waves at the beginning of 
the XX Century [10] but was extended in 1959 to particle 
physics to prove the analyticity of the scattering amplitude 
in potential scattering [9] (see also [29] for more 
details). As we will see, it emerges rather naturally when 
establishing the convergence domain of the scattering 
amplitude to perform a correct analytic continuation to 
arbitrarily large energies. In addition, this technique 
provides the most successful representative of what we 
could call the class of {\it t-channel models}.
 \bigskip
 
 \noindent{\it  I.3.8.1 Convergence of the partial wave 
expansion.}
 \medskip
 
 Our starting point will be the partial wave expansion 
introduced earlier $(I.3.28)$\footnote {We stick, for 
simplicity, to the fictitious case of identical spinless
particles to avoid unnecessary formal complications. 
Similarly, as long as possible, we limit our considerations 
to the two-body elastic reaction $(I.3.3)$.}.
  $$f(p,\theta) =  \sum_{\ell =0}^{\infty} \,
 (2 \ell +1) \, a_{\ell }(p) P_{\ell }(cos \theta).\eqno(I.3.28)$$
 Like every representation of a physical quantity, also 
$(I.3.28)$ (which is, originally, well defined in the physical 
domain of the s-channel $(I.3.51)$) will have a certain 
domain of existence. The questions that arise at this point 
are the following: {\it does the expression $(I.3.28)$ converge 
in a domain of the complex $s$, $t$ and $u$ variables larger 
than just $(I.3.51)$? if so, does it converge in a domain large 
enough to contain also the physical domains of the $t$ and 
$u$ channels? If yes, can we use the same representation to 
study all the physical reactions related one to the other by 
crossing? if so, can we continue it to asymptotic values of 
one (or more) of the Mandelstam variables? if not, can we 
find different representations that allow us to connect these 
various channels one to the other? Can we, ultimately, find 
the proper asymptotic behaviors of $f(s,t)$ in any given 
channel?}
 
 Amazing as it may seem, answering these abstract-seeming 
 questions and using only general properties, one derives 
 enough physical insight to describe the asymptotic behavior 
 of high energy data.
 
 The convergence domain of $(I.3.28)$ in the {\it complex} 
 $\theta$ plane ($\theta = \theta_1 + i \theta_2$), is obtained 
 using the asymptotic expansion of $P_{\ell }(cos\, \theta)$ 
 when $\ell $ is {\it real} and $ \rightarrow \infty$ 
 $$P_{\ell }(cos\, \theta) \approx O[e^{\ell  |\theta_2|}].$$
 As a consequence, the series $(I.3.28)$ converges only if, as 
 $\ell \rightarrow \infty$
 $$a_{\ell }(p) \approx  e^{-\ell  \, \alpha(p)}.\eqno(I.3.63)$$
 If $(I.3.63)$ is satisfied, convergence in the complex $\theta$ plane is 
 insured in {\it a strip parallel to and symmetric with respect 
 to the imaginary $\theta$ axis of width $\alpha(p)$}
 $$|\theta_2| \leq \alpha(p).$$
 Setting $\chi = ch\ \alpha(p)$ (which is always $>1$),  the 
 corresponding convergence domain of the partial wave 
 expansion $(I.3.28)$ in the complex $cos\, \theta = x\, +\, i\, y$ 
 plane, is
 $${{x^2} \over {\chi^2}} +\, {{y^2} \over{\chi^2 -1}} =\, 1.
\eqno(I.3.64a)$$
 This domain is an ellipse of foci $\pm 1$ with semiaxes 
 determined by $\alpha(p)$ known in physics as the {\it 
 Lehmann ellipse}.
 
 The above result implies that the usual partial wave 
expansion converges in a domain which, although larger 
than the simple physical domain $-1 \leq cos\, \theta 
\leq 1$ (to which it reduces if $\alpha(p) \rightarrow 0$), 
never allows to extend$(I.3.28)$ to arbitrarily large values of the 
complex variable $cos\, \theta$. Translated into the 
 language of the Mandelstam variables $(I.3.4)$, this 
means that, for any {\it finite $s$}, the expansion 
$(I.3.28)$ converges in a {\it finite} domain of the 
$|t|$ and $|u|$ variables. Thus, we could not continue the 
amplitude to regions where any of the Mandelstam variables 
$t$ or $u$ can become arbitrarily large\footnote {Owing
 to the constraint $(I.3.5)$, if any of the Mandelstam 
variables becomes unbounded, at least another variable 
must simultaneously do so.}. Thus, had we planned to 
continue our representation to either the $t$ or the $u$ 
channel and let the corresponding energy variable ($t$ or 
$u$ as the case may be) become arbitrarily large, we could 
not do it using the expansion $(I.3.28)$ because, as we just 
found out, this does not allow a continuation to arbitrarily 
large values of either $t$ (or $u$). 
 
 It is straightforward to see how the situation would change 
if the expansion $(I.3.28)$ was not over {\it purely real} 
but, rather, over {\it purely imaginary} values of  $\ell $. 
In this case, in fact, provided 
 $$a_{|\ell |}(p) \approx e^{- |\ell | \beta(p)},\eqno(I.3.63b)$$
 as $\ell \rightarrow   i \infty$, convergence would be 
insured in  {\it in a strip parallel to and symmetric with 
respect to the real axis in the complex $\theta$ plane, i.e.} 
in $|\theta_1 | \leq \beta(p)$. Setting, accordingly, 
$cos\, \beta(p) = \phi$ (which is always $\leq 1$), the 
 convergence domain in the {\it complex} $cos\, \theta =
\, x +\, i\, y$ would be
 $${{x^2} \over {\phi^2}} -\, {{y^2} \over{1-\phi^2}} =\, 
1.\eqno(I.3.64b)$$
 Contrary to the previous case, $(I.3.64b)$ is an {\it open} 
domain (a hyperbola with foci $\pm 1$) and convergence 
is insured outside one of its halves. This hyperbola, in addition, 
overlaps in part with the Lehmann ellipse which guarantees 
that, if we can continue $(I.3.28)$ to imaginary $\ell $ values, 
the new expansion represents the {\it same analytic function
i.e. the same scattering amplitude} in a domain where $|t|$ 
(and/or $u$) can become arbitrarily large. 
 
 The problem, therefore, is reduced to continuing $(I.3.28)$ to 
 imaginary (or complex) values of $\ell $ even though, obviously, 
only {\it real} angular momenta have direct physical meaning.
 
 \bigskip
 \noindent{\it  I.3.8.2 Continuation to complex angular momenta}
 \medskip
 
 The program outlined at the end of the previous Section requires 
that sufficient properties of the partial wave amplitudes be obtained
 to continue $(I.3.28)$ to complex $\ell $ values. 
 
 This can be done explicitly in ordinary (nonrelativistic) Quantum 
 Mechanics where, owing to the properties of the Schr\"odinger 
 equation, for suitable classes of the interaction potential we can
 gather sufficient information to perform an explicit analytic 
 continuation.
 
 We will not go into these mathematical developments but simply 
 refer the interested reader to Refs. [9] and [29] where the problem 
 is analyzed in great details for the case of generalized Yukawa 
 potentials (see also [5,7]. For these classes of interactions one can prove:
 \begin {itemize}
 \item {i)} $(I.3.63a)$ holds with $ch\ \alpha = 
1 +{\mu^2 \over 2 p^2}$
 where $\mu$ is the lowest mass that can be exchanged. This 
allows the analytic continuation of the scattering amplitude 
$f(p, \theta)$ to complex  $\ell $ values since one can write 
 $$f(p,\theta) = - (1/2p) \oint_C d\lambda 
 {{\lambda} \over {cos\, \pi\, \lambda}}
 \, P_{\lambda -1/2}(-cos\, \theta)\, [S(\lambda, p) - 1]
\eqno(I.3.65a)$$
 where $\lambda = \ell +1/2$ and $S(\lambda, p)$ is the 
analytic continuation of $S_{\ell}(p)$. The contour of 
integration is shown in Fig.9.
 \item {ii)} The analytic continuation of $S(\lambda, p)$ in 
$(I.3.65a)$  is unique.
 \item {iii)} For $Re \lambda > 0$, the only singularities of 
 $S(\lambda, p)$ are a {\it finite} number N of {\it simple poles} 
which lie in the upper quadrant of the complex $\lambda$ plane. 
These poles are called {\it Regge poles} in the literature. We will 
denote by $\lambda_n(p) = \ell_n(p) + 1/2 $ ($n = 1, 2, ...N$ the 
positions of these poles (which will, in general, be functions of $p$) 
and by $S_n(p)$ their residues.
 \item{iv)} There exists one pole whose real part is the largest; 
let $\alpha(s)$ be its real part, {\it i.e.}
 $$\alpha(s) = Max \{Re \lambda_n(p)\}.\eqno(I.3.66)$$
 \item{v)} The integrand in $(I.3.65a)$ tends to zero faster than 
 $\lambda$ along any direction in the complex $\lambda$ plane 
 when Re $\lambda >0$.
 \end {itemize}
 \medskip
 
 Fig.9. {\it The contour of integration of eq. $(I.3.65a)$.}
 \medskip
 
 Using the above properties, general theorems on analytic 
 functions allow us to open the contour of integration in Fig. 9 
as shown in Fig. 10. This doing, we pick up the contributions 
of all singularities in the complex angular momentum plane (in 
our case simple poles). When we let the radius of integration of 
such a contour go off to $\infty$, this term vanishes and we 
are left with an {\it integral along the imaginary $\lambda$ 
axis} plus the residues of the N simple poles in the complex 
$\lambda$ plane, {\it i.e.}
 $$f(s,t)=-(1/2p) \int_{-i \infty}^{i \infty}  d\lambda \, \lambda \, 
 P_{\lambda -1/2}(-cos\, \theta) \, {[e^{2i\delta(\lambda, p)} \, -\, 1] 
 \over {cos\, \pi \lambda}}$$
 $$ + i (\pi /p) \sum_{n=1}^N  S_n(p) P_{\ell _n(p)}(- cos\, \theta) \, 
 {{2 \ell_n(p) + 1}\over {sin\, \pi \ell_n(p)}}.\eqno(I.3.65b)$$ 
 
 \medskip
 Fig. 10. {\it Opening of the contour of integration to obtain eq. 
$(I.3.65b)$.}
 \medskip
 
 The above expression was obtained long ago [10] and is known 
as the {\it Watson Sommerfeld transform.} It was rederived by 
Regge [9] as a new representation of the scattering amplitude 
through the explicit analytic continuation outlined above of the 
partial wave expansion to a domain within which we can allow 
$t$ (and $u$) to go to $\infty$. 
 
 \bigskip
 \noindent{\it  I.3.8.3 Asymptotic form of the 
Watson-Sommerfeld representation.}
 \medskip
 
 The crucial point, we recall, is that, aside from the finite sum 
(which poses no problems of convergence), the integral in 
$(I.3.65b)$, as a function of the complex $\theta$ variable, is 
well defined in an open domain. This implies that, inside its 
convergence domain, we can take $|cos\, \theta| \rightarrow 
\infty$ while $s$ remains constant. Keeping only the leading 
contribution, we find 
 $$f(s,t) \approx\, \beta(s)\,  {{|cos\, \theta|^{\alpha(s)}} \over
 {sin\, \pi {\alpha(s)}}}\eqno(I.3.67a)$$
 where we have lumped into $\beta(s)$ all the s-dependent
 factors coming from the leading contribution in $(I.3.65b)$ 
and where $\alpha$ was defined previously $(I.3.66)$.
 
 It will always be understood that $(I.3.67a)$ holds up to 
logaritmic factors\footnote {Here we are simplifying things 
considerably....}.
 
 The obvious objection to all we did so far is that 
in potential scattering, $|cos\, \theta|$ cannot 
exceed unity so that this whole exercise of letting it become 
unrestrictedly large may appear rather academic.
 
 \bigskip
 \noindent{\it  I.3.8.4 Asymptotic expansion in the relativistic 
case.}
 \medskip
 
 The relativistic case differs from ordinary Quantum Mechanics 
since, as discussed earlier, one has three coupled channels (which 
we have called $s$, $t$ and $u$). We will assume that {\it the 
developments of Section I.3.8.2 hold}\footnote {This is, {\it a 
priori}, far from obvious since the {\it explicit analytic 
continuation} of $(I.3.28)$ leading to $(I.3.65b)$ and, then, to 
its asymptotic expansion $(I.3.67a)$ was derived under the 
specific condition that the interaction be (any) superposition 
of Yukawa potentials and it is far from obvious that this 
should be true in the relativistic case. More than likely, in 
fact, it doesn't; there is nowadays sufficient evidence that the 
relativistic case is indeed considerably more complicate, (for 
instance other singularities, like cuts are present beside poles 
in the complex angular momentum variable).} {\it i.e.} that the 
analytic continuation to complex angular momenta can be 
performed in the relativistic case and that properties {\it i) - 
v)} of Section I.3.8.2 are still valid.
 
 In this case, recalling the relation $(I.3.6b)$ between $t$ and 
$cos \theta$, the limit $|cos \theta| \rightarrow \infty$ can
 be viewed as the limit $|t| \rightarrow \infty$ and $(I.3.67a)$ can 
 be reinterpreted as the asymptotic behavior when $|t|, |u| \rightarrow 
 \infty$ {\it at fixed $s$}\footnote {This brilliant idea is originally 
 due to G. Chew and collaborators [7] and it is what converts this 
 mathematical trick into a useful physical approach.}. In this case, 
 we rewrite $(I.3.67a)$ as the limit $|t| \rightarrow \infty$ ({\it at 
 fixed $s$}) of the scattering amplitude {\it i.e.}
 $$\lim_{t \rightarrow \infty} T(s,t,u) \sim \, \beta(s) {t^{\alpha(s)}
 \over {sin\, \pi \alpha(s)}}\eqno(I.3.67b)$$
 (up to subasymptotic contributions).
 
 In $(I.3.67b)$, $T(s,t,u)$ is the same invariant scattering 
amplitude used previously.
 
 Eq. $(I.3.67b)$ is a fundamental result which can be 
reinterpreted as follows: {\it The leading singularity in the 
complex angular momentum plane ({\it i.e.} the singularity 
with the largest real part) in the scattering amplitude in a 
given channel determines the asymptotic behavior in the 
crossed channels}. Thus, eq.$(I.3.67b)$ means that {\it the 
asymptotic behavior $t \rightarrow \infty$ at fixed $s$ is 
determined by the rightmost complex angular 
momentum singularity in the crossed s-channel}. Similarly,
{\it the asymptotic behavior $s \rightarrow \infty$ at 
fixed $t$ will be determined by the rightmost complex angular 
momentum singularity in the crossed t-channel}. It was shown 
by Gribov that, when all complications coming from having three 
channels are properly taken into account, one obtains for the 
invariant scattering amplitude the following asymptotic 
behavior as {\it $s$ tends to $\infty$ at constant $t$ }
 $$T(s,t,u) \lim_{|s| \rightarrow \infty, |t|= const} \sim \beta(t)
 \, s^{\alpha(t)} {{1 + \xi \, e^{-i \pi \alpha(t)}} \over{ sin\, \pi 
 \alpha(t)}}\eqno(I.3.68)$$
 (which is nothing but eq $(I.3.67b)$ properly readjusted to 
take relativistic complications into account). $\beta(t)$ takes 
the name of {\it residue function} which we know essentially 
nothing about (may even be a complex quantity); $\xi$ can be 
either $+1$ or $-1$ and is known as {\it signature} (and arises 
precisely because of having three channels inherently 
linked together); its role will now be briefly discussed
together with the implications of this result for understanding 
high energy data\footnote {Notice that, according to $(I.3.4)$, 
increasing $s$ (to positive values),
 means increasing the CM momentum $p$; in this 
case $t$ and $u$ are definite negative. If, however, $t$ has to 
remain constants, as $p$ increases, the scattering angle $\theta$ 
must become smaller and smaller {\it i.e.} closer and 
closer to the forward direction. This is why most of the events at 
high energy are concentrated near the forward direction. This, 
in turn, is the motivation for the so-called {\it eikonal}
(from the Greek $\epsilon \iota \kappa \omega \nu$ for 
{\it image}) or {\it impact parameter} representation.}.
 
Again a word of caution; the reinterpretation of $(I.3.67b)$ as
the asymptotic behavior in the crossed channel is a very
brilliant idea. It should, however, be kept in mind that this
means using crossing {\it after} having taken an asymptotic 
limit. It is not at all obvious that these two
operations are interchangeable.

 We already mentioned it but let us stress it again: eq.$(I.3.67b)$ 
(or $(I.3.68)$, more properly), represents only the {\it dominant} 
contribution to the asymptotic behavior of T(s,t,u) (and,
in addition, we have disregarded logarithmic corrections). 
At subasymptotic energies, other contributions in $(I.3.65b)$ 
may be non-negligible and we may then have to retain a 
few additional next-to-leading terms. Before we discuss 
(very briefly) this point, however, we need to interpret 
physically the results obtained so far.

 \bigskip
 \noindent{\it  I.3.8.5 Regge trajectories}

 \medskip
 
 In order to understand the physical consequences of 
$(I.3.68)$, let us notice that its denominator vanishes 
whenever $\alpha(t)$ crosses an integer. For $t$ physical, 
a singularity in the complex angular momentum plane is 
in general complex and $\alpha(t)$ will become integer 
only at some non-physical (complex) value of $t$. 
 
 Suppose, then, that for some {\it real} $t_0$, we have
 $$\alpha(t_0) = \ell  + \epsilon(t_0) + i \eta(t_0)$$
 and that $\epsilon(t_0)$ and $\eta(t_0)$ are {\it small}
 (compared with unity). In this case, the denominator in eq.
$(I.3.68)$. will take the form
 $$\propto {1\over{t-t_0 -i \Gamma}}$$ 
 (where $\Gamma$ is proportional to $Im \, \alpha(t_0) = 
\eta(t_0)$). This form has the typical structure of a
Breit-Wigner resonant term. This result will hold whenever 
the real part of $\alpha(t)$ is close to an integer. Thus, 
if $\alpha(t)$ grows with $|t|$, a pole in the complex 
angular momentum plane interpolates an 
{\it a priori} arbitrary number of resonances of increasing 
angular momentum ({\it i.e. spin}) $\ell $. When such a 
situation occurs for real and positive values of $t$ ({\it i.e. 
when $t$ is a mass squared)}, the denominator in $(I.3.68)$
blows up every time $\alpha(t)$ crosses an integer.
In this case we say that we have a family or a {\it trajectory}
interpolating many bound states (or resonances) whose spin 
increases by one unit at the time. In the case of $(I.3.68)$, 
however, owing to the signature factor $( {1 + \xi \, e^{-i \pi 
\alpha(t)}})$ also the numerator vanishes at {\it every other 
integer} value of $\ell $. As a consequence, a trajectory with 
{\it positive signature} ($\xi =1$) interpolates between {\it 
even angular momentum} resonances whereas a {\it negative 
signature} ($\xi =-1$) trajectory interpolates between {\it odd 
 angular momentum} resonances.
 
A simple way to visualize these trajectories, is to expand 
$\alpha(t)$ in power series around $t=0$. In this case, for 
$t$ small enough, we can write
 $$\alpha(t) = \alpha_0 + \alpha' t.\eqno(I.3.69).$$
Quite unexpectedly, when interpolating resonances with the 
same quantum numbers (other than the spin), one finds that 
the expansion $(I.3.69)$ which was {\it a priori} justified only 
for {\it small $t$'s} holds actually for rather large values of $t$
(up to several units of $GeV^2$). In addition, this is true for
both mesonic and fermionic trajectories ({\it i.e.} trajectories
which interpolate between integer and half-integer spin
respectively). The situation is exemplified in Fig. 11 where the 
leading {\it mesonic} trajectories\footnote{Each trajectory has the
quantum numbers, isospin, C and G parity, strangeness etc. of
the first recurrence of which takes the name.More specifically,
we have for \pom (and $f_2$) $[P=+, C=+, G=+, I=0, \xi=+]$, for
$\rho$ $ [P=-, C=-, G=+, I=1, \xi=-]$, for $\omega$ $[P=-, C=-, G=-, I=0, 
\xi=-]$ and for $a_2$ $[P=+, C=+, G=-, I=1, \xi=+]$.} the $\rho$, the 
$f_2$, the $a_2$ and the $\omega$ are shown (they are all 
superimposed) together with the Pomeron (much flatter). 
The main fermionic trajectories, not shown, would have a very 
similar slope ($\alpha'$) but a considerably lower intercept 
$\alpha_0$ (at $t=0$). All other trajectories not shown in Fig. 11 
(for instance those interpolating strange particles) also have 
lower intercepts. This, notice, will be extremely relevant when 
selecting the trajectories to retain as the leading (or subleading) 
contributions to a given reaction. Eq. $(I.3.68)$, in fact, shows 
clearly that the larger the intercept, the more important the 
contribution will be as $s$ increases.
\medskip
 
Fig. 11. {\it The main mesonic trajectories, the $\rho$, the $f_2$, 
the $a_2$, the $\omega$ (all superimposed) and, much flatter
but with higher intercept at $t=0$, the Pomeron.} 
\medskip
 
 Many comments are in order.  First, notice the 
unexpectedly large interval of masses for which 
the trajectories are basically linear. Next, for all 
trajectories other than the Pomeron, the slope 
$\alpha'$ is essentially universal (and of order 
$ 1 (GeV/c^2)^{-1}$). This universal value is also 
called the {\it string tension} in the language of 
string theories. The third point is that all (mesonic) 
trajectories are essentially {\it degenerate} in that 
they all lie one on top of the other (see Fig.11). This 
property is, in the literature, called {\it exchange 
degeneracy}. Thus, even though, in principle, each 
trajectory interpolates among {\it even} or {\it odd} 
angular momenta according to whether it has positive 
or negative signature, in practice, $\xi=+1$ trajectories 
(the $f_2$ and the $a_2$) are undistinguishable from 
$\xi=-1$ trajectories (the $\rho$ and the $\omega$).
 
The last and perhaps most important comment concerns 
the one trajectory which appears so markedly different 
from all the others, the Pomeron. The latter, (to be briefly 
discussed later on from the point of view of being the 
dominant trajectory for diffraction), {\it a priori} is 
expected to behave differently because its recurrences 
should be {\it glueballs} rather than conventional 
resonances. In Fig. 11, a tentative candidate shows 
that the trajectory is indeed much flatter than the 
others. This flatteness, incidentally, is perfectly
compatible with the determination of the Pomeron
slope obtained fitting elastic high energy scattering 
data. These scattering data will also explain why 
the Pomeron trajectory is expected to have the larger 
$t=0$ intercept $\alpha_0 =1$. This is considerably 
higher than those of the other trajectories which, 
by contrast, are determined by the first resonance 
we encounter on the trajectory\footnote {A careful 
analysis of the various mesonic trajectories would 
reveal that the exchange degeneracy displayed in 
Fig. 11 is only approximate; the $\rho$ intercept, for 
instance, is slightly higher than that of the $\omega$. 
This reflects the fact that the $\rho$ mass is slightly 
lower than that of the $\omega$ (and similarly for 
$f_2$ and $a_2$). As a consequence, one often prefers 
to talk of $\rho - f_2$ and $\omega - a_2$ degeneracies.}. 
Roughly speaking, from Fig. 11 we see that all the 
leading mesonic trajectories have intercept $\alpha_0 
\approx 1/2$). 
 
Going back to the general discussion about Regge 
trajectories, the important message to learn is that 
the behavior $(I.3.68)$ is due to the {\it exchange 
of a family of resonances in the crossed channel}. 
Understanding this point  is fundamental since it 
clarifies the nature and the role of a singularity in 
the complex angular momentum. Notice, incidentally, 
that this is in line with the original message from the 
Yukawa conjecture about the existence of the meson. 
The novelty, is that this r\^ole of dominance of the 
exchange in the crossed channels is now transferred 
to the asymptotic behavior. Summarizing our findings, 
we can state that {\it the asymptotic behavior in a given 
channel is provided by the exchange of an interpolation 
of resonances in the crossed channels}. These, in turn, 
build up a singularity in the complex angular momentum. 
This would be an extraordinarily powerful result if 
unrestricdedly correct. As it turns out, however, the 
situation is tremendously more complex; for instance, 
as we have already mentioned, the {\it dominant} 
asymptotic trajectory (the Pomeron) escapes the general 
rule and more complicated singularities than simple poles 
are presumably also present in the complex angular 
momentum plane such as, for instance, the cuts induced 
by unitarity corrections. Only the advent of QCD, has made 
this issue somewhat more clear\footnote {Qualitatively, 
are the gluons which make up the Pomeron in QCD.} but 
we are still a long way even from a semi-quantitative 
understanding of the Pomeron. 
 
Trying to pin down the very physical meaning of a complex 
angular momentum trajectory, we saw that it corresponds to 
the interpolation of families of resonances which have all the 
same quantum numbers but spin. Thus, different reactions will, 
in general, receive different contribution from the various 
trajectories, (or, stated the other way around, the various 
trajectories contribute to different reactions in different 
combinations. These, according to our general discussion will 
depend on the quantum numbers that the given reaction 
exhibits in the {\it crossed channels}) but the precise ways
these combinations are formed will depend on the crossing 
matrices relating the various crossed channels. Elastic and 
diffractive processes, however, share one common feature, {\it 
the same trajectory} with the quantum numbers of the vacuum 
gives the dominant contribution. It is this trajectory, known as 
the {\it Pomeron} which has less simple an origin than the 
other trajectories: it does not result from interpolating 
ordinary resonances in the crossed channel but from the 
interpolation of glueballs. 

A different way of looking at the Pomeron follows from the
fact that one could also argue that the origin of the Pomeron 
must be related to the unitarity constraint. Although, 
unfortunately, the relative argument can only be made 
qualitatively, it is of sufficient importance that we discuss
it briefly here. We postpone, however, this point to after 
examining how Regge poles account for high energy data.
 
 \bigskip
\noindent{\it  I.3.8.6 Regge poles and high energy data.}
\medskip

We can not discuss in details how combining a handful of Regge 
trajectories one can reproduce the bulk of all high energy 
elastic and diffractive reactions. For this, we refer the reader
to the literature already quoted (Refs. [5-7]). 

We recall (Section I.2.7) that high energy hadronic data are 
concentrated near the forward direction\footnote{Let us 
remark that this is exactly why the 
complex angular momentum analysis is so very relevant for 
the discussion of high energy hadronic reactions. As we have 
seen, in fact, the proper asymptotic behavior of a complex 
angular momentum pole obtains when $s \rightarrow \infty$ 
at $t-fixed$. Remembering eq.$(I.3.6b)$, however, the only way 
$t$ can remain constant as the energy (therefore $p$) increases, 
is that $cos \theta \, \rightarrow 1$ ( {\it i.e.} 
$\theta \rightarrow 0$).}.

As an example, we give the various combinations that 
provide the contributions of the leading Regge trajectories 
to the most important elastic reactions, those induced on
nucleons by an initial particle and by its antiparticle. 
$$\pi^-p = \pom + f_2 + \rho  \eqno(I.3.70a)$$
$$\pi^+p = \pom + f_2 - \rho  \eqno(I.3.70b)$$
$$K^-p = \pom + f_2 + \rho + a_2 + \omega \eqno(I.3.70c)$$
$$K^+p = \pom + f_2 - \rho + a_2 - \omega \eqno(I.3.70d)$$
$$pp = \pom + f_2 - \rho + a_2 - \omega \eqno(I.3.70e)$$
$$p \bar p = \pom + f_2 + \rho + a_2 + \omega \eqno(I.3.70f)$$
$$np = \pom + f_2 + \rho - a_2 - \omega. \eqno(I.3.70g)$$

The above formulae follow from what has been called the 
property of {\it line reversal}. The two processes 
$a+b \rightarrow c+d$ and $a+\bar c \rightarrow \bar b+d$ 
have the same quantum numbers in the crossed $t$-channel 
so that they can exchange and receive contributions from
the same Regge poles; there will just be a change of sign 
for negative signature trajectories when reversing a boson 
line or a change of sign for the $C-$negative terms when 
reversing a nucleon line. 

One should notice how much simpler in terms of Regge 
poles are the charge exchange reactions
since the dominant term, the Pomeron, does not 
contribute to them. In particular,
if one neglects any other singularity (like branch cuts), 
from $(I.3.70)$ we find that the following differences 
of total cross sections should become asymptotically 
equal
$$[\sigma (K^- p)-\sigma (K^+ p)] \approx  2(\omega + 
\rho)\eqno(I.3.71a)$$
$$[\sigma (\bar p p)-\sigma (p p)] \approx  2(\omega + 
\rho)\eqno(I.3.71b)$$
$$[\sigma (p n)-\sigma (p p)] \approx  2(\rho - a_2).
\eqno(I.3.71c)$$
and the cleanest reaction is
$$[\sigma (\pi^- p)-\sigma (\pi ^+ p)] \approx 2 \rho 
\eqno(I.3.71d)$$
which, as we will see, leads to very interesting and 
peculiar suggestions (the residues can, of course,
be different for the various couplings to $\pi$ or
$K$ or to nucleons).

It is interesting to compare the above relations $(I.3.71)$
to those that follow from $SU(6)$ invariance in the forward
direction as $s \rightarrow \infty$
$${{1} \over {2}} [\sigma (K^- p)-\sigma (K^+ p)] \approx 
[\sigma (K^0 p)-\sigma (\bar K^0 p)] \approx 
[\sigma (\pi^- p)-\sigma (\pi ^+ p)].$$
Not only one finds $\alpha_{\rho}(0) =\alpha_{\omega}(0)$
(in agreement with exchange degeneracy) but one finds also
relations between the various coupling constants. 
The experimental comparison is quite good [7a]. Assuming isospin 
invariance, the following prediction follows from the previous
relation
$${{1} \over {2}} [\sigma (K^- p)-\sigma (K^+ p)] \approx 
[\sigma (K^- n)-\sigma (K^+ n)].$$ 

Assuming that the only singularities in the complex 
angular momentum plane are simple poles and 
neglecting interference terms, if one uses the 
combinations $(I.3.70)$ for the amplitudes, each 
individual Regge pole (let $a_i(t)$ be the trajectory) 
will contribute a term proportional to
$$ s^{2 (\alpha_i(t) -1)} = e^{b_i(s) t} \, s^{2 (\alpha_i(0)-1)}$$
where the slope $b_i(s)$ associated to the Regge pole in 
question
$$b_i(s) = 2 \, \alpha'_i(0) \, ln s\eqno(I.3.72)$$
is energy dependent and leads to logaritmic shrinking 
of the forward peak in agreement with the trend of the 
data (Fig.3). Before we discuss the energy behavior of a 
Regge pole, let us see what happens for total cross sections.

These equalities follow also from 
$SU(6)$ invariance.

Each Regge pole contributes a term 
$$  {s^{ (\alpha_0 -1)} \over {sin \pi \alpha_0}}  \times 
Im [\beta(0) (1 +\xi \, e^{-i \pi \alpha_0})]\eqno(I.3.73)$$
to the total cross section $\sigma_{tot}$. If we recall that
total cross sections stay nearly constant as $s \rightarrow 
\infty$, we see that this demands the leading trajectory
to have $\alpha_0 \approx 1.$
The only trajectory for which we are at liberty to assume
an intercept close to unity is the Pomeron; all the others 
have their intercept determined by the recurrences on the 
trajectories and we have already stated that they all have
$$\alpha_0 \approx {{1} \over {2}}.\eqno(I.3.74)$$
Thus, we will assume from now on that the Pomeron 
intercept is $=1$. In this case, however, owing to the fact 
that $\xi =+1$ for the Pomeron, we must also demand
the Pomeron residue to be purely (or predominantly)
imaginary. With these choices, however, not only the
total cross sections but also the optical point of elastic
angular distributions ({\it i.e.} ${{d \sigma} \over {dt}}|_{t=0}$)
would remain strictly constant with energy whereas, as 
discussed in Section I.2.7 they both seem to increase 
logarithmically. The usually accepted way out of this 
dilemma is that poles are not the only singularities in 
the complex angular momenta but more complicated 
singularities can be present. As a matter of fact, it was 
Mandelstam [11] the first to prove that unitarity 
corrections to a simple pole in the complex angular 
momentum plane generate a cut and that this grows
like some power of $ln s$. Since the maximum growth 
allowed by the Froissart-Martin bound is indeed $\propto 
ln^2 \, s$, everything seems to be quite consistent.

As an example of how things work, in Fig. 3 we have shown how 
the combination $(I.3.70e)$ of the mesonic trajectories plotted in 
Fig. 11 reproduce the high energy $pp$ data near the {\it forward}, 
({\it i.e.} $t \approx 0$) direction. Similarly, in Fig. 12 we show how 
the few basic {\it fermionic} trajectories would reproduce the data 
in the {\it backward}, ({\it i.e.} $u \approx 0$) direction.
\medskip

Fig. 12. {\it $\pi^{+} p$ elastic backward high energy data as 
reproduced by the leading fermionic Regge trajectories.}
 \medskip
 
Let us just add that this seemingly very good description 
of the data displayed in Figs.3 and 12, fails when one tries 
to incorporate polarization data in the fit. Even though 
they are just a small fraction of the data, this proves 
that the agreement is just only qualitative. To make it more 
quantitative, one must, once again, enlarge the picture to 
include other singularities in the complex angular momentum 
plane (cuts) besides the poles. This brings about the appearance 
of logarithms which are indeed a prominent feature of high 
energy physics.

 \bigskip
\noindent {\it I.3.8.7 Duality}
\medskip

The notion of duality in high energy physics was 
introduced by Dolen, Horn and Schmidt [30] who 
showed how, in some particularly simple case, {\it 
the superposition of direct channel resonances is 
averaged by the corresponding Reggeon(s) exhanged 
in the crossed channel}
$$<\sum_i Res^{(s)}_i(s,t)> = R^{(t)}(s,t)\eqno(I.3.75)$$
where we denote by $R^{(t)}(s,t)$ the Reggeon(s) with
the proper quantum numbers.

The key to the proof of $(I.3.75)$ is the choice of the 
appropriate rection. Ideal, in this sense is the pion - nucleon 
charge exchange process $\pi^- p \rightarrow \pi^0 n$ whose
(direct channel) isospin content is the same as that of the
difference $\pi^- p - \pi^+ p$ and which, according to 
$(I.3.71d)$ has just $\rho$ exchange in the crossed channel.
The result of this comparison is exhibited
in Fig. 13 and shows how nicely the $\rho$ trajectory 
interpolates among all the resonances that contribute to
$\sigma_{tot}(\pi^- p) - \sigma_{tot}(\pi^+ p)$.

\medskip
Fig. 13 {\it The comparison of direct channel resonances and 
crossed $rho$-Reggeon for $\pi^- p - \pi^+ p$.}
\medskip

Several other examples of {\it duality} in the sense used 
here have been worked out but none as clean and convincing as the 
case of $\pi^- p - \pi^+ p$. We will just mention, without any 
elaboration, that the word duality in high energy physics has
been extended to cover the case of quark-diagrams and, in due
time, has led to the birth of an entirely new field known as
{\it string theory}.

 \bigskip
\noindent{\it  I.3.8.8 Unitarity constraints}
\medskip

Very simple conditions on the parameters of the trajectories,
intercept $\alpha_0$ and slope $\alpha'$ follow by combining 
the asymptotic form $(I.3.68)$ with the (two-body) unitarity 
equation $(I.3.47a)$. We rewrite the two-body contribution
$E(s,t)$ in the asymptotic limit $s \rightarrow \infty$ as
$$E(s,t) \, = \, {{1} \over {64 \pi^2 }} \int d{\Omega _{k_1}} \, 
T(s, t_1) T^*(s, t_2)\eqno(I.3.76) $$
whereas, recall, about the inelastic part I(s,t) we know only
that $I(s,t) > 0$. This, however, is sufficient to prove that the 
parameters must satisfy rather constraining conditions. One 
finds
$$\alpha_0 \leq 1\eqno(I.3.77)$$
and
$$\alpha' >0.\eqno(I.3.78)$$
If one of the previous two relations is not obeyed, $I(s,t)$ 
could become negative (these conditions have been derived 
first by Froissart [31]). Needless to say, the above constraints 
apply only to the leading complex angular momentum 
singularity {\it i.e.} to the Pomeron.

 \bigskip
\noindent{\it  I.3.8.9 The Pomeron and the data}
\medskip

In spite of their simplicity, conditions $(I.3.77, 78)$ derived 
from unitarity on the parameters of the leading complex 
angular momentum trajectory (the Pomeron) are extremely
useful.

Let us recall that total cross sections are, experimentally,
growing very mildly (Sction I.2.7) and, theoretically, 
cannot grow faster than $ln^2 \, s$ (Froissart bound, eq. 
$(I.3.53)$). As a consequence, the Pomeron intercept 
$\alpha_0$ can, at most, saturate the unitarity bound 
$(I.3.77)$ {\it i.e.}
$$\alpha_0\, =\, 1.\eqno(I.3.79)$$

Next, to kill unwanted physical singularities, the Pomeron
must have positive signature
$$\xi\, =\, 1.\eqno(I.3.80)$$

Finally, being the total cross section proportional to the 
imaginary part of the (forward) amplitude, the residue 
function must be essentially imaginary. From the data,
on $\rho$ $(I.2.10)$ (Fig. 7) this is found to be true to 
within 10-15 $\%$. Thus, we can write the Pomeron in 
the form
$$\pom (s,t) \approx -i\, s\, |\beta(t)| \, 
e^{\alpha'\, ln\, s \, t} \, (ln^{\gamma} \, s)\eqno(I.3.81)$$
where the power of the logarithmic factor(s) $\gamma$
cannot exceed $2$ and we will agree that their appearance
originates from unitarity corrections (\`a la Mandelstam 
[11]) but about which we will not elaborate any further. 
The r\^ole of the Pomeron, as we have seen, is, therefore, 
that of providing the dominant contribution as $s \rightarrow 
\infty$ for both $\sigma_{tot}$ as well as for 
${{d \sigma} \over {dt}}$ in the forward cone. Owing to the 
positivity of $\alpha'$, however (which, as we have discussed, 
is a consequence of unitarity),  as we move away from the 
forward direction ($t=0$) towards increasingly (negative) 
$t$ values, the Pomeron contribution to the angular 
distribution drops exponentially. Differently stated, the 
Pomeron {\it predicts} a (diffractive) peak as we 
move away from $t=0$. To have (more or less) the right 
order of magnitude to reproduce the data, the Pomeron 
slope turns out to be, roughly
$$\alpha' \approx\, 0.25\, (GeV/c)^{-2}\eqno(I.3.82)$$
(notice, incidentally, that with this slope, there are no 
particles made of quarks which could reasonably claim a 
right to be Pomeron recurrences; yet another way of saying 
that the Pomeron is an atypical trajectory, see Fig. 11).

In conclusion, {\it qualitatively}, the Pomeron has the 
right properties to reproduce the gross features exhibited
by the data. As already mentioned it, however, it is only
when combining together the few leading Reggeons 
discussed in Sections I.3.7.6,7 that the high energy data are
well reproduced\footnote{This will characterize also
the attempts to reproduce diffractive data at HERA. A subleading
contribution will be necessary in addition to the Pomeron
to fit the data (see Section III.3).}. For instance, the minimum displayed
by the total cross sections (Fig. 5) is, strictly, an effect of
the subleading trajectories and so is the maximum in $\rho$.
The two effects are, in fact, related [32].

Before we end this Section, however, a word of caution is 
perhaps necessary. The continuation to complex angular 
momenta has led us to select the trajectory which supposedly 
gives the dominant contribution as $s \rightarrow \infty$. 
This we have called the {\it Pomeron}. The literature, 
however, is by now full of somewhat different things 
that different people call {\it Pomeron}.

First of all, from a phenomenological and semiempirical 
viewpoint, the Pomeron we are talking about is what one uses 
in {\it soft} physics and could be substantially different from 
the one used in {\it hard diffraction} ({\it i.e.} the physics 
studied presently at HERA, see Chapter III) which produces
large rapidity gaps between jets [see 18, 19, 20]. Of course, 
the hope is that in all these instances it is always the same 
object which is at work. In the next Section(s) we will give 
some arguments to support such a universal seeming role 
of the Pomeron. From yet another point of view (which, again, 
goes to prove that many objects are sold on the market under 
the general name of {\it Pomeron}  we simply list: i) the Low - 
Nussinov Pomeron [15] (a prototype of what is known today 
as the Pomeron for hard diffraction and the first instance in 
which an assimilation of two-gluons exchange to the Pomeron 
was proposed; ii) the BFKL (or perturbative) Pomeron [16],
a valiant attempt to sum ladders of gluons to find out their 
behavior even though the result conflicts with unitarity 
exhibiting a power of the form $s^\delta$ (with $\delta 
\approx 0.2 - 0.3$); iii) the empirical Donnachie - Landshoff 
Pomeron, whose growth $s^\epsilon$ with $\approx 0.08$ 
we have already discussed (Fig. 6) and whose conflict with 
unitarity is expected to become observable at energies 
just above the Tevatron. 

More (and in fact more complex) varieties of Pomeron exist
which we shall not even mention here.

Like in all cases when many different forms or interpretations 
of some quantity exist, what all the above goes to prove, I fear, 
is that there is no general consensus on what the Pomeron really 
is or, in other words, no one really {\it knows} what the Pomeron 
is. As we will discuss in the next Section, this is not really so 
surprising since we will end up with the following somewhat 
disturbing suggestion: {\it the Pomeron is probably the effect
of the constraint imposed at high energy by unitarity on any 
hadronic reaction where vacuum quantum numbers can be 
exchanged}.

 \bigskip
 \noindent{\it  I.3.8.10 The Pomeron as a consequence of 
unitarity .}
 \medskip
 
We will establish ({\it alas}, qualitatively only) the 
following logical sequence of statements:

\begin {itemize}
\item{a)} High energy unitarity demands elastic (or 
diffractive) reactions to exhibit a rapid fall as we move 
away from the forward direction (although we have no 
way to establish the analytic form of this fall).
\item{b)} As we have seen in the previous Section, the 
Pomeron we have obtained from the analytic continuation 
to complex angular momenta suggests a {\it very specific} 
form of diffractive peak for the elastic two-body reactions 
with constraints on its parameters coming from unitarity.
\item{c)} To the extent that complete unitarity is the most 
general mechanism capable of predicting a diffraction peak 
(see point a) above), it is very plausible that what we  
call Pomerom may indeed be related to unitarity.
\item{d)} As we will see, the argument is very general and 
predicts that similar situations should arise in a host of other 
(more general) cases. Also for these situations, we predict 
that asymptotic behaviors similar to that of elastic two-body 
reactions should occur. Unfortunately, only one of these other 
more general cases has some chances of being measurable 
and seems indeed to confirm our intuitive conclusions.
\end{itemize}

Unitarity of the Scattering Matrix S $(I.3.30)$ reads
$$S\, S^\dagger\, =\, S^\dagger\, S\, =\, 1.\eqno(I.3.30)$$
We introduce the Transition Matrix T defined by $(I.3.10)$
$$ S\, =\, 1\, +\, i\, T.\eqno(I.3.10)$$
Taking $(I.3.10)$ between an initial ($ |i> $) and a final 
($ |f> $) state and using the completeness of the set $|n>$ 
of the physical states, we get the following set of
{\it symbolic} equations\footnote{Recall that the symbol 
$\si$ implies summation over discrete and integration 
over continuous variables.}
$$Im\, T_{if} \, =\, \si_n  c_n T_{in} T^*_{fn}\eqno(I.3.43)$$
where the index $n$ on the {\it r.h.s.} runs over all physically 
accessible states (in practice, the higher the energy, the larger 
the number of accessible states which becomes infinitely 
large at infinite energies) and where $c_n$ is a positive 
coefficient that includes also an energy momentum 
conservation $\delta^{(4)}(p_i - p_f)$.

Notice that eqs. $(I.3.43, 44)$ provide an interesting 
connection with the ordinary phenomenon of diffraction
since they state that {\it the sum over the (nearly) 
infinite number of inelastic collision} builds up the 
{\it total cross section} (in the case of $(I.3.44)$)) or
that the {\it shadow} of all inelastic channels gives 
rise to diffraction.

Eqs.$(I.3.43)$ represent a system of (infinitely many, in 
practice)  non linear integral equations constraining 
the imaginary part for all $i \rightarrow f$ transitions.
 Of these, the usual optical theorem is just the first and 
the simplest obtained for $|i> \equiv |f> = |2>$. 
In this case, eq. $(I.3.43)$ reduces to $(I.3.44)$ while
everytime $|f> \equiv |i> \neq 2$ we have a generalization
of $(I.3.44)$.

As already discussed (Section (I.3.5)), eq.$(I.3.44)$ contains the 
optical theorem as that special case when the initial state is 
two body. In this case, given that the final state has to be 
identical to the initial one, the final particles must fly off in 
the same direction as the initial ones; differently stated, we 
have the case of forward elastic collisions; the {\it l.h.s.} is the 
imaginary part of the forward elastic amplitude while the 
{\it r.h.s.} is the sum over all integrated cross sections; as we
just recalled, this is exactly the traditional optical theorem. 

As nothing restricts (conceptually) the initial state to be two-body, 
if we stick to $|i> = |f>$, unitarity gives a sum rule of the same 
kind as the optical theorem for {\it all} $ i \rightarrow f$ 
reactions for which the final state coincides with the initial one. 
In practice, there are very infinitesimal chances, 
unfortunately, that we can ever use $(I.3.44)$ for $i>2$
but, if we stick to the two body initial case,
 we can release slightly our assumptions, {\it i.e.} 
we can demand
$$|f> \approx |i>.\eqno(I.3.83)$$
For a two body reaction, condition $(I.3.83)$ means a {\it quasi - 
forward elastic collision i.e.} a situation in which we are {\it almost} 
(but not quite) in the forward direction. In this case, the coherence 
in sign we have among all the terms on the {\it 
r.h.s.} of eq.$(I.3.84)$ is {\it almost} (but not quite) as absolute
as in the case of eq.$(I.3.44)$. Nevertheless, being  the {\it l.h.s.} still
a {\it real quantity}, the sum over {\it an infinitely large} number of
complex quantities on the {\it r.h.s.} must again end up cancelling 
all the imaginary contributions. If we assume, (which is probably 
not exactly the case) that the {\it phases} of the complex numbers 
entering in each term in the sum on the {\it r.h.s.} are randomly 
distributed, the only way this cancellation can occur is, precisely, 
a rapid drop of the angular distribution dictated from unitarity. 
This is exactly the origin of the diffraction peak which we have 
mentioned among the properties of elastic reactions. Notice that
the same mechanism must be at work if we have a {\it diffractive} 
rather than an elastic reaction: so long as no quantum numbers 
are exchanged, the same general results apply and we must 
expect a diffractive reaction to exhibit a forward peak very
similar to the elastic one (which is, indeed the case). It is 
precisely in this sense that the shadow of all inelastic contributions
gives rise to diffraction.

It is most unfortunate that our simple and qualitative 
argument does not lend itself to be cast in an even
slightly more quantitative form.

We can actually imagine other situations in which the final 
state is {\it almost} (but not quite) identical to the initial state 
and random cancellations must occur once again as we move 
away from the forward direction\footnote{This is situation is 
encountered in the case of the so-called {\it leading hadron} in 
inclusive reactions and which, in particular, is presently being 
studied at HERA; we shall return briefly to this case later on. 
For this case, the unitarity constraint takes a special form 
which was first investigated by Mueller [12] (Section I.3.10).}.

In conclusion, we have given intuitive arguments suggesting 
that the diffraction peak in elastic reactions arises from unitarity. 
To the extent that the r\^ole of the Pomeron is precisely to 
reproduce the diffraction peak, we feel authorized to conjecture 
that {\it the roots of the Pomeron are in the unitarity  
constraint}. 

The fact that the unitarity constraint acts not just on elastic 
amplitudes (for which we have only two independent variables) 
but also on amplitudes involving an arbitrary number of particles 
in both the initial and the final state, tells us that what we call 
{\it Pomeron} may well be a {\it very} complicated object or, in 
fact, that it may cover a large variety of analogous situations 
in which the same physical constraint, unitarity, is at work. These 
different situations will, generally speaking, correspond to kinematical 
configurations with many variables which, {\it a priori} may look
very unlike the simple ones discussed here. If our conjecture
is correct, it may be very difficult indeed to come to terms 
with {\it the Pomeron} unless we learn how to handle unitarity.

\bigskip
\noindent{\bf I.3.9 s-Channel models.}
\bigskip

As we mentioned, the technique of extension 
to complex angular momenta provides the most 
successful $t-$channel model for which a number of 
attempts have succeeded at incorporating some 
unitarity. Other classes of models 
that have been used are the so-called {\it eikonal}
(or $s-$channel) [33, 34] approach and the geometrical 
model.

Although these models (especially the eikonal), have
been (and still are) very much used, they have been
especially useful for discussing unitarity preserving
approaches complementary to the one using complex
angular momenta. Given that the present focus is
rather on the implications of the continuation of Regge
poles into the field of modern diffraction, we feel that
we should skip this discussion altogether. A {\it dynamically}
motivated approach is the classic and beautiful paper by 
Glauber [34 a)] while a systematic approach between the
partial wave expansion and the impact parameter
representation is found in [34 b)]. A somewhat more
ample introduction to these problems can be found in
[33]; see also [5, 7].

\bigskip
\noindent{\bf  I.3.10 Inclusive processes.}
\bigskip
 
We have already defined and exemplified in eqs. $(I.2.1)$ 
and $(I.2.2)$ what exclusive and inclusive reactions are. 
In particular, we have already commented on the fact that 
the single inclusive reaction $(I.2.3)$ or 
 $$a(p_1)\, +\, b(p_2)\, \rightarrow a^*({p'}_1)\, +\, X(p_X),
\eqno(I.3.84)$$
is diffractive according to our general 
definition (no exchange of quantum number different from 
those of the vacuum). In this case, we 
have one extra variable as compared with the 
elastic case, where two variables, ($p $ and $\theta$ or $s$ 
and $t$),  describe entirely the process. A third variable 
often used is the {\it missing mass}
$$M_{b^*}^2 \equiv M_X^2 = (p_1+p_2-{p'}_1) \, =
\, E_X^2 - {\vec p}_X^2.\eqno(I.3.85)$$
 
 Other sets of variables routinely used instead of $s, t$ 
and $ M_X^2$ are, for instance, Feynman's $x_F$ defined
as
$$x_F \equiv {{|p^{a^*}_{long}|}\over {p^{a^*}_{long}}} \approx
1 - {{M_X^2} \over {s}}\eqno(I.3.86)$$
where the last equality follows from $p_{long} \approx \sqrt s /2$.
Another commonly used variable is the {\it rapidity}
$$y \, = \, ln {{E^{a^*} +  p^{a^*}_{long}}\over {E^{a^*} -  
p^{a^*}_{long}}}\eqno(I.3.87)$$
or the {\it pseudorapidity}
$$\eta \, = \, - \, ln \, tg \theta_{a^{*}}\eqno(I.3.88)$$
(the latter, especially used in cosmic ray physics in the past, 
has recently been widely used to describe jets and high 
energy accelerator data). Remarkable is that a
 rapidity difference is invariant under a Lorentz
transformation and grows, roughly, as $\Delta y \approx 
\Delta \eta \approx ln{{s}\over {M_X^2}}$.

The complete kinematics of inclusive processes (included the
definitions od cross sections in terms of the scattering 
amplitudes and included sum rules) is outside the scope of 
the present lectures. The interested reader is referred to the 
existing literature (see, for instance, Refs. [33, 34]).

 \bigskip
\noindent{\it I.3.10.1 Inclusive processes and the triple 
Pomeron vertex.}
\medskip

Using once more the property that an {\it outgoing particle} 
can be viewed as an {\it incoming antiparticle}, reaction 
$(I.3.84)$ can be written as
 $$a(p_1) \, + \, b(p_2) \,+ \bar a^*(-p'{_1}) \,  \rightarrow  \, X.
 \eqno(I.3.89)$$
 Graphically, the above property is depicted in the first line of 
Fig. 14. Doing so, we have, formally, arrived at a {\it generalized
form of the optical theorem} [12] and, following the logical 
developments of the second line of Fig. 14 we can say that 
{\it the inclusive reaction $(I.3.89)$ can be 
viewed as the imaginary part}, (or, more correctly, the {\it 
discontinuity} over the variable $M_X^2$) of the elastic {\it three -
body amplitude} $a \, + \, b \, + \, \bar a^* \,  
 \rightarrow  a \, + \, b \, + \, \bar a^* \,$ (where the initial
 and final states {\it must be identical}). 
\medskip
 
Fig. 14 . {\it Generalized optical theorem \`a la Mueller.}
\medskip
 
 If we now select a kinematical configuration  for which
 $$s \, >> \, M_X^2 \, >> \, t,\eqno(I.3.90)$$
the same asymptotic expansion obtained when performing the 
analytic continuation to complex angular momenta can be used. 
Thus, if $R_i$ are the Reggeons with the right 
quantum numbers to be exchanged in the crossed channel, the 
inclusive cross section for $(I.3.85)$ looked \`a la Mueller 
$(I.3.89)$ can be written, symbolically, as the first step of 
Fig. 15. In mathematical terms, this means 
$${{d^2 \sigma} \over {{dM_X^2}{dt}}} \approx \sum_i G_i(t) \, 
{{M_X^2} \over {s^2}} \, (s/M_X^2)^{ 2\, \alpha_i(t)} \, 
\sigma_i(M_X^2,t)\eqno(I.3.91)$$
where $\alpha_i(t)$ and $G_i(t)$ are the trajectory and the 
residue of $R_i$ and $\sigma_i(M_X^2,t)$ is the cross section 
for a Reggeon of mass $t$ and hadron $b$ to yield a hadronic 
state of mass $M_X^2$.
If $(I.3.90)$ hold, we can also assume $\sigma_i(M_X^2,t)$ to 
be dominated by the exchange of a Reggeon of intercept 
$\alpha_k(0)$ 
$$M_X^2 \, {{d^2 \sigma} \over {{dM_X^2}{dt}}} \approx  
\sum_{i,k} \beta_i^k(t) \, (s/M_X^2)^{ 2\, \alpha_i(t) \, - \, 1} \, 
(M_X^2)^{\alpha_k(0) \, - \, 1},\eqno(I.3.92)$$ 
where $\beta_i^k$ is the appropriate residue function.
 
On the other hand, we could develop directly the $| \,|^2$ in the 
second diagram of Fig. 15 which means double exchange of 
Reggeons $R_i$ and $R_j$ or, in the greatest generality,
$$x \, {{d^2 \sigma} \over {{dx}{dp_T^2}}} \approx  \sum_{i,j,k} 
\beta_{i,j}^k(t) \, ({1 \over {1-x}})^{ \alpha_i(t) \, + \alpha_j(t) 
- \, \alpha_k(0)} \, s^{\alpha_k(0) \, - \, 1}\eqno(I.3.93)$$ 
(where, as often done, we have written the inclusive cross section 
in terms of $x \approx 1 - {{M_X^2} \over {s}}$ and $p_T^2$  instead of 
$t \approx \, - p_T^2/x \, - m^2 \, (1-x)^2 /x \approx \, -p_T^2/x$).
\medskip
 
 Fig. 15. {\it Triple Pomeron contribution to diffraction dissociation.}
\medskip
 
If, finally, we make use of the fact that the reaction is 
diffractive, Pomeron exchange dominates asymptotically and gives 
the leading contribution to $(I.3.93)$. If we neglect all the other 
contributions, {\it diffractive dissociation} is dominated by the 
so-called {\it triple Pomeron} (or \pom \pom \pom) vertex which 
obtains when we choose 
$i = j = k =$ \pom. ({\it i.e.} $\alpha_i(0) = \alpha_j(0) = 
\alpha_k(0) =1$). In this case, we have the prediction
$$ {{d \sigma} \over {dM_X^2}} \propto 1/M_x^2\eqno(I.3.94)$$
which is in reasonable agreement with conventional hadronic 
data (see Fig.16). As far as the Tevatron data are concerned,  
basically the same behavior is encountered (Fig. 17)
(see, however, Ref. [35] for details). In Chapter III we 
discuss also the HERA data.

\medskip
Fig. 16. {\it $1/M_X^2$ behavior of the inclusive $\bar p p$ 
conventional data.}
 
\medskip
Fig. 17. {\it $1/M_X^2$ behavior of the inclusive $\bar p p$ 
data at the Tevatron.}

\medskip

Before closing this subject, it should be stressed that the 
asymptotic behavior $(I.3.94)$ of an inclusive diffractive
process follows only if the triple Pomeron vertex 
 $\pom \pom \pom$ dominates and all subasymptotic contributions 
\pom \pom $R$ can be neglected. When this is so, one talks 
of {\it factorizability} since in $(I.3.93)$ the dependence on 
all the variables factorizes. The present day data, however,
do not appear to be already asymptotic and this explains why 
the issue of factorizability has been so controversial in 
recent times.

 \bigskip
\noindent{\it I.3.10.2 Inclusive processes and the leading particle 
effect.}
\medskip

A final point which should be briefly mentioned concerns 
the {\it leading particle effect}. By this term one denotes 
a well known property typical of high energy hadronic 
interactions. When a highly energetic hadron strikes a 
target and we do not look for the complete final state 
({\it i.e.} we look at inclusive processes) a non
negligible portion of the events (say $\approx 10$\% 
of the total) has a rather peculiar configuration
which is especially simple in the Lab system: the same
incident particle flies off essentially unscathed in the (near)
forward direction leaving behind a stream of slow particles
produced. Experimentally [36], the resulting inclusive cross
section has most of the features of elastic distributions.
The situation is clearly understood [37] in terms of diffraction.
The configuration we have depicted is a special case meeting
the requirement of our general criterion: between the initial
colliding particle and the final fast emerging one, there has 
been no changes of quantum numbers (the particle is the
same) therefore the reaction is diffractive and has all the
properties we have seen for these processes. In addition,
what goes on can easily be explained [37] in terms of our
intuitive argumentof Section I.3.8.10: the final state is, indeed,
extremely similar to the initial one from which it differs 
only by the fact that a stream of slow particles is being 
radiated by the fast particle flying off; no change of quantum
numbers and a very minor loss of momentum to produce 
particles. 

It should be stressed, however, that this leading particle effect
(which, as we said, represents $\approx 10$\% of the events)
requires the fast particle to be {\it exactly the same} as the
initial one. For instance, if a proton strikes, a proton must
come out, {\it not a neutron}. If a neutron comes out as the
fast particle, charge is being exchanged. The process, accordingly,
 cannot be attributed to {\it the exchange of zero quantum 
numbers} ({\it i.e} to {\it diffraction}) but, for instance, to pion
exchange. Such an event must be expected to be depressed
further and cannot be more than few \% and, should not be
confused with what we have called {\it leading particle 
effect}. Of course, 
nothing prevents anybody from extending this notion to
cover a larger variety of cases but one should then not marvel
at finding so different responses if the
fast particle detected is a proton or a neutron (as has been
the case recently when analyzing data from HERA). 

\bigskip
\noindent{\bf  I.3.11 Diffraction and large rapidity gaps.}
 \bigskip

We are now equipped to make one further step which will
be seen to be especially relevant in Part III.

If we reconsider the case of a leading particle (say $a^*$)
diffractively produced with the mechanism discussed above, 
precisely because the process is diffractive, we must expect 
the quasiparticle $X$ of reaction $(I.3.84)$ to be produced at
the opposite end of the rapidity spectrum or, as people say, 
with a {\it large rapidity gap} from $a^*$. The reason for this 
is fairly evident. By our definition of diffraction,
$a^*$ has the same quantum numbers as $a$ and no quantum
number can be exchanged until $X$ is produced. If a particle, say
a pion, is produced in between, however, this corresponds 
automatically to exchanging some quantum number and the process 
is no longer diffractive. Thus, ideally, no particle whatsoever 
can be produced between $a^*$ and $X$ and the two are therefore
separated by a large rapidity gap (equal to  $\Delta y \approx 
\Delta \eta \approx ln{{s}\over {M_X^2}}$).

This observation appears to be due to Bjorken [18] and is 
nowadays widely used as a definition of diffraction itself.
As we saw, it is contained into (and, basically, contains) our
definition. In addition, it is quite clear that, exactly like
our definition, it applies to all cases, from single to double
diffraction.

\bigskip
\noindent{\bf I.3.12 Conclusion to Part I.}
 \bigskip

We have spent a considerable portion of these lectures trying
to introduce the unexperienced reader to as large variety as
possible of different subjects related to hadronic diffraction. 
What we could not discuss in detail has been referred to as 
best as we could. Other publications trying to provide an
analogous service have appeared recently [38] which are
to some extent complementary to the present one.

\bigskip
\centerline {\large \bf PART II}
\bigskip

\centerline {\large \bf ELEMENTS OF DEEP INELASTIC SCATTERING (DIS).}

\bigskip

\noindent{\bf Preliminaries to Part II.}
\medskip

Deep Inelastic Scattering (DIS) is a subject that nowadays 
can be found in many textbooks (see Ref. [6] for a 
comprehensive treatment and a large literature). 
For this reason, in this paper, we will confine ourselves 
to a sketchy derivation of the basic formulae of the 
leptoinduced DIS on nucleons (Section II.1) and to a brief 
discussion of Bjorken scaling [13] as the best evidence that 
hadrons are composite systems made of (presumably) 
elementary, spin 1/2 constituents (called, indifferently, 
partons or quarks). Part II, as a consequence, 
is extremely concise, compared to 
Part I. The point, however, is that all we need about DIS
here, is merely introducing the terminology. We shall not
even mention the problems connected with polarized 
structure functions which are, presently, the most
interesting subject in the field.

\medskip

\noindent {\large \bf II.1 Basic kinematics.}
 \bigskip

\noindent {\bf II.1.1 Towards DIS.}
 \bigskip

Let us consider the (fully) inclusive process
$$\ell (k)\, h(p) \rightarrow \ell'(k')\, X\eqno(II.1)$$
where $\ell \ (\ell ')$ is an incoming (outcoming) charged 
lepton (electron or muon), 
$h$ is the hadronic target and $X$ represents the hadronic debris. 
Aside from the incoming lepton energy ($E$, in the hadron's rest 
frame), the other two independent variables needed to  describe 
reaction (II.1) will be chosen among the following 
ones (the notation is explained in Fig. 18)
$$\nu =p\cdot q/m=E-E'\eqno(II.2a)$$
$$Q^2=-q^2=-(k-k')^2>0\eqno(II.2b)$$
$$x=Q^2/2p\cdot q=Q^2/2m\nu\eqno(II.2c)$$
where $q=k-k'$ is the fourmomentum transferred from the 
incoming lepton to the hadronic target. The last equalities in 
eqs.(II.2a) and (II.2c) refer to the hadron's rest frame 
(the lab. system).

\medskip
Fig.18 {\it Diagram and kinematics for reaction $(II.1)$.}
 \medskip

Up to very large $Q^2$ values ({\it i.e.} sufficiently
 below the $Z^0$ threshold), reaction (II.1) is dominated 
by one-photon exchange (the validity of this approximation 
has been checked experimentally). In this case, $Q^2$ is 
the photon virtuality (we will never enter into the 
complications that arise when $Q^2$ becomes so large that 
$Z^0$ exchange can not be neglected anymore, for this, see
[6]).

What is called Deep Inelastic Scattering is the regime in which 
both $\nu$ and $Q^2$ can became arbitrarily large but $x$ 
(eq. (II.2c)) remains finite and is kept fixed; as a consequence
$$\nu \, >> |Q|\, >> 0.\eqno(II.3)$$

 \bigskip
\noindent {\bf II.1.2 A pedagogical exercise}
 \bigskip

Let us investigate the elastic collision of two non identical, 
elementary, spin $1/2$ particles like $e\mu\rightarrow 
e\mu$; such a reaction is not (yet) experimentally
accessible. Neglecting
the masses of the electron $m_e$ and of the muon $m$ 
(compared to the incident energy), standard evaluation of 
the proper Feynman graphs leads to the double differential 
cross section [see eq. $(15.1.18)$ of Ref. 6] 
$${d^2\sigma\over d\Omega\ dE'}={\alpha^2\over 4\nu E^2 
\sin^4{\theta\over 2}}\ \biggl(\cos^2 {\theta\over 2}+{Q^2\over 
2m^2}\sin^2{\theta\over 2}\biggr)\ \delta (1-{Q^2\over 
2m\nu}).\eqno(II.4)$$

Integrating over $E'$ one gets what could be called the Mott 
cross section for the scattering of {\it two} elementary spin $1/2$ 
particles.

Notice that in the exercise: i) the coefficients of $\cos^2{\theta \over 
2}$ and $\sin^2{\theta \over 2}$ in (II.4) do not go to zero as 
$Q^2\rightarrow\infty$; ii) the only dependence on $\nu$
(aside from an overall factor) 
is entirely contained in the Dirac delta function and its
effective dependence appears in the combination defining $x$ 
(eq.(II.2c)).

 \bigskip
\noindent {\bf II.1.3 DIS: Elastic electron proton scattering}
 \bigskip

If we consider the (realistic) case of electron-proton scattering, 
{\it i.e.} the collision of {\it one elementary} spin $1/2$ particle 
(the electron) against {\it one composite} spin $1/2$  hadron 
(say a nucleon), general invariance principles lead to the form
$${d^2\sigma\over d\Omega\ dE'}
={\alpha^2\over 4\nu E^2 \sin^4{\theta\over 2}} \times$$
$$ \biggl(\biggl[ 
F_1^2(Q^2)+{\kappa^2 Q^2\over 4m^2}F_2^2(Q^2)\biggr]\ \cos^2{{\theta} 
\over {2}}+\biggl[F_1(Q^2)+\kappa F_2(Q^2)\biggr]^2{Q^2\over 
2m^2}\sin^2{\theta 
\over 2}\biggr)\ \delta (1-{Q^2\over 2m\nu})\eqno(II.5)$$
which differs from the case of {\it elementary particles}
$(II.4)$ because of the appearance of the Dirac form factors 
of the nucleon $F_{1,2}(Q^2)$, normalized so that $F_1(0)=
F_2(0)=1$; $\kappa$ is the anomalous magnetic moment
of the nucleon. A third form factor appears when parity 
is not conserved like in $\nu$-collisions or when W-exchange 
becomes sizeable. We will ignore all this here, see [6]).

Notice that the only, but major, difference as compared with the 
previous case $e\mu\rightarrow e\mu$ is the appearence of the 
terms in square brackets in eq. (II.5). Differently stated, eq. (II.5) 
reduces to the $e \mu$ elastic case if $F_1\equiv F_2\equiv 1$ 
and $\kappa=0.$

We recall at this point that, experimentally, the nucleon form 
factors fall very quickly with $Q^2$. Fits to the data, (see 
Fig.19) suggest a $Q^{-4}$ behavior (also known as a {\it dipole}
behavior).

\medskip
Fig. 19 {\it Nucleon form factors as function of $Q^2$.}
\medskip

Thus, the main differences between the two cases analysed
here ($(II.4)$ and $(II.5)$) are that the cross section for the 
collision of one elementary and one composite spin $1/2$ particle: 
a) is not function of just the combination $x=Q^2/2m\nu$ (as 
it was the case for collisions of {\it two} elementary spin $1/2$ 
particles) but depends on both $Q^2$ and $\nu$ (or $Q^2$ 
and $x$) and, b) the coefficients of $\cos^2{\theta\over 2}$ and 
$\sin^2{\theta\over 2}$ drop very rapidly with increasing $Q^2$ 
at fixed $x$ (whereas they were $Q^2$-independent in the 
collision of elementary spin $1/2$ particles).

Again, integrating over $E'$, standard results are obtained, 
{\it i.e.} the Rosenbluth formula. 
\bigskip

\noindent {\large \bf  II.2 Basic properties of DIS.}
 \bigskip

\noindent {\bf II.2.1 Deep Inelastic $\ell  N\rightarrow\ell' X$ 
collision; the structure of the nucleons and Bjorken scaling.}
 \bigskip

We consider now the case of interest: inelastic $\ell 
N \rightarrow \ell' X$ collision where $\ell$ is the incoming 
and $\ell'$ the outgoing lepton (electron or muon) of 
fourmomenta $k$ and $k'$, respectively. In the already 
mentioned one photon approximation it is found that eq. 
$(II.5)$ is replaced by
$${d^2\sigma\over d\Omega\ dE'}={\alpha^2\over 
4E^2\sin^4{\theta\over 2}}\biggl[2W_1(\nu,Q^2)\sin^2{\theta\over 
2}+W_2(\nu,Q^2)\cos^2{\theta\over 2}\biggr],\eqno(II.6)$$
where, instead of the two form factors encountered previously
(functions of the single variable $Q^2$), we have now two {\it 
structure functions} $W_1$ and $W_2$ that, {\it a priori} are 
functions of both $Q^2$ and $\nu$ (or $Q^2$ and $x$).

If eq. $(II.6)$ is compared with $(II.5)$, {\it i.e.}
to the case of elastic scattering $ep\rightarrow ep$,
the structure functions reduce to the following 
combinations of the two form factors $F_1(Q^2)$ 
and $F_2(Q^2)$ 
$$W_1^{(elast.)}(\nu,Q^2)={Q^2\over 4m^2\nu}(F_1+\kappa F_2)^2\ 
\delta(1-{Q^2\over 2m\nu})\eqno(II.7a)$$
$$W_2^{(elast.)}(\nu,Q^2)={1\over \nu}(F_1^2+\kappa^2{Q^2\over 
4m^2}F_2^2)\  \delta(1-{Q^2\over 2m\nu}).\eqno(II.7b)$$
On the other hand, comparing with the elastic collisions 
of elementary spin $1/2$ particles, the correspondence 
would be 
$$W_1^{(e)}(\nu,Q^2)={Q^2\over 4\nu m^2}\ 
\delta(1-{Q^2\over m\nu})= {x\over 2m}\ \delta (1-x)
\eqno(II.8a)$$
$$W_2^{(e)}(\nu,Q^2)={1\over \nu}\ \delta (1-{Q^2\over2m\nu})
={1\over \nu}\ \delta (1-x).\eqno(II.8b)$$

{\it A priori}, we know nothing about $W_{1,2}(\nu, Q^2)$, 
but they must be functions of the two variables 
they depend on. Looking to the SLAC data, however, back 
in 1969, Bjorken was led to the very far reaching conjecture
that {\it hadrons are composed of elementary spin $1/2$
components}. To see the rationale for this, we will suppose 
that at large $\nu$ and $Q^2$ the incoming lepton sees the 
target hadron as made of point-like (elementary) spin $1/2$ 
constituents (which we will call indifferently quarks or 
partons) and, furthermore, that in the prescribed conditions 
($\nu$ and $Q^2$ both large), these constituents are seen by 
the incident lepton as essentially free particles. If $ Q_i$ 
is the charge of the $i$-th constituent, the structure functions 
$W_1(\nu,Q^2)$ and $W_2(\nu,Q^2)$ should then be given, in
strict analogy with the free-elementary case, by 
$$W_{1,2}(\nu,Q^2)=\sum_i Q_i^2 W_{1,2}^{(i)},\eqno(II.9)$$
where the functions on the {\it r.h.s.} are the analog of  
$(II.8)$\footnote{These functions will be called the 
{\it parton distribution functions}.}.

If the above hypotesis is correct 
the structure functions of the 
inelastic reaction $\ell \,N\rightarrow \ell'\,  X$ in 
the large $\nu, Q^2$ limit, {\it i.e.} in the deep inelastic 
regime, should {\it i)} be function of just $x=Q^2/2m\nu$ 
(and not of both $Q^2$ and $\nu$ (or $Q^2$ and $x$) 
separately) and {\it ii)} they should not vanish 
in the DIS limit (remember $x$ is kept fixed as $\nu,  
Q^2\rightarrow\infty$). More precisely, we should have 
$$\displaystyle\lim_{Bj}\ mW_1(\nu,Q^2)=
F_1(x)\eqno(II.10a)$$
$$\displaystyle\lim_{Bj}\ \nu W_2(\nu,Q^2)=
F_2(x).\eqno(II.10b)$$
 In addition, from $(II.8)$, we also have
$$2xF_1(x)=F_2(x)\eqno(II.11)$$
which is known as the Callan-Gross relation.

In eqs. $(II.10)$, $\displaystyle\lim_{Bj}$ is shorthand for
the deep inelastic limit
$$\displaystyle\lim_{Bj} \Longrightarrow 
\quad  Q^2\rightarrow \infty, \quad \nu \rightarrow \infty, 
\quad Q^2/2m \nu =x \quad \rm fixed.\eqno(II.12)$$

The reason why the limit $(II.10)$ is named 
$\displaystyle\lim_{Bj}$ is to remind the reader that this 
conjecture was first made by Bjorken [13] who named it the 
{\it scaling hypothesis} (since the dependence of $W_{1,2}$ 
is now only on the scaling variable $Q^2/2m\nu$ and not on 
$Q^2$ and $\nu$ separately). 

That, indeed, in the DIS limit, Bjorken's scaling conjecture 
was well supported by the original data (see ref. [6]), 
is still the best experimental evidence that an 
elementary probe sees the proton (like any other hadron) 
as made of elementary spin $1/2$ constituents which were 
named (in different contexts) partons by Feynman and 
quarks by Gell-Mann.

While it is conceivable that the structure functions 
$W_{1,2}(\nu,Q^2)$ may be very complicate functions of 
$\nu$ and $Q^2$ at low energies, it is not surprising that 
they should {\it become simple}, {\it i.e.} scale in the DIS 
regime. Not only it is to be expected of any function of two 
variables that it should become function of their ratio when
they both go to infinity (unless it either vanishes or becomes
unbounded), but it is to be expected on physical grounds 
if the target is made of elementary subconstituents. Let us 
stress the similarity between Bjorken conjecture and 
the discovery by E. Rutherford and collaborators (60 years earlier!) 
that atoms are made of {\it elementary constituents}, the nucleus 
and the electrons. Just as Rutherford was only able to prove that 
the atomic nucleus must be smaller than $\approx 10^{-13} m$ (it 
took some time before one could prove that nuclear radii are, 
actually, of the order 
$\approx 10^{-15} m$), similarly, to date, we only know that 
partons (like leptons!) are to be considered {\it elementary, i.e. 
pointlike} down to $\approx 10^{-18} m$ and it is anybody's 
guess what is going to happen to smaller distances and/or whether 
partons (and leptons) will keep appearing pointlike at much smaller 
distances or whether they will turn out to be, once more, made of 
even smaller (more {\it elementary}) constituents [39].

Several comments are in order. First, Bjorken scaling can be  
connected to the dilatational invariance of the hadronic tensor 
under the operation: $p_\mu\rightarrow\alpha p_\mu, 
q_\mu\rightarrow\alpha q_\mu$.

Secondly, it can be argued that partons and quarks are similar  
entities (in the more recent literature, the term {\it parton} 
appears to be used in a more general contest, {\it i.e.} to include 
all hadronic constituents such as {\it quarks} and {\it gluons}). 

Thirdly, throughout this short summary, we have tacitly assumed 
that spin plays no role. It turns out that this is not at all 
so\footnote{Contrary to an old prejudice that spin physics 
should become irrelevant at high energies, spin physics has,
on the contrary, always given great surprises and invariably 
led to new interesting information.}. If spin is taken into account, 
two new structure functions appear and the data rise very 
intriguing questions (see [6, 25e)].

Last, and perhaps most important, within the prevailing 
theory of strong interactions, QCD, several complications 
arise which go beyond the simple parton picture, for instance 
gluons must be considered in the game and this leads (among 
other things) to theoretically predicting a violation of 
Bjorken scaling {\it i.e.} a mild $Q^2$ dependence. This 
violation has, by now, been observed and the (large)
consensus is that the $Q^2$ dependence which is found
experimentally conforms to the QCD predictions. 
In this paper we will not open this Pandora box and, once 
again, refer the interested reader to the literature [6].

We end this section by noticing that, so far, we have no 
way to actually {\it derive} parton distributions and 
structure functions; they can, however, be determined 
from the data using appropriate phenomenological forms
and this has indeed been
done extensively in the recent past.  As a consequence, a 
large number of more or less empirical structure functions 
have been suggested in the past 30 years and can be found 
in the literature. These structure functions cover by now all 
needs (from the valence quarks to the sea-quarks to the gluons). 
While most of the recent proposals coincide over large 
kinematical ranges, the danger one should not overlook is 
that, {\it being all these attempts phenomenological ones}, 
extrapolations with different explicit forms may differ 
arbitrarily when continued to kinematical ranges sufficiently
far away from those where they have been determined.

For instance, if one collects the various structure functions 
that have been used in the past 25 years and continues 
them to very low values of Bjorken $x$ (where the new 
interesting physics lies), they may become uncontrollably
different one from the other. An interesting (albeit, by 
now, somewhat old) collection of structure functions 
extrapolated to smaller ${{1} \over {x}}$ values than 
those for which the parametrizations had been used 
[40], makes the point quite beautifully (see Fig. 20).

\medskip

Fig. 20 {\it Collection of extrapolations of structure functions
to values of ${{1} \over {x}}$ smaller than used by the original 
authors to fit the data.}

\medskip

With the above {\it caveat},  however, the fact remains that, 
at least in principle, the data from the purely inclusive 
reaction we have so far considered, {\it i.e.} deep inelastic 
$ep \rightarrow eX$ scattering are sufficient to derive with 
an {\it a priori} arbitrary precision the hadronic structure 
functions.

\bigskip
\noindent{\bf II.2.2 Seminclusive deep inelastic}
 \bigskip

\noindent{\it II.2.2.1 Generalities}
\medskip

Whereas inclusive DIS is in principle sufficient to 
reconstruct completely all structure functions, more 
information comes when we go to the next step, {\it i.e.} 
when we move to {\it seminclusive} DIS whereby one of 
the final hadrons is also measured 
$$\ell (k) \, h(p) \rightarrow \ell'(k')\, h'(p') \, X.\eqno(II.13)$$
In this case, a total of 5 particles (or quasiparticles, one of 
them, $X$ is off its mass shell) take part in the process and, 
therefore, we need altogether 6 independet variables (3 more 
than in the fully inclusive case considered previously) and 
two additional structure functions (again, we limit ourselves  
to the unpolarized case). For instance, we could, in addition 
to the previous variables, use also
$$z \, =\, p \cdot p' / p\cdot q\eqno(II.14a)$$
$$p_T \, = \, |\vec p_T|\eqno(II.14b)$$
$$ cos \phi \, = \, {{(\vec q \times \vec k) \cdot (\vec q \times \vec p)}
\over {  |\vec q \times \vec k| |\vec q \times \vec p|}}\eqno(II.14c)$$
and we could define the 5-fold differential cross section in
$x, y, z, p_T^2$ and $\phi$ for producing the hadron $h$ as
$${{d \sigma^h} \over {dx \, dy \, dz \, dp_T^2 \, d\phi}} = {{8 \pi 
\alpha^2 M E} \over {Q^4}} [x \, y^2 \ H_1^h(x,Q^2,z,p_T)+$$
$$(1-y)\, H_2^h + {{2 p_T}\over {Q}}(2-y)(1-y)^{1/2} \, cos \phi \, 
H_3^h +{{p_T^2}\over {Q^2}} cos 2\phi \, H_4^h]\eqno(II.15)$$
where all $H_i^h$ are functions of $x, Q^2, z$ and $p_T$ 
if we have azimuthal invariance. In this case, we can
integrate over $\phi$ and $H_4$ drops out. Often, however, one 
is satisfied with much less information; for instance, one can
integrate over $p_T^2$) and eq.(II.15) reduces to
$${{d\sigma^h}\over{dx \, dy\, dz}}= {{8 \pi \alpha^2 M E}\over{Q^4}} 
[x \, y^2 \bar H_1^h(x,y,z)+(1-y) \bar H_2^h(x,y,z)]\eqno(II.16)$$
where, by comparison with (II.6), we would have
$$F_i(x,y)= \sum_h \int_0^1 dz \, z\, \bar H_i^h(x,y,z)\eqno(II.17)$$
and $\sum_h$ denotes the sum over all kinds of final 
hadrons detected.

If, to begin with, we focuse on the kinematical region 
of quark fragmentation, gluons can presumably be neglected
 and we can write
$$H_i^h(x,Q^2,z) \, = \, Q_i^2 x [q_i(x,Q^2) \, D_{h/i}(z,Q^2)\, +\,
\bar q_i \, \bar D_{h/i}]\eqno(II.18)$$
where $q_i(x,Q^2)$ are the parton distribution fanctions (in 
terms of wich the structure functions $F_i$ are defined 
within the parton model) and $D_{h/i}(z, Q^2)$ are the 
fragmentation functions for parton $i$ to produce a hadron 
$h$ with a longitudinal momentum $z$ at a given $Q^2$.

For an isoscalar target and reasonably large x ($x \geq 0.2 - 
0.3$) and if heavy flavors are neglected, we find
$${1 \over {\sigma_{tot}}} 
{{d\sigma(\ell  h \rightarrow \ell' h' X)}\over {dz}} 
\approx {1 \over 5} [4 D_{h/u}(z,Q^2)+D_{h/d}(z,Q^2)].\eqno(II.19)$$ 
The message contained in $(II.19)$ is that seminclusive 
measurements are necessary to go one step beyond the
determination of structure functions {\it i.e.} when we 
want to determine also the fragmentation functions.

\bigskip

\noindent{\it II.2.2.2 Seminclusive Diffractive DIS (DDIS).}
\medskip

Much more interesting from the point of view of the present
paper is another configuration of seminclusive DIS, the one
where diffraction is at work.

Consider the special case of reaction $(II.13)$ when the
hadron detected $h'(p')$ is exactly the same as the incident
hadron $h$. To be more precise, consider the seminclusive
process
$$\ell (k) \, N(p) \rightarrow \ell'(k')\, N(p') \, X\eqno(II.20)$$
where $N$ is a nucleon (say a proton). In this case, the diagram
of Fig. 18 is replaced by the one of Fig. 21 and the process is
manifestly diffractive because no quantum numbers are 
exchanged between the virtual photon and the lower vertex.
In this case, we have in fact $J_X^{PC} = 1^{--}$ (like for the 
incoming $\gamma^*$) .

By the same argument, by the way, also the production af {\it any 
vector meson}\footnote{$\rho, \omega, \phi, J/\Psi, \Upsilon$.}
 at the upper vertex is equally a diffractive process, {\it i.e.}
$$\ell (k) \, N(p) \rightarrow \ell'(k')\, N^*(p') \, V\eqno(II.21)$$
because, in this case, we have chosen a configuration in which it 
is at the upper vertex of Fig. 21 that the quantum numbers
$J_X^{PC} = 1^{--}$ are those of the virtual photon.
In this case, $N^*$ is a quasiparticle with the same
quantum numbers of the initial proton and the latter, 
incidentally, is a leading particle according to the definition
given in Section (I.3.10.2)\footnote{One can appreciate
the fact that the production, for instance, of a fast neutron, or
more generally, of a fast neutral particle instead of the fast
proton, would not be the same thing. The latter is {\it not}
diffractive and is therefore considerably reduced since is
comes from pion exchange.}. A special case of either $(II.20)$
or $(II.21)$ is the exclusive reaction
$\ell (k) \, N(p) \rightarrow \ell'(k')\, N(p') \, V$
which is especially interesting for many reasons.

\medskip
Fig. 21 {\it The (diffractive) case of reaction $(II.20)$.} 
\medskip

Referring to the configuration of Fig. 21, in addition to the 
variables introduced earlier ($(II.2)$), it is convenient
to introduce a set of auxiliary variables which are the
following
 $$ W^2 = (p+q)^2 = m^2 + Q^2 (1/x -1)\eqno(II.22a)$$
 $$t = (p-p')^2\eqno(II.22b)$$
 $$x \equiv x_{Bj} ={{Q^2} \over {Q^2 + W^2 - m^2}} \approx {{Q^2} 
\over {Q^2 + W^2}}\eqno(II.22c)$$
 $$x_{\pom} = {{q\cdot (p-p')} \over {q\cdot p}} =
{{Q^2 +M_X^2-t}  \over {Q^2 + W^2 - m^2}} \approx {{Q^2 +M_X^2} 
 \over {Q^2 + W^2}}\eqno(II.22d)$$
 $$ \beta = {{Q^2} \over {2 q\cdot (p-p')}} = {{Q^2} \over {Q^2 + 
 M_X^2 -t}} \approx {{Q^2} \over {Q^2 + M_X^2}}\eqno(II.22e)$$
$$M_X^2 = {{1 - \beta} \over {\beta}} Q^2\eqno(II.22f)$$
 $$ x = \beta x_{\pom}\eqno(II.22g)$$
 $$s= (p+k)^2\eqno(II.22h)$$
 $$x_F = 1- x_{\pom}\eqno(II.22i)$$
 (needless to say, not all the previous relations are mutually 
 independent). 
$x, x_{\pom}, x_F $ and $\beta$ all vary between 0 and 1.
 $x_{\pom}$ is the fraction of momentum carried from the 
upper to the lower vertex, {\it i.e.} in the slang we have 
been using, by the {\it Pomeron}. $\beta$ is the fraction of 
momentum carried by the parton directly coupled to the 
virtual photon. $t $ is the conventional four-momentum 
transfer of hadronic reaction (this will always be taken to 
be very small\footnote{Like in all hadronic processes, the
$t-$distribution will drop very rapidly with $|t|$, see
Section III.3.}). 

We postpone to Part III the discussion of the dynamical 
implications of the previous definition of Diffractive 
Deep Inelastic Scattering (DDIS). Let us just note that the 
region of interest will be that of a small as possible $\beta$
(and $x_{\pom}$) and, therefore, as small as possible Bjorken
$x$; exactly the region complementary to that considered
in Section (II.2.2.1) and, incidentally, exactly the region
particularly suited to the investigation at HERA.
It is precisely because of these properties that HERA has
turned out to be so valuable to study diffraction.

\bigskip

\centerline {\large \bf PART III}

\bigskip

\centerline {\large \bf  MODERN HADRONIC DIFFRACTION.}
\bigskip

\noindent {\bf Preliminaries to Part III.}
\medskip

As we have mentioned repeatedly, recent times have witnessed
a proper resurrection of diffractive physics after the FNAL 
Tevatron on the one hand (hadronic physics), and the HERA 
Collider on the other hand ($ep$ physics), have come in operation.
This last Part will be devoted to a brief discussion of the 
perspectives opened by these new data.

 \bigskip
\noindent {\large \bf  III.1 Diffraction with hadrons.}
\bigskip

 Diffraction at hadronic machines (of which, waiting for 
LHC, the FNAL Tevatron is presently the most prominent 
example) does not differ, basically, from what we have 
analyzed as {\it conventional diffraction} in Part I. If 
anything, going so much higher in energy than any other 
previous hadronic accelerator, the Tevatron makes all 
signals a lot cleaner. Various points are worth mentioning.

\medskip

{\it i) $M_X^2$ behavior.}
\medskip

The first point concerns the triple Pomeron or
large-$M_X^2$ behavior. We have already shown (Fig. 17)
the Tevatron CDF data in Part I (see Ref. 
[35] for analogies and problems with these data).

\medskip 
{\it ii) Large rapidity gaps.} 
\medskip

The cleanest signal for diffractive hadronic physics at very 
high energies, comes from the {\it large rapidity gaps}
anticipated by Bjorken [18] and discussed briefly in Section
I.3.11. Indeed, these gaps in rapidity have been found, [19]
and several topologies have been uncovered [41]. Some of 
them are illustrated in Fig. 22 in the {\it pseudorapidity 
vs azimuthal angle} diagram {\it i.e.} in the $\eta - \Phi$ 
lego-plot. Fig.22 a) shows the diagrams for soft and hard
processes; the actual data (from $D0$) are shown in Fig. 22 b). 

Proceeding from top to bottom of Fig. 22, first we have 
a typical (non-diffractive) event in which particles 
are showered all over the $\eta - \Phi$ plot. The second 
event corresponds to the case of {\it  hard single diffraction 
(HSD)} when the upper vertex is undisturbed ($a^*$, in eq. 
$(1.1,12)$, coincides with $a$, an antiproton, in this case) 
while $b^*$ is a diffractive excitation of $b$ (a proton) which 
dissociates into two jets (both well visible in the plot). The 
third event corresponds to the case in which each hadron 
emits a Pomeron ({\it hard double Pomeron (HDP))}; these 
two Pomerons {\it collide} giving rise to two 
jets half way between the two vertices {\it i.e.} in the middle
of the lego plot. Finally, the bottom diagram is the case of 
 double diffraction; each hadron is excited and deexcites 
emitting one jet and exchanging a Pomeron ({\it HDS}). 

Several other topologies exist but the above examples 
illustrate the point {\it ad abundantiam}: large rapidity gaps 
are well evident in hadronic machines. 

\medskip

{\it iii) Energy dependence, Fraction of events, Color effects etc.}
\medskip

The comparison of the $ 1800$ and $630 GeV$ data 
has been performed by both $D0$ and $CDF$ but 
the analysis in terms of the various models proposed
so far ({\it color-singlet, soft color rearrangement,
$U(1)$ gauge boson, colored Pomeron, flavored
Pomeron} etc.), definitely deserves more attention.

\medskip
 
 Fig. 22. {\it Selected topologies of large rapidity gaps at 
the Tevatron. a) Diagrams, b) data from D0 dijet events 
(from Ref. [41]).}
\medskip

We conclude this extremely brief survey of diffraction
at hadronic machines with two general comments. As
anticipated, diffraction is established at the highest
energies beyond any doubt and we are perhaps
witnessing a transition in this field from soft to
hard diffraction. Its phenomenology, 
however, remains somewhat controversial in its
physical interpretation signalling on the one hand 
the traditional difficulties inherent in hadronic 
physics (which nowadays make QCD such a tough
customer to cope with) and, at the same time, its
great possibilities of opening new paths to our
understanding of strong interaction.

 \bigskip
\noindent {\large \bf III.2 Diffraction at HERA.}
\bigskip

HERA, a Collider where $30 \, GeV$ electrons strike
$800 \, GeV$ protons, was originally conceived as the
machine which would have entirely cleared the field 
of DIS. An accelerator at which the highest lepton -
hadron collisions were to be attained had in itself the 
richest potentiality to better explore the structure of
strong interacting particles. As it turned out, the
latter promise was indeed amply fulfilled and HERA
has become the number one device to investigate
diffraction.

\medskip

\noindent {\bf Back to DIS.}
\medskip

Let us go back to traditional (fully inclusive) DIS 
(Section II.2.1). This, we recall, was originally applied
in the DIS limit $(II.12)$ of large $Q^2$ and large $\nu$
but the kinematics leading to $(II.6)$ was quite general
and one can extend it down to as small $Q^2$ as one
wants\footnote{The limit $Q^2 \rightarrow 0$ corresponds 
to photoproduction.}. If we neglect  $F_1(x, Q^2)$ (it is 
really not necessary but it makes things slightly simpler 
and it also gives a small contribution), eq. $(II.6)$ takes
the form
$$\sigma^{\gamma^* p} = {{ 4 \pi^2 \alpha \, (1+4m^2 x^2/Q^2)}
\over {Q^2 (1-x)}} \, F_2(x, Q^2)\eqno(III.1)$$
where $\sigma^{\gamma^* p}$ is the virtual-photon 
Compton scattering. Eq. $(III.1)$ tells us that the structure
function $F_2(x, Q^2)$ must vanish like $Q^2$ in the
limit $Q^2 \rightarrow 0$.

\medskip

\noindent {\bf Transition from low to large $Q^2$ values.}

\medskip

At low $Q^2$, good agreement (Fig. 23) was found between 
the data and the behavior advocated in Ref. [14] which, 
we recall, appeared to give a good qualitative account of 
hadronic total cross sections (Fig. 6) while violating Froissart
bound $(I.3.53)$. As we move away from the domain of 
very small $Q^2$, however, the agreement worsens (Fig. 24)
and the data exhibit a progressively more marked growth
with $1/x$. This sudden growth was generally unexpected 
even though a number of authors [21] had anticipated it
as Fig. 25 shows. This increase with $1/x$ is attributed to
the growth of the gluon component and, according to many
[22] marks the transition between the so-called {\it soft} 
and the {\it hard} diffraction. Many authors have shown, 
however, that the transition between these two regimes
can be interpolated by very smooth functions [23] as
Fig.26 shows [42]
\medskip

Fig. 23 {\it The photoproduction cross section 
$\sigma^{\gamma^* p}$ fitted \`a la Donnachie and 
Landshoff [14].}

\medskip
Fig. 24 {\it Discrepancy of the curve [14] fitting the data 
at $Q^2=0$ when compared with the $Q^2 = 8.5 GeV^2$ data.}

\medskip
Fig. 25 {\it Prediction of the rate of growth of the glue
distribution (Ref. 21 b) with $1/x$ for various inputs compared
with data.}
 
\medskip
Fig. 26 {\it Curves [42] and data for $\sigma^{\gamma^* p}$ 
as function of $W^2$ (eq. $(II.22a)$) for various bins in $Q^2$.}

\medskip

\noindent {\bf Perturbative (BFKL) Pomeron.}

\medskip

A valiant attempt to actually sum ladders of gluon 
contributions led [16] to predicting a power like 
behavior of the Pomeron in perturbation theory known 
as the BFKL Pomeron of the form $({{1} 
\over {x}})^{\alpha_{\pom}(0)}=({{1} \over {x}})^{1+\Delta}$
where, in the one loop approximation
$$\Delta={{3}\over {4}} {{\alpha_s} \over {\pi}} \, ln 2
\approx 0.2-0.3.\eqno(III.2)$$ 

The algebra necessary to derive $(III.2)$ is far from
trivial but some very good review paper can help [43].

The last paper of Ref. [16] reports about very recent 
progresses in this perturbative approach. The disturbing
feature of these new results, however, is that, as expected
and feared by many, the calculation of $\Delta$ in $(III.2)$
does not appear to converge yet. While these developments
are of the utmost interest, other approaches are also
promising [44] and should be pursued.

 \bigskip

\noindent {\bf III.2.1 Diffractive DIS (DDIS).}
\bigskip

As mentioned already (Section II.2.2.2), the reaction
$$\ell (k) \, N(p) \rightarrow \ell'(k')\, N(p') \, X
\eqno(III.3)$$
is diffractive if the hadron detected in the final state 
coincides with the incident one. Following the outcome 
of the incident proton is, therefore, of the utmost
importance and this was the motivation for the LPS
({\it Leading Proton Spectrometer}) which has been
operating for some years at the ZEUS experiment at
HERA.

To implement the kinematics of Section II.2.2.2, let us
define the cross section relevant for the Diffractive DIS
(for this it will be necessary that we are in the appropriate
kinematical range, see below, eq. $(III.6)$). We denote by
 $$K ={{4 \pi  {\alpha}^2} \over {Q^4 x}} (1 - y +{{y^2} \over 
{2(1+ R)}}) \eqno(III.4)$$
the overall factor that appears in the definition of the 
differential $e \gamma \rightarrow e X$ cross 
section\footnote {$K$ is not, really, a kinematical factor 
since $R$ is defined as $R= 2x{{F_1} \over {F_2}} -1$ where 
$F_1$ and $F_2$ are the structure functions defined earlier 
(eq.$II.10)$). This quantity, however, is known to be very 
slowly varying in DIS and believed to do so also in DDIS.}. 
With this notation, we can define various multi - 
differential cross sections for reaction $(III.3)$ 
beginning with the four-dimensional one 
 $${{d^4\sigma} \over {dx dQ^2 dt dx_{\pom}}} = 
K  F_2^{D(4)}(x, Q^2, t, x_{\pom})\eqno (III.5a)$$
or, with a slight abuse of notation,
$${{d^4\sigma} \over {d\beta dQ^2 dt dx_{\pom}}} = 
K  F_2^{D(4)}(\beta, Q^2, t, x_{\pom}).\eqno (III.5b)$$
 Next, we can, for instance, integrate over $t$ 
(the statistics is rather 
low). This defines $F_2^{D(3)}(x, Q^2, x_{\pom})$. If we 
further integrate over $x_{\pom}$ we have 
$F_2^{D(2)}(x, Q^2)$ which is nothing but the 
diffractive part of $F_2(x. Q^2)$, the conventional 
structure function of DIS . The affix $D(4)$ 
 (or $D(3)$ or $D(2)$) marks the qualitatively different 
situations of diffraction we wish to explore. As
remarked earlier, (Section II.2.2.2), necessary
condition for the process to be diffractive is that 
$1/x$ be as large as possible.

 \bigskip

\noindent {\it III.2.1.1 Mueller theorem in DDIS, 
factorization and the Pomeron structure function.}
\medskip

 The old weapon for this new situation is the same 
generalized optical theorem developed by Mueller 
for inclusive processes and described earlier (see 
Section I.310 and Fig. 14). It is, graphically, quite 
evident that the diagram of eq. $(III.3)$ (Fig. 21) 
depicts a situation which, in a profoundly different 
physical context (we have now two and possibly three 
legs off-shell) is, qualitatively, the same as in Fig.15. 

We can, therefore, repeat, essentially, the same 
procedure used previously and come to very similar 
predictions. 

The so-called {\it Regge limit} in which we can 
presumably apply the asymptotic results found 
previously in the continuation to complex angular 
momenta requires
 $$s >> W^2 >> (M_X^2 , Q^2) >> (|t| , m^2).\eqno(III.6)$$
Pretending we can apply the full machinery of
 Section I.3.10, let us rewrite eq. $(I.3.93)$ in the 
following schematic form  
$$ \sigma = \pom \pom \pom + 
\pom \pom R\eqno(III.7)$$  
where $\sigma$ is any of the diffractive DIS cross
sections $(III.5)$ defined earlier ($D(4), D(3)$ 
or $D(2)$). $\pom \pom \pom $ is the leading {\it 
triple Pomeron} term and $\pom \pom R$ is the first
subleading contribution (with one {\it Reggeon} $R$).

The question one can ask is whether the leading
term $\pom \pom \pom $ is sufficient to account for 
the DDIS data. In principle it should be if we were 
already fully immersed in Asymptopia. Forty years of 
hadronic physics, however, have tought us that 
the present hadronic data are not yet asymptotic 
and a back of the envelope calculation shows that we 
are no more asymptotic at HERA than at hadronic 
accelerators. Therefore, the guess is that at least one 
subleading contribution $\pom \pom R$ will be 
necessary to have agreement with the DDIS data. 
This, indeed, turns out to be the case and we will come 
back to this point shortly. Notice that the question 
we just addressed ourselves is not academic since, 
had it been sufficient to retain only the leading term
$\pom \pom \pom $, the next relevant observation
would have been that, in the regime $(III.6)$ we
would have expected factorization of the $1/x$
term as in the hadronic case {\it i.e.} an asymptoyic
Regge-like behavior.
Consider the case $D(3)$ to be specific, {\it i.e.}
the case when the cross section is proportional to
$F_2^{D(3)}(\beta, Q^2, x_{\pom})$. 
At present, there is a rather heated debate about 
whether or not the triple Pomeron ($\pom \pom \pom $)
can be factorized into a function of $x_{\pom}$ only
$f(x_{\pom})$ (sometimes called the {\it Pomeron 
flux times}) times a remaining $F_2^{\pom}(\beta, Q^2)$
sometimes called the {\it Pomeron structure function}
{\it i.e.} whether or not one can write
$$\pom \pom \pom = f(x_{\pom}) \, F_2^{\pom}(\beta, 
Q^2).\eqno(III.8)$$
The point is that the answer to this question
is purely model dependent and cannot be 
solved in unambiguous terms. Many people, however,
mantain that it is inherently impossible to define 
something like a Pomeron structure function since the
Pomeron is not a particle. We will not
even try to tackle this question and refer the interested
reader to the existing literature [35,38].

\bigskip
\noindent {\bf III.2.2 The basis for diffraction at HERA.}
\bigskip

Real and virtual photons are known to exhibit hadronic
properties as they can fluctuate into $q \bar q$ pairs (the 
first evidence for this was the vector meson dominance 
[VMD] of old). The {\it larger the photon virtuality 
$Q^2$, the smaller the size of the $q \bar q$ pair} and, 
therefore, the {\it greater the power resolution} $1/Q^2$
of the photon itself\footnote{This is, incidentally, the 
very basis that made DIS the right instrument to explore
the hadronic structure.}. On the other hand, the {\it 
smaller Bjorken x}, the {\it larger the hadronic lifetime 
of the hadronic fluctuation of the photon} (which is
proportional to $1/x$) and, accordingly, the better the
virtual photon can explore its target. In the case of
reaction $(III.3)$ (Fig. 21), these conditions, large as
possible $Q^2$ and small as possible $x$ make the
virtual photon as pointlike as possible and give it the
best chance to explore tha structure of the Pomeron 
{\it i.e.} of diffraction. Thus, one expects that as the
distances probed decrease, we should see a transition
between the soft VMD regime (when the photon
fluctuates into vector mesons) and the hard regime
where perturbative QCD should apply. This gives us,
in principle at least, the best chance to see the transition 
between the soft and the hard behavior of the Pomeron.
In practice, if one believes in the existence of these two
Pomerons (soft and hard) [22] this could mean seing the 
transition between $\alpha_{\pom}(0) \approx 1.08$
(as in [14]) in the soft regime to a 
$\alpha_{\pom}(0) \approx 1.3$ in the hard regime
[16] (see eq. $(III.2)$). No evidence of this
transition has been seen so far and it is quite difficult
that it will be seen in a near future (this, obviously, is
just the personal viewpoint of this author).

\bigskip
 \noindent {\bf III.2.3 Evidence for diffraction at HERA.}
\bigskip

Several signatures of diffraction at HERA have been
reported. Here we will analyze a few of them.

\bigskip
 \noindent {\it III.2.3.1 Quasi-elastic (leading proton).}
\medskip

If a leading (quasi-elastic) proton is detected, this, as
discussed previously, is a clean evidence for diffraction.
Fig. 27 shows the (1994) ZEUS data [20] reported as 
function of
$$x_{long} \equiv x_L = 1- x_{\pom}.\eqno(III.9)$$
The acceptance of the LPS is rather limited and so is,
correspondingly, the statistics, but the events are very
clean and the diffractive peak at $x_L \approx 1$ is
uncontroversial.

\medskip

Fig.27 {\it Number of events due to a scattered proton
with $E'_p \approx = E_{beam}$}.

 \bigskip
 \noindent {\it III.2.3.2 Large rapidity gaps at HERA.}
\medskip

The two experiments at HERA, H1 and ZEUS have different
and complementary ways for searching for large rapidity
gaps. They appear as events with no particles in the 
(forward) proton direction {\it i.e.} as large angle events
(in accord with the definition $(I.3.88)$ of pseudorapidity)

\bigskip
\noindent {\it III.2.3.3. $1/M_X^2$ behavior.}
\medskip

According to the triple Pomeron formula, diffractive 
events are expected to behave as $1/M_X^2$ whereas
non diffractive events should exhibit an exponential
fall-off. A set of events at $Q^2 =14 \, GeV^2$ and $W 
\approx 150 \, GeV$ is shown in Fig. 28. The straightline
(in the log scale) accounts for the exponential drop {\it
i.e.} for the non-diffractive component. The events in
excess of the straightline are the diffractive ones. A
complete display of the data shows this diffractive 
component to become more and more important the
larger $W$ compared with $\sqrt Q^2$ (when they
comparable, the diffractive component is almost
absent).

\medskip

Fig.28 {\it  Example of a fit for the determination of the
non-diffractive background in the W interval $134-164
\, GeV$ at $Q^2=14 \, GeV^2$ with its diffractive excess.}.

\bigskip
\noindent {\it III.2.3.4 Triple Pomeron fit.}
\medskip

A fit by H1 with the form $(III.7)$ to the small $Q^2$
data shows that no acceptable agreement can be obtained
without a subleading $\pom \pom R$ term.  This leads to the 
determination of a Pomeron intercept of the soft kind
(as expected in this $Q^2$ interval
$$\alpha_{\pom}(0) = 1.068 \pm 0.016 (stat.) \pm
0.022 (syst.) \pm 0.041 (model)\eqno(III.10)$$
compatible with the determination from fixed target
(Fig. 29).

\medskip

Fig.29 {\it  Fit of the small $Q^2$ data to the form $(III.7)$.}

\bigskip
 \noindent {\it III.2.3.5  Inclusive slope (small $Q^2$).}
\medskip

The small-$|t|$ data (at small $Q^2$) have been fitted 
to the traditional exponential fall-off
$${{d\sigma_{inc}} \over {dt}} \propto e^{bt}.\eqno(III.11)$$
The result
$$b=7.3 \pm 0.9 (stat.) \pm 1.0 (syst.) \, GeV^{-2}\eqno(III.12)$$
is perfectly compatible with the hadronic slopes (Fig. 30)

\medskip

Fig.30 {\it  Fit of the near forward, small $Q^2$ data to 
the exponential form $(III.11)$.}

\bigskip
\noindent {\it III.2.3.6  Inclusive slope (large $Q^2$).}
\medskip
 
Also the small-$|t|$ data at large $Q^2$ have been fitted 
to the  same exponential fall-off $(III.11)$. The result
$$b=7.2 \pm 1.1 (stat.) \pm 0.8 (syst.) GeV^{-2}\eqno(III.13)$$
is, once more, perfectly compatible with the 
hadronic expectation.

\medskip

Fig.31 {\it  Fit of the near forward, large $Q^2$ data to 
the exponential form $(III.11)$.}

 \bigskip
\noindent {\it III.2.3.7 Leading particle effect.}
\medskip

A clean leading particle effect has been also reported
as a signature for diffraction [45]. As a matter of fact,
while the leading proton events amounts to some 
10 \% of the data, asking for a fast outgoing neutron
(rather than a proton), one finds only few \% of the 
events. This, as already mentioned , is perfectly in
accord with the fact that neutron production occurs
via pion (or $\rho$) exchange and is {\it not 
diffractively produced}.

\medskip

\bigskip
\noindent {\bf III.2.4 Vector meson production.  }
 \bigskip

A new important chapter of diffraction at HERA opened
with the search of another diffractive reaction 
whereby one looks not for a leading proton in the 
proton hemisphere but for a vector meson produced 
at the $\gamma^* \pom$ vertex (eq. $(II.21)$)
$$\ell (k) \, N(p) \rightarrow \ell'(k')\, N^*(p') \, 
V.\eqno(III.14)$$
As already remarked, this reaction is diffractive 
since the quantum numbers of the hadronic system $X$ 
are now those of the virtual $\gamma^*$. In this case, to
increase the statistics, instead of a proton we have a
hadronic system with the quantum numbers of the
proton at the lower vertex. What is exciting about
reaction $(III.14)$ is the flexibility one has of
choosing any of the known vector mesons from the
lightest (the $\rho$) to the heavier (the $J/\Psi$ and
the $\Upsilon$) whereby one hopes to see a trnsition
 from a (presumably) soft to a (presumably) hard regime.
Fig. 32 shows the collection of all data on the elastic
vector meson photoproduction cross sections [46]. 
The change in regime when one
gets to the $J/\Psi$ may indeed be indicative of this
transition but it can also be due to some kind of 
heavy flavor threshold and more evidence is certainly
called for before one can draw any conclusion.

\medskip
Fig.32 {\it  Elastic cross sections for vector meson production. }.

\medskip
 
Many more details can be found in the literature [46].

\bigskip 
\noindent {\large \bf III.3 Concluding remarks and 
perspectives.}
 \bigskip

There are no conclusions in the strict sense of the world. 
Decades of somnolent interest in diffraction have gone by.
As usual, it is far simpler trying to hide the difficulties
under the carpet. Presently, however, both the theoretical 
and the experimental communities of high energy physics have 
realized not only the great interest of diffractive physics 
but also the unescapable necessity of understanding it 
better if we wish to come to terms with hadrons. 
The hope is, of course, that all these efforts with meet
with success.

\vfill \eject
 \centerline{REFERENCES}
 \bigskip
 \begin {enumerate}
 \item{} L. D. Landau and I. Y. Pomeranchuk, {\it Zu. Eksper. Teor. Fiz. } 
 {\bf 24} (1953) 505.
 E. Feinberg, I. Y. Pomeranchuk, {\it Nuovo Cimento Suppl.}
 {\bf 3} (1956), 652; A. I. Akhiezer and I. Y. Pomeranchuk, {\it Uspekhi, 
 Fiz. Nauk.} {\bf 65} (1958), 593; A. Sitenko, {\it Uspekhi, Fiz. Nauk.} 
 {\bf 67} (1959), 377; V. N. Gribov, {\it Soviet Jetp} {\bf 29} (1969), 377.
 \item{} Born and Wolf, {\it Principles of Optics}, Pergamon Press,
London (1959), p. 397. 
 \item{} T. T. Wu, CERN preprint TH-95-238, August 1995.
 \item{} M. L. Good and W. D. Walker, {\it Phys. Rev.} {\bf 120} (1960),
\item{} a) P. D. B. Collins and A. D. Martin, {\it Hadron interactions},
Adam Hilger Ltd 1984, Bristol. b) P. D. B. Collins, {\it Regge theory 
and high energy physics}, Cambridge University Press (1977). 
c) H. M. Pilkhun, {\it Relativistic particle physics}, Springer Verlag,
N.Y. (1979). d) The TOTEM Collaboration, CERN/LHCC 97-49 (August 
1997). 
\item{} E. Leader and E. Predazzi, {\it An introduction to gauge 
theories and modern  particle physics: }, Cambridge University 
Press, (1996).
\item{} a) R. J. Eden, {\it High energy collisions of
elementary particles}, Cambridge Press (1967). b) V. Barone and
E. Predazzi, {\it High energy particle diffraction}, in preparation.
 \item{} G. Veneziano, {\it Nuovo Cimento} {\bf 57A} (1968), 190.      
 \item{} T. Regge, {\it Nuovo Cimento} {\bf 14} (1959) 951 and {\bf 18} 
 (1959) 947 .
 \item{} Earlier derivations are discussed by A. Sommerfeld in {\it Partial 
 Differential Equations in Physics}, Academic Press, New York (1949)
where extended reference to previous work is to be found.
 \item{} S. Mandelstam, {\it Nuovo Cimento} {\bf 30} (1963) 1127 
and 1148.
 \item{} A. H. Mueller, {\it Phys. Rev.} {\bf D2} (1970) 2963.
 \item{} J. D. Bjorken, {\it Phys. Rev.} {\bf 179} (1969) 1547.
 \item{} A. Donnachie and P. V. Landshoff, {\it Phys. Lett.} {\bf B 191} 
 (1987), 309; {\it Nucl. Phys.} {\bf B 303 } (1988), 634.
 \item{} F. E. Low, {\it Phys. Rev.} {\bf D 12} (1975) 163; S. Nussinov, 
 {\it Phys.  Rev. Lett.} {\bf 34} (1976), 1286; N. N. Nikolaev and B. G.
Zakharov, {\it Zeit. f. Physik} {\bf C 49} (1991), 607; {\it Zeit. f. Physik} 
 {\bf C 53} (1992), 331.
 \item{} E. A. Kuraev, L. N. Lipatov and V. S. Fadin, {\it Z. Eksp. Teor.
Fiz.} {\bf 72 } (1977), 199; Ya. Ya. Balitsky and L. N. Lipatov, {\bf 28} 
 (1978), 1957 ({\it Sov. J. Nucl. Phys.} {\bf 28} (1978), 822); {\it Sov. 
 Phys. JETP} {\bf 63} (1986), 904. For very recent updating, see, V. S. 
 Fadin and L. N. Lipatov, {\it Proceedings of the LISHEP International
 Conference}, Rio de Janeiro, February 1998 (to be published by
 World Scientific).
 \item{} G. Ingelman and P. Schlein, {\it Phys. Lett.} {\bf B 152} 
 (1985), 256.
 \item{} J. D. Bjorken, {\it Phys. Rev.} {\bf D47} (1993) 101.
 \item{} S. Abachi {\it et al.} (the D0 Collaboration), {\it Phys. Rev.
Lett.} {\bf 72} (1994) 2332; The CDF Collaboration, FERMILAB Pub-
 94/194-E CDF (July 1994), F. Abe {\it et al.}, {\it Phys. Rev. Lett.} 
{\bf 74} (1995), 855. 
 \item{} See P. Marage, {\it Proceedings of the LISHEP International
 Conference}, Rio de Janeiro, February 1998 (to be published by
 World Scientific).
 \item{} a) M. Gluck, E. Reya and A. Vog, {\it Zeit. f. Phys.} 
 {\bf C 5} (1992), 127; b) V. Barone, M. Genovese, N. N. Nikolaev, E. 
 Predazzi and B. G. Zakharov, {\it Zeit. f. Phys.} {\bf C 58} (1993), 541.
 \item{} a) P. V. Landshoff, {\it Proceedings of Hadronic aspects of Collider
physics}, Zuoz, Swit\-zer\-land, August (1994); b) A. Don\-na\-chie and P. V. 
Land\-shoff, {\it Small x: Two Po\-me\-rons!}, 
 DAMTP-1998-34; M/C TH 98-9.
 \item{} a) A Capella, A. Kaidalov, C. Merino and Tran Thanh Van, {\it 
 Phys. Lett.} {\bf B 337} (1994), 358;
 b) M. Bertini, M. Giffon and E. Predazzi, {\it Phys. Lett.} {\bf B 349} 
 (1995), 561.
 c) M. Bertini, M. Giffon. L. L. Jenkovszky, F. Paccanoni and E. Predazzi,
 {\it La Rivista del Nuovo Cimento} {\bf Vol. 19, N. 1} (1996), 1.
 \item{} See, for instance, R. Hagedorn, {\it Relativistic kinematics},
 W. A Benjamin, New York (1963); see also S. Gasiorowicz: {\it 
 Elementary particle physics}, Wiley, New York(1967). 
 \item{} See: a) M. Jacob and G. C. Wick, {\it Ann. of Phys.} {\bf 7}  
 (1959); b) A. R. Edmonds, {\it Angular momentum in Quantum
 Mechanics}, Princeton University Press, Princeton (1957); c) M. E. 
 Rose, {\it Elementary theory of angular momentum}, J. Wiley 
 New York (1957); d) E. P. Wigner, {\it Group Theory and its applications
 to Quantum Mechanics of atomic spectra}, Academic Press, New York
 (1957); e) M. Anselmino, A. Efremov and E. Leader, Phys. Reports {\bf 
261} (1995), 1.
\item{} M. Froissart, {\it Phys. Rev.} {\bf 123} (1961), 1053.
\item{} a) A. Martin, {\it Nuovo Cimento} {\bf 42} (1966) 930.
b) A. Martin, {\it Scattering Theory: unitarity, analyticity and 
crossing}, Springer, Berlin (1969); F. Cheung and A. Martin,
{\it Analyticity properties and bounds of the scattering amplitudes},
Gordon and Breach, New York (1970). 
c) S. M. Roy, {\it Physics Reports} {\bf 5} (1972) 125
d) S. W. Mc Dowell and A. Martin, {\it Phys. Rev.} {\bf B 135}
(1964), 960.
\item{} I. Ya Pomeranchuk, {\it Soviet Physics JETP} {\bf 3}
(1956), 306; L. B. Okun and I. Ya Pomeranchuk, {\it Soviet Physics 
JETP} {\bf 3} (1956), 307.
\item{} A. Bottino, A. M. Longoni and T. Regge, {\it Nuovo Cimento} 
 {\bf 23} (1962), 954.
\item{} R. Dolen, D. Horn and C. Schmidt, {\it Phys. Rev.} {\bf 166}
(1968), 1768.
\item{} M. Froissart, {\it Proceedings of the 1961 La Jolla Conference}    
 (unpublished).
\item{} E. Predazzi and A. T. Samokhin, to be published.
\item{} E. Predazzi, {\it Rivista del Nuovo Cimento} {\bf Vol. 6, n. 2}
(1976),217; S. M. Roy, {\it Physics Reports} {\bf 67} (1980), 201.
\item{} a) R. Glauber, {\it Lectures in Theoretical Physics}, New York 
(1985), 315; b) E. Predazzi, {\it Ann. of Physics} {\bf 36} (1966), 228, 
250.
\item{31} D. Goulianos, {\it Proceedings of the LISHEP International
 Conference}, Rio de Janeiro, February 1998 (to be published by
 World Scientific).
\item{} M. Basile {\it et al.}, {\it Phys. Lett} {\bf 92 B} (1980), 367.
\item{} M. Giffon and E. Predazzi, {\it Lettere al Nuovo Cimento}
{\bf 29} (1980),111; {\it Nuovo Cimento } {\bf 62A} (1981), 238.
\item{} a) E. Levin, {\it Everything about Reggeons}, TAUP 
2465-97, DESY 1997-213; b) A. Levy, {\it Low-x Physics at 
HERA}, DESY 97-013, TAUP 2398-96.
\item{} J. J. Dugne, S. Fredriksson, J. Hanson and E. Predazzi,{\it 
Proceedings of the Ouronopolis Meeting}, Greece, (1997); to be published
by World Scientific.
\item{} A Br\"ull, {\it Die Strukturfunktionen des Nukleons bei
kleinen Werten von $x$}, PhD Thesis, University of Freiburg (1993). 
\item{} See, for instance, A. Brandt, {\it Proceedings of the 
LISHEP International  Conference}, Rio de Janeiro, February 
1998 (to be published by World Scientific).
\item{} H. Abramowicz {\it et al.},{\it Phys. Lett. } {\bf B 269}
(1991), 465; see also Ref. [38 b] for more details.
\item{} V. Del Duca, DESY preprint 95-023 (1995)
\item{} R. Ball and S. Forte, see, for instance, {\it Phys. Lett.}
{\bf B 405} (1997), 317 and references 
therein. 
\item{} M.Erdmann {\it Proceedings of the ICHEP98 
International Conference}, Vancouver, July 1998.
\item{} J. Crittenden, {\it Proceedings of the LISHEP 
International Conference}, Rio de Ja\-nei\-ro, February 
1998 (to be published by World Scientific).

 \end {enumerate}
 \end {document}